\documentclass[aps,prb,twocolumn,superscriptaddress,showpacs]{revtex4-2}
\usepackage{amsmath,amssymb,amsfonts,amsthm}
\usepackage{braket}
\usepackage{graphicx}
\usepackage{bm}
\usepackage{bbm}
\usepackage{color}
\usepackage{booktabs,tabularx}
\usepackage{dcolumn}   % needed for some tables
\usepackage{epstopdf}
\usepackage{bbold}
\usepackage[colorlinks=true,linkcolor=blue,urlcolor=blue,citecolor=blue]{hyperref}
\usepackage{commath}
\usepackage[caption=false]{subfig}
\begin{document}
\newcommand{\mycomment}[1]{}

 \newcommand{\breite}{1.0} %  for twocolumn

\newtheorem{prop}{Proposition}
\newtheorem{cor}{Corollary}

\newcommand{\be}{\begin{equation}}
\newcommand{\ee}{\end{equation}}

\newcommand{\bea}{\begin{eqnarray}}
\newcommand{\eea}{\end{eqnarray}}

\newcommand{\Reals}{\mathbb{R}}     % Reals
\newcommand{\Com}{\mathbb{C}}       % Complex #
\newcommand{\Nat}{\mathbb{N}}       % Natural #

\newcommand{\id}{\mathbbm{1}}    

\newcommand{\Real}{\mathop{\mathrm{Re}}}
\newcommand{\Imag}{\mathop{\mathrm{Im}}}

\def\O{\mbox{$\mathcal{O}$}}   % Order epsilon ... 
\def\F{\mathcal{F}}			% FourierTrafo
\def\sgn{\text{sgn}}

\newcommand{\deo}{\ensuremath{\Delta_0}}
\newcommand{\dea}{\ensuremath{\Delta}}
\newcommand{\ak}{\ensuremath{a_k}}
\newcommand{\ad}{\ensuremath{a^{\dagger}_{-k}}}
\newcommand{\sx}{\ensuremath{\sigma_x}}
\newcommand{\sz}{\ensuremath{\sigma_z}}
\newcommand{\spl}{\ensuremath{\sigma_{+}}}
\newcommand{\smi}{\ensuremath{\sigma_{-}}}
\newcommand{\alk}{\ensuremath{\alpha_{k}}}
\newcommand{\bk}{\ensuremath{\beta_{k}}}
\newcommand{\ok}{\ensuremath{\omega_{k}}}
\newcommand{\vd}{\ensuremath{V^{\dagger}_1}}
\newcommand{\vi}{\ensuremath{V_1}}
\newcommand{\vo}{\ensuremath{V_o}}
\newcommand{\zc}{\ensuremath{\frac{E_z}{E}}}
\newcommand{\xc}{\ensuremath{\frac{\Delta}{E}}}
\newcommand{\xd}{\ensuremath{X^{\dagger}}}
\newcommand{\aok}{\ensuremath{\frac{\alk}{\ok}}}
\newcommand{\tpw}{\ensuremath{e^{i \ok s }}}
\newcommand{\tpe}{\ensuremath{e^{2iE s }}}
\newcommand{\tmw}{\ensuremath{e^{-i \ok s }}}
\newcommand{\tme}{\ensuremath{e^{-2iE s }}}
\newcommand{\epls}{\ensuremath{e^{F(s)}}}
\newcommand{\emis}{\ensuremath{e^{-F(s)}}}
\newcommand{\epl}{\ensuremath{e^{F(0)}}}
\newcommand{\emi}{\ensuremath{e^{F(0)}}}
\newcommand{\vs}[1]{\boldsymbol{#1}}
\newcommand{\mkcomm}[1]{{\color{red}MK: #1}}

\newcommand{\lr}[1]{\left( #1 \right)}
\newcommand{\lrs}[1]{\left( #1 \right)^2}
\newcommand{\lrb}[1]{\left< #1\right>}
\newcommand{\nbt}{\ensuremath{\lr{ \lr{n_k + 1} \tmw + n_k \tpw  }}}
\newcommand{\om}{\ensuremath{\omega}}
\newcommand{\dw}{\ensuremath{\Delta_0}}
\newcommand{\wbp}{\ensuremath{\omega_0}}
\newcommand{\dv}{\ensuremath{\Delta_0}}
\newcommand{\vbp}{\ensuremath{\nu_0}}
\newcommand{\vplus}{\ensuremath{\nu_{+}}}
\newcommand{\vminus}{\ensuremath{\nu_{-}}}
\newcommand{\wplus}{\ensuremath{\omega_{+}}}
\newcommand{\wminus}{\ensuremath{\omega_{-}}}
\newcommand{\uv}[1]{\ensuremath{\mathbf{\hat{#1}}}} % for unit vector
\newcommand{\avg}[1]{\left< #1 \right>} % for average

\let\underdot=\d % rename builtin command \d{} to \underdot{}
\renewcommand{\d}[2]{\frac{d #1}{d #2}} % for derivatives
\newcommand{\dd}[2]{\frac{d^2 #1}{d #2^2}} % for double derivatives

% for partial derivatives
\newcommand{\pdd}[2]{\frac{\partial^2 #1}{\partial #2^2}} 
% for double partial derivatives
\newcommand{\pdc}[3]{\left( \frac{\partial #1}{\partial #2}
 \right)_{#3}} % for thermodynamic partial derivatives
\newcommand{\matrixel}[3]{\left< #1 \vphantom{#2#3} \right|
 #2 \left| #3 \vphantom{#1#2} \right>} % for Dirac matrix elements
\newcommand{\grad}[1]{\gv{\nabla} #1} % for gradient
\let\divsymb=\div % rename builtin command \div to \divsymb
\renewcommand{\div}[1]{\gv{\nabla} \cdot #1} % for divergence
\newcommand{\curl}[1]{\gv{\nabla} \times #1} % for curl
\let\baraccent=\= % rename builtin command \= to \baraccent
\definecolor{mjg}{rgb}{.08,.05,.8}
\newcommand{\mjg}[1]{{\color{mjg} #1}}

%%%%%%%%%%%%%%%%%%%%%%%%%%%%%%%%%%%%%%%%%%%%%
% End Definitions
%%%%%%%%%%%%%%%%%%%%%%%%%%%%%%%%%%%%%%%%%%%%%

%Title of paper
\title{Broken unitary picture of dynamics in quantum many-body scars.}

\author{Pierre-Gabriel Rozon}
\email[]{pierre-gabriel.rozon@mail.mcgill.ca}
\affiliation{Physics Department, McGill University, Montr\'eal, Qu\'ebec H3A 2T8, Canada}
\author{Kartiek Agarwal}
\affiliation{Physics Department, McGill University, Montr\'eal, Qu\'ebec H3A 2T8, Canada}

\date{\today}
\begin{abstract}
Quantum many-body scars (QMBSs) are a novel paradigm for the violation of the eigenstate thermalization hypothesis---Hamiltonians of these systems exhibit mid-spectrum eigenstates that are equidistant in energy and which possess low entanglement and evade thermalization for long times. We present a novel approach for understanding the anomalous dynamical behavior in these systems. Specifically, we postulate that QMBS Hamiltonians $H$ can generically be partitioned into a set of terms $O_a$ which do not commute over the entire Hilbert space, but commute to all orders within the subspace of scar states. All states in the scar subspace thus evolve according to a `broken unitary' $U_s(t) = \prod_a e^{-iO_at}$, where $H = \sum_a O_a$; alternatively, the scar subspace may be viewed as being spawned by common eigenstates of all $O_a$. While the scar Hamiltonian $H$ may be non-integrable, operators $O_a$ usually exhibit a trivial spectrum with equidistant eigenergies which gives rise to revivals at an appropriate period. Two classes of scar models emerge in this picture---those with a finite set of $O_a$, as pertaining to, for instance, scars in the AKLT model, and those with an extensive set of such operators, such as eta pairing scar states in the Hubbard model. Besides discussing how well known scar models fit into the above picture, we use this dynamical decoupling to generalize conditions under which scar subspaces can be engendered in the two settings, generalizing constraints satisfied by quasiparticles in the Matrix Product State (MPS) construction of scar eigenstates, and the Shiraishi-Mori approach. We also discuss connections between this approach and others in the literature.   
\end{abstract}
\maketitle
\section{Introduction}
Quantum many-body scars (QMBSs) represent a new paradigm for the breaking of the eigenstate thermalization hypothesis ETH)~\cite{SERBYN2021_ReviewQuantumScars, MOUDGALYA2022_ReviewOfExactResults, PAPIE2021_ScarsFromEntanglement, CHANDRAN2022_QuasiParticlesPictureReview, ALBA2015_ETHinIntegrableSpinChains, ABANIN2019_MBLcolloquium, SHIBATA2020_OnsagerScars, IADECOLA2020_QMBSGlikeModel,RIGOL2008_ThermalizationMechanisms,SCHERG2021_OscillationsInTiltedBoseHubbard,MARK2020_TrueScarsHubbardModel,OK2020_TopologicalScars,LEE2020_ScarsInFrustratedModel,BANERJEE2021_ScarsInLatticeGauge,CHIRSTOFER2022_RaibowScars,PAI2019_FractonScars,GUSTAFSON2023_PreparingScarsOnQuantumComputers,SUN2023_MajoranaScars}. Prior to the discovery of such models, it had been assumed that for many-body systems, all mid-spectrum eigenstates exhibit volume law entanglement (and thus respect the ETH \cite{DEUTSCH1991_StatMechInClosedSystems, SREDIKI1994_BerrysConjectureThermalEigenstates, RIGOL2007_GeneralizedGibbsEnsemble, ALESSIO2016_FullETHReview, NANDKISHORE2015_MBLHuse,MORI2018_ETHReview}), as in the case of a thermalizing system, or area law entanglement, as in the case of many-body localized system. That certain Hamiltonians, such as the PXP model ~\cite{BULL2020_PXPandBrokenLieAlgebra, PAPIE2022_DrivingQuantumScars, WINDT2022_SqueezingScars, BERNIEN2017_RydbergScarExperiment, BULL2019_ExtendedPXPScars, TURNER2018_PXPScarsAnalysis, WILKINSON2020_ExactPXPsolution, IADECOLA2020_PXPautomataDeformations,LIN2019_ExactPXPScars,LIN2020_2DPXPScars,CHOI2019_SU2PXP,DESAULES2022_RandomWalkScars,KHEMANI2019_PXPIntegrability,MIZUTA2020_FloquetPXP,SUGIURA2021_DrivenScars} first uncovered experimentally, support mid-spectrum low-entanglement states came as a surprise. Although a small number of states exhibit this `scar' property, they leave an indelible signature on the dynamics of the system---if the system is prepared in an initial state with large overlap with scar eigenstates, such as the N\'{e}el product states in case of the PXP model, it exhibits many-body revivals for a long time. On the other hand, all other initial states relax on microscopic timescales, as expected for a thermalizing system. 

Since the discovery of the PXP model, much has been understood about Hamiltonians that support such scar states, and further progress has been made to understand systems that support more aggressive forms of Hilbert space `fragmentation' ~\cite{CHENG2020_FragmentationFromConfinement, MOUDGALYA2022_ReviewOfExactResults, MOUDGALYA2022_CommutantAlgebraFragmentation, SALA2020_DipoleFragmentation, KHEMANI2020_LocalizationInFragmentation}. It is understood for instance, that many scar models support an exact or approximate hidden spectrum generating algebra \cite{MARK2020_RestrictedSGA}, which guarantees the presence of eigenstates with equidistant eigenvalues, generated by repeated application of an appropriate raising operator. These operators can also be interpreted as quasiparticle creation operators; here sub volume-law entanglement states in the presence of many quasiparticle excitations arise because of specific scattering properties of the quasiparticles \cite{CHANDRAN2022_QuasiParticlesPictureReview, MOUDGALYA2020_MPSQuasiparticles} which inhibit the growth of entanglement. On a more general footing, scar Hamiltonians can be constructed from projectors, according to the Shiraishi-Mori approach \cite{SHIRAISHI2017_EmbeddedScars}, or using local operators that comprise the bond algebra of operators that commute with the projectors on scar eigenstates \cite{MOUDGALYA2022_BondAlgebra, MOUDGALYA2022_CommutantAlgebraFragmentation}. 

In contrast, the present authors proposed an alternative interpretation of the presence of such scar states by demonstrating a relation between certain scar models, such as the PXP Hamiltonian, and a related Floquet system. In this approach, as discussed in Ref.~\cite{ROZON2022_FloquetAutomataScars}, the Floquet system can be a remarkably simple automaton in which all computational basis states are mapped to other basis states and which exhibit revivals after a certain, finite time of Floquet evolution. Mathematically, Ref.~\cite{ROZON2022_FloquetAutomataScars} considered a two-layer Floquet Unitary $U_s = U_1 U_2 \equiv e^{-iO_1} e^{-iO_2}$, where $U_s$ simply maps computational basis states to other computational basis states besides an overall phase. Although all states in the Floquet system exhibit revivals trivially, the \emph{Hamiltonian} $H = O_1+ O_2$, constructed from the zeroth order Baker-Campbell-Hausdorff (BCH) expansion, exhibits many-body revivals only for a small number of basis states $\ket{\psi_s}$, which satisfy the relation $[O_1^{n_1}, O_2^{n_2}]\ket{\psi_s} = 0$ for arbitrary powers $n_1,n_2$. For such states, it is easy to see that the time evolution by $H$ is identical to that by $U_s$, and revivals are thus natural; for all other states, the Hamiltonian $H$ typically contains matrix elements connecting these to one another and thus leads to their thermalization. Remarkably, the PXP model fits nicely into the above picture, and a timescale for the decay of the many-body revivals can be estimated by the impact of terms truncated in the BCH expansion. Several other models with varying levels of revival were also found using this approach in Ref.~\cite{ROZON2022_FloquetAutomataScars}. 

Although the above work assumed that a simple automaton with low-entanglement Floquet eigenstates can be used as a starting point for the construction of scar Hamiltonians, it is not obvious that most scar Hamiltonians admit such a connection. Instead, in this work, we focus on an alternative and more general interpretation of the above ideas---the presence of even a single low entanglement state $\ket{\psi_s}$ for which some set of operators $O_1, O_2, ...$ act as commuting operators when acting on it (a more precise definition is given in Sec. \ref{sec:GlobalRules}) directly implies the existence of a  subspace $\mathcal{S}$ composed of states $\ket{\epsilon_l}$ that are common eigenstates of the operators $O_1, O_2, ...\;$. The time evolution of states in this subspace $\mathcal{S}$ resulting from the Hamiltonian $H = \sum_aO_a$ can then be described by a `broken unitary' as $e^{-i\sum_aO_at}\mathcal{S} = \prod_ae^{-iO_at}\mathcal{S}$. Provided the operators $O_a$ represent simple (integrable, in practice, as we show that $O_a$ are often composed of spatially disjoint terms) Hamiltonians with eigenvalues separated by integer multiples of an energy width $\Delta E$, the dynamics within the subspace $\mathcal{S}$ of any state $\ket{\psi_s}$ will show revivals with period $t_* = \frac{2\pi}{\Delta E}$. We thus interpret more broadly QMBSs as possessing Hamiltonians which can be partitioned into simpler (often integrable) parts $O_a, ...$, and the scar subspace as composed of common eigenstates of $O_a, ...\;$.

%It is worth noting that prior knowledge of the scar eigenstates $\ket{\epsilon_l}$ is not a requirement here and that nothing is known about the scar subspace other than the fact that it exists and it contains $\ket{\psi_s}$. We refer to any state within the restricted subspace $\mathcal{S}$ (which is not necessarily a common eigenstate of the operators $O_1, O_2, ...$) as $\ket{\psi_s}$. The common eigenstates $\ket{\epsilon_l}$ are then naturally identified as the scar eigenstates of the model as they satisfy additional unexpected symmetries, whilst the states $\ket{\psi_s}$ can be thought of as initial states of the system that will result in anomalous dynamical properties. 

%For instance, if the common eigenstates of $O_1, O_2, ....$ have eigenvalues that are separated by some integer multiple of an energy difference $\Delta E$, then the time evolution of any state $\ket{\psi_s}$ will satisfy $U_s(t_*)\ket{\psi_s} = e^{-iO_1t_*}e^{-iO_2t_*}...\ket{\psi_s} = \ket{\psi_s}$ after a time $t_* = \frac{2\pi}{\Delta E}$, as was the case in the automaton associated with the PXP model. In practice, we find that this turns out to often be the case due to the fact the operators $O_a$ are in general given by sums of spatially decoupled terms and consequently have a trivial spectrum.
The aim of this work is threefold---i) to examine several prominent examples of quantum many-body scars in the literature, and better understand how they fit into the picture proposed above, ii) to use the `broken unitary' picture to generalize conditions under which scar Hamiltonians are realized, and iii) to make clear the distinctions and similarities between this approach and several other approaches that have been outlined previously---namely, the Shiraishi-Mori approach \cite{SHIRAISHI2017_EmbeddedScars}, the quasiparticle viewpoint \cite{MOUDGALYA2020_MPSQuasiparticles,CHANDRAN2022_QuasiParticlesPictureReview}, the bond algebra approach \cite{MOUDGALYA2022_BondAlgebra,MOUDGALYA2022_CommutantAlgebraFragmentation}, the group invariant formalism \cite{PAKROUSKI2020_GIformalism, PAKROUSKI2021_GIformalismApplication} and the tunnels to tower formalism \cite{DEA2020_TunnelsToTowers}. 

A simple illustration of the above points can be obtained by examining the spin-$1$ AKLT model. As we show, the Hamiltonian  can be partitioned into two integrable parts $H_{\text{AKLT}} = O_1 + O_2$ where $O_1$ consists exclusively of even bonds and $O_2$ of odd bonds. These operators themselves are trivially integrable and possess integer eigenvalues. We show that the scar subspace in this system, first studied in Ref.~\cite{MOUDGALYA2018_ExactExcitedStates}, is composed of common eigenstates of $O_1, O_2$. Note that scar eigenstates of $H_{\text{AKLT}}$ have been interpreted as being constructed by adding a particular quasiparticle repeatedly on top of the ground state~\cite{MARK2020_RestrictedSGA}. Mathematically, a state containing $l$ quasiparticle excitations is represented as $\ket{\psi_l} = (Q^{\dagger})^l\ket{\psi_A}$ for some appropriate base state $\ket{\psi_A}$ and quasiparticle creation operator $Q^{\dagger}$. Certain rules \cite{MOUDGALYA2020_MPSQuasiparticles}, described in the Matrix Product State (MPS) picture, have been shown to be satisfied by the MPS tensor representing the quasiparticle excitation $\ket{\psi_l}$ in the spin-1 AKLT model. Satisfaction of these rules was shown to be a sufficient set of conditions to ensure that the states $\ket{\psi_l}$ are eigenstates of the Hamiltonian. Using the broken unitary picture as a guiding principle, thus assuming that scar eigenstates are common eigenstates of partitions $O_a$ of a generic scar Hamiltonian $H$, we derive a more general set of local conditions (local in the sense that the rules only need to be verified on a subset of sites of the full wavefunction), that quasiparticle MPS tensors must satisfy, and which encompass those observed in the spin-$1$ AKLT model, besides a domain wall preserving scar Hamiltonian. This set of local MPS rules is referred to as Type I rules in this work. Importantly, the local conditions we derive necessitate that the quasiparticle creation operator is represented in the MPS language by a sum of single-site operations on the MPS wavefunction. However, the physical operator creating this excitation may be composed of operators acting on multiple sites; for instance, the physical operator that creates quasiparticle excitations is a sum of single-site operators for the AKLT model but a sum of three-site operators in the domain wall conserving model studied. Nevertheless, both of them admit quasiparticle excitations that can be constructed in the MPS language from an operator $Q^{\dagger}$ composed of single-site operations, and so Type I rules apply to both. We discuss these generalized conditions in Sec.~\ref{sec:localRulesFewExtensive}.   

Similarly, we discuss eta-pairing scar eigenstates in the generalized Hubbard model, as discussed in Ref.~\cite{MOUDGALYA2020_GeneralizedHubbard}. In this, and related models, scar eigenstates can be viewed as common eigenstates of an extensive set of local operators, along with some global operators. We consider this class of models more generally and describe the simplest set of local conditions that allow for states in the scar subspace $\mathcal{S}$ to exhibit  `broken unitary' time evolution. This set of conditions is referred to as type II rules in this work. We find that the principles behind type II rules allow us to generalize the Shiraishi-Mori (SM) conditions for obtaining scar Hamiltonians, as discussed in Sec.~\ref{sec:LocalRulesExtensive}. We additionally study the spin-$1$ XY magnets which are also known to exhibit scar subspaces, and as we show can be understood via type II rules.
%To the first end, we derive simple sets of strictly local rules that, if satisfied, ensure that the broken unitary picture holds and we proceed to show that the rules apply to the spin-1 AKLT model, a domain wall conserving model, the Hubbard model, and the spin-$1$ XY magnet. 
Notably, in certain cases, such as the spin-1 XY magnet, there exists an alternate scar tower comprising of two-site quasiparticles---for this tower, it is not obvious how to partition the Hamiltonian into eigenoperators of the scar states. However, mapping to a spin-$1/2$ model allows one to define appropriate partitions, as we show. It is also worth mentioning here, that although the type I rules, and models which satisfy them, only demand that a finite set of global operators commute in the scar subspace, an extensive number of local operators commute on the extremal states (containing, for instance, no quasiparticles) of the scar subspace. This in constrast to models which follow type II rules in which an extensive number of local operators commute on the entire scar subspace.  %We then turn to a discussion of other existing approaches in the literature and we elaborate on the similarities and distinctions with the broken unitary picture presented in this work. 
%In particular, we discuss how this approach generalizes the Shiraishii Mori approach, and also, how it provides a simpler recipe to the more general bond algebra approach. 

This manuscript is organized as follows. In Sec.~\ref{sec:SummaryOfIdeas}, we provide an overview of all the models studied, the partitions for which the scar states are common eigenstates, and connections between the broken unitary picture and various approaches to obtaining scar Hamiltonians in the literature. In Sec. \ref{sec:GlobalRules}, we discuss the most general requirements to ensure the existence of a scar subspace $\mathcal{S}$ that evolves in time according to a `broken unitary' operator $U_s(t)$, which is followed by Sec. \ref{sec:localRules} in which we discuss specific cases for which simpler sets of strictly local requirements for the broken unitary picture to hold can be found. In Sec.~\ref{sec:ExplicitModels}, we describe in detail the various models studied and show that the scar eigenstates are common eigenstates of appropriate partitions. 
%In Sec.~\ref{sec:nonpartitionexamples}, we discuss scar states that do not fall into the common eigenstate construction, namely the Arovas A and B states in the AKLT model, and the two-site quasiparticle scar tower in the spin-1 XY magnet (which requires a mapping to a spin-1/2 model to fall into this construction). 
In Sec.~\ref{sec:Methods}, we discuss the distinctions and similarities between the approach proposed in this work and formalisms introduced in the literature such as the Shiraishi Mori approach, the bond algebra picture, the group invariant formalism as well as the tunnel to tower formalism. We end in section \ref{sec:Discussion} with a discussion on possible extensions of our work.  

\section{Summary of main ideas and results}\label{sec:SummaryOfIdeas}

The main ideas and findings in this work can be summarized in the following bullet points---

\begin{itemize}

\item Our main observation is that scar Hamiltonians can usually be partitioned into terms $H = \sum_aO_a$ and commutators of arbitrary powers of products of these operators vanish on the low entanglement mid-spectrum subspace $\mathcal{S}$ of scar states. This leads to time evolution on any state $\ket{\psi_s}$ in the scar subspace by a `broken unitary': $e^{-i(\sum_aO_a)t}\ket{\psi_s} = \prod_ae^{-iO_at}\ket{\psi_s}$ for any time $t$. Alternatively put, these operators commute over a subspace even though they do not commute over the entire Hilbert space; the scar subspace is then spanned by common eigenstates of all the $O_a$. 

Each term $O_a$ is usually composed of few-body operators  that pairwise commute, and are often even spatially decoupled. Thus, the spectrum of these operators (which individually can be integrable) is usually trivial, even though the spectrum of the full Hamiltonian (which is not integrable) is not. Provided the common eigenstates are separated in energy by some multiple of a constant energy shift $\Delta E$, any linear combination of the scar eigenstates exhibits perfect many body revivals with period $t_* = \frac{2\pi}{\Delta E}$. Since scar eigenstates form a small subspace of common eigenstates of the operators $O_a$ they often exhibit sub-volume law entanglement entropy and equidistant eigenvalues. 

\item In Sec.~\ref{sec:localRules}, we derive two sets of simple local rules to achieve the fracturing of the unitary as discussed above. The first set of rules (type I rules) is a direct generalization of a formalism designed to build quantum scars in the language of MPSs and is appropriate for understanding scar states that are common eigenstates of a few global operators that are extensive sums of local operators. The core idea is that given a Hamiltonian $H = O_1 + O_2 + ...$, where each $O_a$ are assumed to be spatially decoupled commuting few-body operators, if one finds states that are common eigenstate of all the operators $O_a$, the time evolution of any linear combination of such states (in general, the superposition may not be an eigenstate of the operator $O_a$) will be characterized by $\prod_ae^{-iO_at}$ instead of the more complicated unitary $e^{-iHt}$. We derive a set of local conditions that, if satisfied for any state $\ket{\psi_s}$, imply the above decomposition of the unitary operator $e^{-iHt}$ over the subspace spanned by the states $\ket{\psi_s}$. The rules are formulated in a way that directly applies to excited states on top of a base MPS state $\ket{\psi_A} = \ket{AAA...}$.

The second set of rules (type II rules) is closely related to a generalized Shiraishi-Mori formalism and relies on the fact that many QMBS models are built from Hamiltonians of the form $H = \sum_j a_jh_j + v_j$ with arbitrary coefficients $a_j$ for which there exists a subspace $\mathcal{S}$ of states $\ket{\psi_s}$ that vanish under the action of each individual $h_j$ and such that $\sum_j v_j$ preserves $\mathcal{S}$. The resulting time evolution of the states $\ket{\psi_s}$ is then described by $e^{-i\sum_j v_j t}\ket{\psi_s}$. In the broken unitary picture, if we are to obtain many-body revivals for arbitrary amplitudes $a_j$, we must find a time $t_*$ such that all operators $e^{-ih_j t_*} = \mathbb{1}$ on the scar subspace. This is easiest to achieve if $h_j$s annihilate the scar subspace. Additionally, we identify a set of requirements necessary to achieve the simplified time evolution of the states $\ket{\psi_s}$. These two requirement when taken together form the type II rules. We show in Sec. \ref{sec:Methods} that they can be used to generalize the original Shiraishi-Mori construction. 
%Note that in the broken unitary picture, if we are to obtain many-body revivals for artbirary amplitudes of $h_j$, we must find a time $t_*$ such that all operators $e^{-ih_j t_*} = \mathbb{1}$ when operating on scar states. This is easiest to achieve if the $h_j$

Finally, we discuss a set of local rules (rules of type III) that were introduced in a previous work \cite{ROZON2022_FloquetAutomataScars} of the present authors. Such rules are most appropriate whenever the Hamiltonian $H = \sum_jh_j$ hosts a subspace $\mathcal{S}$ composed of common eigenstates of the $h_j$ operators even though the $h_j$ do not pairwise commute. Importantly, we note that most QMBS models host such a subspace, but it is often the case that more scar states can be found that instead satisfy either type I or type II rules but not the type III rules. For instance, this is the case in the AKLT model for which only $2$ out of the $L/2$ scar eigenstate (not accounting for the $SU(2)$ degeneracy) satisfy the rules of type III, where $L$ is the system size. As discussed in \cite{ROZON2022_FloquetAutomataScars}, such rules can be leveraged to construct exact as well as approximate quantum scars. Furthermore, when certain conditions are met, they can be interpreted in the context of quantum cellular automata, see Ref. \cite{ROZON2022_FloquetAutomataScars} for more details.

As a testament of the usefulness of these rules, we show in Sec.~\ref{sec:ExplicitModels} how they can be used to uncover with ease scar states in the Spin-1 AKLT model, the generalized Hubbard model, Spin-1 XY magnets and a domain-wall conserving model. These rules also generalize conditions presented in previous works for constructing scar states in the MPS quasiparticle picture, and the Shiraishi-Mori picture. We also show that many exact methods for constructing QMBSs can be recast in terms of the broken unitary picture discussed above. 
\subsubsection{Explicit models}
\item For the spin-1 AKLT model, $H_{\text{AKLT}} = \sum_j P_j$, where $P_j$ is a projector onto the spin-2 subspace of nearest neighbor spins $j, j+1$. The scar states are common eigenstates of operators $O_1 (O_2) = \sum_{j = \text{odd (even)}} P_j$. The operators $O_1,O_2$ are a sum of decoupled projectors and thus trivially exhibit an integer eigenvalue spectrum; thus in the broken unitary picture, where evolution by $H$ can be broken up into that by $O_1$ and $O_2$ individually, revivals occur for multiples of $t_* = 2\pi$. We show that all scar eigenstates are indeed common eigenstates of $O_1$ and $O_2$. Importantly, one can construct scar towers using an MPS-based quasiparticle picture for this system. Certain conditions are known to be satisfied by the tensors of the MPS which allow for multiple quasiparticle states to still be exact eigenstates of the AKLT Hamiltonian. As we show, the broken unitary picture allows us to generalize the conditions on these tensors by demanding that these are common eigenstates of the partitions $O_1,O_2$. 

\item The domain-wall conserving model is given by the Hamiltonian $H_{\text{DWC}} =\sum_{j=1}^{L}\left[\lambda\left(\sigma^x_j-\sigma_{j-1}^z \sigma^x_j \sigma_{j+1}^z\right)+\Delta \sigma^z_j+J \sigma^z_j \sigma_{j+1}^z\right] 
$ where $\sigma^{x/y/z}_j$ are the usual Pauli matrices on site $j$ and $\sigma^z_j\ket{0}_j = -\ket{0}_j$ $\sigma^z\ket{1}_j = \ket{1}_j$. This model preserves the total number of domain wall configurations, where a domain wall corresponds to a transition between a $\ket{0}$ and a $\ket{1}$ on any pair of adjacent sites, e.g $\ket{111000111}$ contains two domain walls. It was shown in Ref. \cite{IADECOLA2020_QMBSGlikeModel} that a tower of scar eigenstates can be constructed by applying a blockade-free spin-raising operator  to the base state $\ket{00...0}$. The resulting excited states contain no blockade configurations $\ket{11}_{j,j+1}$ for any pair of adjacent sites. By representing the tower of scar states in the MPS language and by finding an appropriate partition of the Hamiltonian, we find that the scar states satisfy the rules of type I. Importantly, the physical operator that creates the quasiparticle excitations in this model is composed of a sum of overlapping operators that have support on three adjacent site, but it turns out that the MPS quasiparticle creation operator $Q^{\dagger}$  can be written out in terms of single-site operations which allows us to apply the type I rules. We then show that this results in the fact that the scar states vanish under the action of the Hamiltonian term proportional to $\lambda$ and that the remaining magnetic terms fully characterize the time evolution of the scar subspace.

\item The generalized Hubbard model consists of local hopping terms, local chemical potential and local Hubbard-U interaction terms which are all weighted by arbitrary coefficients. In subsection \ref{sec:GeneralizedHubbard}, we show that the broken unitary picture can be used to derive the eta-pairing operator which generates the scar eigenstates in this model. To do so, we make use of the type II rules derived in this work. In the Hubbard model, the local hopping terms annihilate the scar subspace (they act as the $h_j$ in the $H = \sum_jh_j + v_j$ discussed above)  while the local chemical potential and local Hubbard-U form the $v_j$. This directly leads to a set of restrictions which, if satisfied, ensures the existence of quantum scars; the conditions derived of course agree with those of Ref. \cite{MOUDGALYA2020_GeneralizedHubbard}. The dynamics within the scar subspace is determined solely by the interaction and chemical potential terms, which allows for revivals. 

\item For the spin-1 XY magnet, two separate scar towers exist which comprise of single-site and two-site quasiparticle excitations, respectively. We show that the scar states in this model follow directly from the set of rules derived in this work. For the single-site case, as has been shown, the exchange terms vanish onto the scar-subspace whilst the magnetic field and Ising aniostropy preserves this subspace. %This directly shows that the dynamics of the scar subspace is determined by the spatially decoupled magnetic terms resulting in trivial time evolution. 
For the two-site case, the connection to the rules is only apparent after a mapping of the Spin-1 model on to an effective spin-1/2 chain. After this mapping, there again exists a decomposition of $H_{\text{XY}}$ for which the rules we identify apply, thus resulting in a simplified time evolution within the scar subspac
\subsubsection{Connection to other constructions}

\item We next turn to the relation that exists between the broken unitary picture and other methods for building exact QMBS models. We start with the Shiraishi-Mori approach which describes a Hamiltonian $H_{\text{SM}} = \sum_j P_j K_j P_j + H'$ for which the $P_j$ annihilate a set of scar states $\ket{\psi_s}$, the $K_j$ are arbitrary Hamiltonian operators and $H'$ is such that $[H',P_j] = 0$ for all $j$. With these conditions in place, it is straightforward to show that the time evolution of the states $\ket{\psi_s}$ is determined by the unitary operator $e^{-iH't}$. 
%We show that this decoupling can be understood via the broken unitary picture and follows from . We further discuss the main differences between the Shiraishi-Mori condition and our rules. It is found that 
The local rules derived in this work are less restrictive than the original SM constructions, but result in the same type of dynamical decoupling. Technically, the rules we identify relax the condition $[H',P_j] = 0$ to commutators of artbirary powers of $H', P_j$ vanishing only on the scar subspace. Note that this is already true of many of the models that were built with the SM construction as a guiding principle. 

\item The bond algebra approach~\cite{MOUDGALYA2022_BondAlgebra} is complete in the sense that it seeks a basis for the algebra of all possible operators $O_a$ that commute with projectors onto the scar eigenstates. The scar Hamiltonian is then given by $H_{\text{BA}} = \sum_a O_a$ . While completely general, it is worth noting that in the approach we put forth, it is not imperative to know the scar eigenstates apriori. Instead, we simply demand that for some states $\ket{\psi_s}$, which may not be an eigenstate of the scar Hamiltonian at all, one achieves the unitary decoupling $e^{-i(O_1 + O_2 + ...)t}\ket{\psi_s} = e^{-iO_1t}e^{-iO_2t}...\ket{\psi_s}$. The scar eigenstates are indeed superpositions of the states $\ket{\psi_s}$ over which this unitary decoupling occurs, which are further common eigenstates of the operators $O_a$. The existence of a single state $\ket{\psi_s}$ satisfying the decoupling conditions guarantees that operators $O_a$ have such common eigenstates which we interpret as scar eigenstates of the scar Hamiltonian $H_{\text{BA}} = \sum_a O_a$. Thus, we may choose $\ket{\psi_s}$ to be simple low-entanglement state, possibly a computational basis state, to start the construction of putative scar Hamiltonians. 
\item Another approach for constructing exact QMBS is the so called tunnels to tower formalism \cite{DEA2020_TunnelsToTowers}. The idea is to start with a Hamiltonian $H_{\text{sym}}$ that has a non-abelian symmetry $G$, which results in multiplets of degenerate eigenstates. One then proceeds to add an operator $H_{\text{SG}}$ that breaks the degeneracy between the multiplets (a standard example would be $H_{\text{sym}} = \sum_j\vec{S_j}\cdot\vec{S}_{j+1}$, $H_{\text{SG}} 
 = \sum_jS_j^z$, where the non-abelian symmetry $G$ in this case is SU(2)). Up to this point, there are no quantum scars since the presence of fully decoupled subspaces of the Hamiltonian can be explained by the presence of global symmetries. To generate quantum scars, one then further adds to this model a Hamiltonian term $H_{\text{A}}$ that vanishes when acting on a multiplet (or a set of multiplets). We denote the states that vanish under $H_{\text{A}}$ by $\ket{\psi_s}$ and they form the scar subspace $\mathcal{S}$ of the model. This procedure generates quantum scars in general since the Hamiltonian $H_A$ is chosen such that it mixes all the multiplets that are not included in $\mathcal{S}$. This means that $H_{\text{sym}}$, $H_{\text{SG}}$ and $H_{\text{A}}$ do not mutually commute in general, and the existence of the scar subspace $\mathcal{S}$ cannot be explained via global symmetries of the Hamiltonian. It is clear in this case that the subspace $\mathcal{S}$ is a common invariant subspace of $H_{\text{sym}}$, $H_{\text{SG}}$ and $H_{\text{A}}$. Furthermore, these operators commute when restricted to $\mathcal{S}$. As a consequence, one obtains the following broken unitary dynamics $e^{-i(H_{\text{sym}} + H_{\text{SG}} + H_{\text{A}})t}\mathcal{S} =  e^{-iH_{\text{sym}}t}e^{-iH_{\text{SG}}t}\mathcal{S}$ in which $H_A$ does not appear since it is identically $0$ on $\mathcal{S}$.

\item Finally, the group invariant formalism~\cite{PAKROUSKI2020_GIformalism} (GI), which uses generators of a Lie group $T_a$ to build scar Hamiltonians of the form $H_{\text{GI}} = \sum_a K_aT_a + H_0$, with the condition $[H_0,C_{G^2}] = WC_{G^2}$, where $C_{G^2}$ is the Casimir operator of the Lie group, and $K_a, W$ are arbitrary operators, also assumes an interpretation in terms of the broken unitary picture. The unitary operator describing the dynamical evolution within the scar subspace may be shown to be $e^{-iH_0t}$ and the scar subspace is given by $\mathcal{S} = \{\ket{\psi_s} \text{ : } C_{G^2}\ket{\psi_s} = 0\}$. This is due to the fact that the imposed conditions guarantee that $\sum_aK_aT_a\ket{\psi_s} = 0$ and that $H_0\mathcal{S}\subset \mathcal{S}$, see Ref. 
 \cite{MOUDGALYA2022_BondAlgebra} and \cite{PAKROUSKI2020_GIformalism} for more details, which as discussed above directly implies that the time evolution within the scar subspace is characterized by the unitary operator $e^{-iH_0t}$.
\end{itemize}

\section{Global rules}\label{sec:GlobalRules}
We first consider the most general situation in which the time evolution unitary due to the Hamiltonian can be broken down into independent pieces over some subspace of the full Hilbert space, as our broken unitary picture demands. Consider a Hamiltonian $H$ that admits the form 
\begin{equation}
H = \sum_a^kO_a
\end{equation}
for some oeperators $O_a$. Then, the following theorem applies, which constitutes a generalization of a result originally derived in Ref.~\cite{SHEMESH1984_CommonEigenstates}\\\\
\textbf{Theorem:} The operators $O_1$, $O_2$, ..., $O_k$ admit a non-vanishing subspace of common eigenvectors if and only if there exist at least one state $\ket{\psi_s}$ such that 
\begin{equation}\label{eq:PolynomialConditions}
[p(O_1,O_2,...,O_k),q(O_1,O_2,...,O_k)]\ket{\psi_s} = 0
\end{equation}
for any $p,q \in \mathrm{P}$ where $\mathrm{P}$ is the set of all possible complex monomials in $k$ noncommutative variables. 
\\\\
\textbf{Proof:} First, assume there exists a state $\ket{\epsilon}$ that is a common eigenstate of $O_1, O_2, ..., O_k$ with respective eigenvalues $\epsilon_1, \epsilon_2, ..., \epsilon_n$, then clearly $f(O_1,O_2,...,O_k)\ket{\epsilon} = f(\epsilon_1,\epsilon_2,...,\epsilon_n)\ket{\epsilon}$ for $f = p, q$. Thus, $[p(O_1,O_2,...,O_k),q(O_1,O_2,...,O_k)]\ket{\epsilon} = [p(\epsilon_1,\epsilon_2,...,\epsilon_k),q(\epsilon_1,\epsilon_2,...,\epsilon_k)]\ket{\epsilon}$ = 0. 
\\\\
For the reverse implication, consider the vector-space $\mathcal{N}_{\ket{\psi_s}} = \{ m(O_1,O_2,...,O_k)\ket{\psi_s} \; \text{such that} \quad m \in \mathrm{P}\}$ for which the state $\ket{\psi_s}$ satisfies Eq.~(\ref{eq:PolynomialConditions}). This vector space is invariant under the action of the operators $O_a$ in the sense that $O_a\mathcal{N}_{\ket{\psi_s}} \subset \mathcal{N}_{\ket{\psi_s}}$. Now consider any state  $\ket{y} \in \mathcal{N}_{\ket{\psi_s}}$. We have that $O_aO_b\ket{y} = O_bO_a\ket{y} = \sum_{n_1,n_2,...,n_k}c_{n_1,n_2,...,n_k}O_1^{n_1}O_2^{n_2}...O_k^{n_k}\ket{\psi}$ for some complex numbers $c_{n_1,n_2,...,n_k}$ and integers $n_1,n_2,...,n_k$ which follows directly from the fact that $\ket{\psi_s}$ satisfies Eq.~(\ref{eq:PolynomialConditions}). This implies that $[O_a,O_b]\ket{y} = 0$ for any state $\ket{y}$ in $\mathcal{N}_{\ket{\psi_s}}$. In other words, the operators $O_a,O_b$ pairwise commute on a common invariant subspace, which implies that the set of operators $O_a$ can be simultaneously diagonalized in the subspace $\mathcal{N}_{\ket{\psi_s}}$. This guarantees the existence of common eigenstates of the operators $O_a$.

Next, the time evolution of states in this subspace $\mathcal{N}_{\ket{\psi_s}}$ satisfying Eq.~(\ref{eq:PolynomialConditions}) is governed by the broken unitary $U_s (t)= \prod_ae^{-iO_at}$. This follows straightforwardly from the fact that arbitrary powers of $O_a$ commute on this subspace of states and thus, the usual time evolution operator $e^{-iHt}$ can be expanded in terms of powers of $O_a$, and re-arranged to yield $U_s (t)$. 

As a final note, it can be verified that when there are only two eigenoperators $O_1,O_2$, Eq. (\ref{eq:PolynomialConditions}) reduces to 
$[O_1^{n_1},O_2^{n_2}]\ket{\psi_s} = 0$ for arbitrary integers $n_1,n_2$. 

%The time evolution of any state $\ket{\psi_l}$ that satisfies Eq.~(\ref{eq:PolynomialConditions}) then evolves in time according to the broken unitary operator $U_s(t) = \prod_je^{-iO_jt}$. 

%Finally, since any state $\ket{\psi}$ satisfying Eq.~(\ref{eq:PolynomialConditions}) is contained in a subspace inside which all the operators $O_j$ can be simultaneously diagonalized, one has that the states $\ket{\psi}$ can be written as a linear superposition of common eigenstates of the $O_j$, in which case it is clear that the broken unitary operator $\prod_je^{-iO_jt}$ properly captures the time evolution.

\section{Local rules}\label{sec:localRules}
We now introduce three sets of conditions that ensure the existence of a subspace $\mathcal{S}$ for which the full-time evolution operator $e^{-iHt}$ can be broken up into a product of simpler unitary operators. The first set of conditions, which we call type I rules, is most appropriate for describing scar Hamiltonians for which the eigenoperators are extensive local operators. A striking example of such a model is the spin-1 AKLT model which, as we will show, admits a scar subspace described by the broken unitary $e^{-iO_1t}e^{-iO_2t}$ where $O_1$ is a sum of two-site projectors $P_{j}$ for odd $j$, and $O_2$ is a sum of two-site projectors for even $j$. 
The second set of conditions, which we call type II rules, is akin to a generalized Shiraishi-Mori construction~\cite{MOUDGALYA2022_BondAlgebra} and is most appropriate for describing systems for which an extensive number of local operators vanish on the scar subspace. The local eigenoperators are usually accompanied by a single extensive operator constructed as a sum of local terms which gives dynamics to the scar subspace. 
Finally the last set of conditions, which we call type III rules, was originally derived in a previous work of the present authors \cite{ROZON2022_FloquetAutomataScars}, and is most appropriate for describing QMBS Hamiltonians $H = \sum_jh_j$ that can be decomposed into an even/odd partition  $H = O_1 + O_2$ where both $O_1$ and $O_2$ are composed of commuting local terms, but such that full Hamiltonian $H = O_1 + O_2$ does not satisfy this property. One can then derive a set of local rules in terms of the $h_j$ (and sometimes in terms of a related unitary operator $e^{-ih_j}$) that if satisfied for some given state $\ket{\psi_s}$ directly implies the existence of a common invariant subspace $\mathcal{S}$ inside of which the $h_j$ act as pairwise commuting operators. This leads to the broken unitary dynamics $e^{-iHt}\ket{\psi_s} = e^{-iO_1t}e^{-iO_2t}\ket{\psi_s} = \prod_je^{-ih_jt}\ket{\psi_s}$ for any state $\ket{\psi_s}$ in $\mathcal{S}$.

\subsection{Type I rules; few (global) eigenoperators}\label{sec:localRulesFewExtensive}

We now consider the situation where scar subspaces are common eigenstates of just a few extensive operators (made up of a sum of local terms) $O_a$ for $a = 1, 2, ...$. In particular, we consider the situation where the Hamiltonian can be partitioned as follows

\begin{align}
H = \sum_a O_a, \; \; O_a = \sum_{j \in P_a} o_{a,j}
\label{eq:Hsecondrules}
\end{align}

where $O_a$ are extensive operators constructed out of a sum of local operators $o_{a,j}$ centered at site $j$, selected from the subset of sites $P_a$. Now, as per the broken unitary picture, we demand that there are states $\ket{\psi_s}$ for which 

\begin{align}
e^{-i H t} \ket{\psi_s} = \prod_a e^{-i O_a t} \ket{\psi_s}
\end{align}

Note that in this case, there is no extensive set of local eigenoperators $h_j$ which annihilate the scar subspace. Usually, models like the above have a base scar state $\ket{\psi_A}$ with local entanglement onto which one may construct a tower of quasiparticle excitations that will correspond to the scar eigenstates of the model. It is more natural to then formulate the base scar state and related scar states with additional quasiparticle excitations in the Matrix Product State (MPS) formalism whereby the state can still be expressed in terms of local tensors. We first briefly review this construction before elucidating the condition for broken unitary dynamics on these MPS states. 

\subsubsection{MPS Quasiparticle Approach}

Consider a system of $L$ qudits of dimension $d$, with a Hilbert space spanned by states  $\ket{m_1,m_2,..., m_L}$, and $m_i = 1,...,d$. Assuming periodic boundary conditions, arbitrary states $\ket{\psi}$ be represented using the MPS formalism as
\begin{equation}\label{eq:MPS}
    \ket{\psi} = \sum_{\{m_j\}}\operatorname{Tr}\left[A_1^{[m_1]}A_2^{[m_2]}... A_L^{[m_L]}\right]\ket{m_1,m_2,...,m_L}.
\end{equation}
Here $A_{j}^{[m_j]}$ are $d$ $\chi \times \chi$ matrices and where $\chi$ is the bond dimension of the MPS which limits the entaglement; alternatively put, $A_j$ represents a $d\times \chi \times \chi$ tensor at site $j$. We assume periodic boundary conditions. % The trace appearing in equation \ref{eq:MPS} indicates that periodic boundary conditions are assumed. 

In what follows, we denote the above state $\ket{\psi}$  ~\cite{MOUDGALYA2020_MPSQuasiparticles}
\begin{equation}
    \ket{\psi} = \ket{[A_1A_2...A_L]},
\end{equation} for brevity. 
%which is sensible as the matrices $A_j$ contain all the necessary information to reconstruct $\ket{\psi}$ using equation \ref{eq:MPS}. 
It will also be useful to refer to segments of the wave function. Given a subset of consecutive lattice sites $D(j,n) = \{j+1, j+2, ..., j+n \}$ for some given lattice site $j$ and integer $n$, we define a reduced wavefunction on these sites as
\begin{widetext}
\begin{equation}\label{eq:MPSLocal}
\begin{aligned}
    \ket{\psi}_{D(j,n)} = \ket{A_{j+1}A_{j+2}...A_{j+n}}  \equiv %\ket{A_{i+1}A_{i+2}...A_{i+n}} =
     \sum_{m_{j+1},m_{j+2}, ...,m_{j+n}}A_{j+1}^{[m_{j+1}]}A_{j+2}^{[m_{j+2}]}...A_{j+n}^{[m_{j+n}]}\ket{m_{j+1},m_{j+2},...,m_{j+n}}
\end{aligned}
\end{equation}
\end{widetext}
%where the absence of square brackets in $\ket{A_{i_1}A_{i_2}...A_{i_n}}$ indicates that the state is not traced over 
Note that in Eq.~(\ref{eq:MPSLocal}), the $d^n$ `coefficients' in front of the states are all $\chi\times \chi$ matrices---the internal bonds between the sites have been summed over, while the physical site indices $m_p$, $p \in \{j+1, ..., j+n\}$ and the two internal bonds at the ends of the subset of sites, remain.

Any translationally invariant state $\ket{\psi}$ can be represented by an MPS with site-independent matrices $A_{j} = A$ $\forall$ $j$. We denote this state as 
\begin{equation}
    \ket{\psi_A} = \ket{[AA...A]}.
\end{equation}
Next, we introduce the local map $Q_j^{\dagger}$ acting on the uncontracted matrices $A_j$  
\begin{equation}
    (Q_j^{\dagger})^n: A_j\rightarrow B_j^n
\end{equation}
where $B_j^n$ is a shorthand for the matrices $(B_j^n)^{[m_j]}$. In other words, for each $n$, $(Q_j^{\dagger})^n$ represents a mapping from the tensor $A_j$ to the tensor $B_j^n$. We may then define a quasiparticle creation operator as 
\begin{equation}
Q^{\dagger} = \sum_{j=1}^Lb_jQ_j^{\dagger}
\end{equation}
for some scalars $b_j$ (which for instance may be momentum factors $b_j = e^{ik j}$). Then, a state composed of $l$ quasiparticles is given by
\begin{widetext}
\begin{align}
\ket{\psi_l} = (Q^{\dagger})^l\ket{[AA...A]} \equiv
\sum_{j_1,j_2,...,j_l}\left(\prod_{j = j_1,j_2,...,j_l}b_j\right)\ket{[A...A\stackrel{j_1}{B}A...A\stackrel{j_2}{B}A...A\stackrel{j_l}{B}A...A]}
\end{align}
\end{widetext}
where the indices $j_m$ run between $1$ and $L$. It is understood that whenever multiple indices $j_m$ are equal, say $n$ of them, then the symbol $B$ at site $j_m$ is replaced with $B^n$.
As a simple example, the fully polarized state in a one-dimensional lattice of $L$ spin-$1$s, given by $\ket{m_1 = -1, m_2 = -1,...,m_L = -1}$, has the MPS representation
\begin{equation}
\begin{aligned}
\ket{m_1 = -1, m_2 = -1,...,m_L = -1} = \ket{[AA...A]}, \\
    A_j^{[m_j]} = \mathbb{1}_{1\times 1}\delta_{m_j = -1}, 
\end{aligned}
\end{equation}
The states $\ket{\psi_l} = (\sum_{j = 1}^{L}(-1)^jS_{j}^{+})^l\ket{\psi_A}$, where $S^+$ is the usual Spin-1 raising operator, can be written in the MPS language by defining the mapping
\begin{equation}
    (B_j^n)^{[m_j]} = \sum_{m_p = -1,0,1}((S_j^{+})^n)_{m_j,m_p}A_j^{[m_p]}. 
\end{equation}
where $((S_j^{+})^n)_{k,l}$ corresponds to the $(k,l)$ matrix element of $(S_j^{+})^n$.

The state $\ket{\psi_A}$ and quasiparticle states built on top of $\ket{\psi_A}$ by acting with $Q^\dagger$ repeatedly will be considered as the scar states in this system. Ref.~\cite{MOUDGALYA2020_MPSQuasiparticles} for instance figures out certain conditions satisfied by the $A, B$ tensors which allow for the states above to be exact low-entanglement eigenstates of the Hamiltonian. In what follows, we will first derive a more general set of conditions for these states to be exact eigenstates with respect to a given local Hamiltonian, by assuming that these states are in fact eigenstates of operators representing a partition of the Hamiltonian, such as the $O_a$ in Eq.~(\ref{eq:Hsecondrules}). We then discuss how conditions derived in Ref.~\cite{MOUDGALYA2018_ExactExcitedStates} are a special case of these more general conditions. 

\subsubsection{Towers of single-site quasiparticles and broken unitary formalism}
%Given an operator $Q^{\dagger}$ creating quasiparticles on top of a base state $\ket{\psi_A}$, one may make use of the MPS representation to devise rules that ensure that the states $(Q^{\dagger})^l\ket{\psi_A}$ are eigenstates of an underlying target Hamiltonian $H = \sum_{i=1}^Lh_i$. Here, we diverge from the rules introduced in Ref.~\cite{MOUDGALYA2020_MPSQuasiparticles}, and we derive instead a more general set of rules. In this new form, the connection to the broken unitary formalism is transparent.

%Assume the existence of a base state $\ket{\psi_A}$, a set of local Hamiltonian's $h_j$ with support on a neighborhood $\mathcal{N}(j)$ of the site $j$ and a map $Q^{\dagger} = \sum_{j=1}^{L}b_jQ_j^{\dagger}$, where the operators $Q_j^{\dagger}$ are single-site maps of the tensors $A_j$ as discussed above. Further,
Let us partition $H$ as shown in Eq.~(\ref{eq:Hsecondrules}), in terms of operators $O_{a} = \sum_{j\in P_a}o_{a,j}$ satisfying the property $\sum_aO_a = H$. Here the $o_{a,j}$ are assumed to be non-overlapping terms with non-trivial support on the set of sites $\mathcal{N}_{a}(j)$ and $P_a$ is a subset of sites such that $\cup_j\mathcal{N}_{a}(j) = \{ 1,2,...,L\}$. The $O_1$, $O_2$ bipartition for the AKLT model is an example of such partitioning---the $O_1$ partition is composed of terms on odd bonds, while the $O_2$ partition contains the even bonds; the operator $O_1$ has non-trivial support on all the sites, and similarly for $O_2$. Finally, $O_1 + O_2$ is indeed equal to $H_{\text{AKLT}}$.  
\\ \\
\textbf{Theorem:} The states $(Q^{\dagger})^l\ket{\psi_{A}}$ are eigenstates of the operators $O_a$ $\forall l$ \textbf{provided}

\begin{itemize}
\item The $o_{a,j}$, where $j\in P_a$, are such that 
\begin{align}
    o_{a,j}\left(\sum_{p \in \mathcal{N}_a(j)}b_pQ_p^{\dagger}\right)^{n_j}\ket{\psi_A}_{\mathcal{N}_a(j)} =\nonumber\\ \epsilon_a (j,n_j)\left(\sum_{p \in \mathcal{N}_a(j)}b_pQ_p^{\dagger}\right)^{n_j}\ket{\psi_A}_{\mathcal{N}_a(j)}
    \label{eq:MPScondition}
\end{align}

for some scalars $\epsilon_a(j,n_j)$, see Fig. \ref{fig:FewExtensiveRules}.
\item For any $\{n_{j }\ \ge 0\}$ satisfying $\sum_{j \in P_a} n_j = l$
\begin{align}
\sum_{j\in P_a}\epsilon_a(j,n_j) \equiv \mathcal{E}_a(l)
\label{eq:IntegerCondition}
\end{align}
depends only on $l$. 

\begin{figure}
\centering
\includegraphics[width=0.46\textwidth]{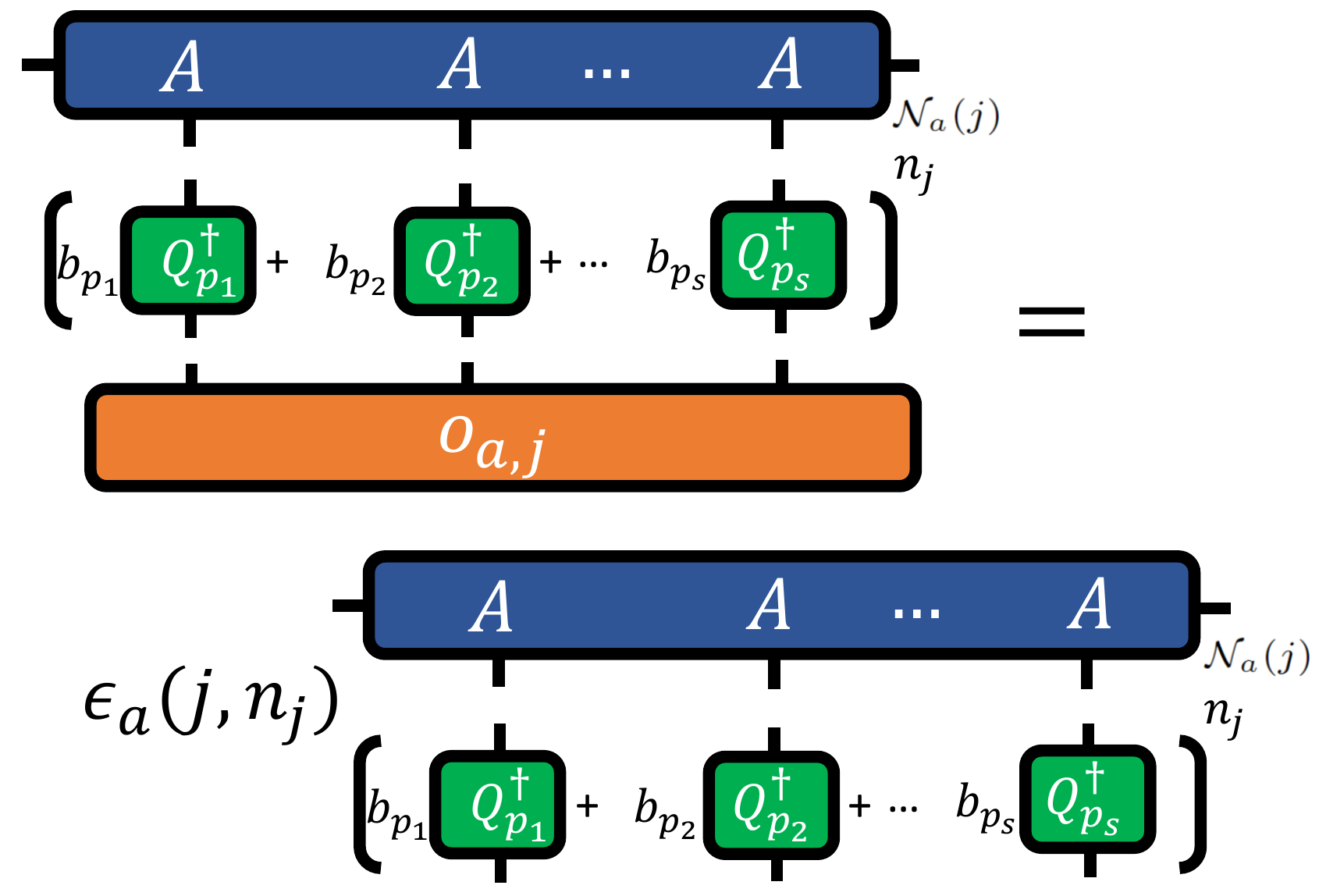}
\caption[]{Visual representation of the rules in Eq. (\ref{eq:MPScondition}). The local operator $o_{a,j}$ acts on the tensor $(\sum_{p \in \mathcal{N}_a(j)}b_pQ_p^{\dagger})^{n_j}\ket{\psi_A}_{\mathcal{N}_a(j)}$ which yields the same tensor multiplied by the scalar $\epsilon_a(j,n_j)$. To get quantum scars, note that the $\epsilon_a(j,n_j)$ must also satisfy Eq. (\ref{eq:IntegerCondition}).}
\label{fig:FewExtensiveRules}
\end{figure}

\end{itemize}
%which is evaluated for any set of integers $\{n_j\}, j\in P_a$, $0 \leq n_j$ satisfying 
\textbf{Proof:}
Let us act with the operator $O_a$ on the state $(Q^{\dagger})^l\ket{\psi_A}$. 
%First, note that by construction, any partition $P_a$ must be composed of sites $j$ such that $\cup_{j \in P_a}\mathcal{N}(j) = \{1,2,...,L\}$, but such that the distinct $\mathcal{N}(j)$ do not overlap. 
We may write the operator $Q^{\dagger}$ into groups of local operators acting onto $\mathcal{N}_a(j)$, that is 
\begin{equation}
    Q^{\dagger} = \sum_{j \in P_a}\sum_{p \in \mathcal{N}_a(j)}b_pQ_{p}^{\dagger}.
\end{equation}
Then, $(Q^{\dagger})^l$ are explicitly given by 

\begin{equation}\label{eq:creationOpPower}
   (Q^{\dagger})^l =  \sum_{\{n_j\}}a_{\{n_j\}}\prod_{j \in P_a}\left(\sum_{p \in \mathcal{N}_a(j)}b_pQ_{p}^{\dagger}\right)^{n_j}.
\end{equation} 
The sum over $\{n_j\}$ is the sum over all possible integers $n_j \ge 0$ such that $\sum_{j \in P_a}n_j = l$, and $a_{\{n_j\}}$ are the usual multinomial coefficients. Acting on the state $(Q^{\dagger})^l\ket{\psi_A}$ with the operator $O_a$ then  yields 
\begin{widetext}
\begin{equation}
\begin{aligned}
O_a(Q^{\dagger})^l\ket{\psi_A} =\sum_{\{n_j\}}\left(\sum_{q\in P_a}\epsilon_a(q,n_q)|_{\{n_j\}}\right)a_{\{n_j\}}\prod_{j \in P_a}\left(\sum_{p \in \mathcal{N}_a(j)}b_pQ_{p}^{\dagger}\right)^{n_j}\ket{\psi_A} = \\
\mathcal{E}_a(l)\sum_{\{n_j\}}a_{\{n_j\}}\prod_{j \in P_a}\left(\sum_{p \in \mathcal{N}_a(j)}b_pQ_{p}^{\dagger}\right)^{n_j}\ket{\psi_A} =\mathcal{E}_a(l)(Q^{\dagger})^l\ket{\psi_A}.
\end{aligned}
\end{equation}
\end{widetext}
%Note that provided the $\mathcal{E}_n$ are some multiple of some constant energy $\mathcal{E}$, this shows that the individual $O_i$ all satisfy a spectrum generating algebra. 

It is important to note that Eq.~(\ref{eq:MPScondition}) is \emph{not} equivalent to demanding that the wavefunction on the entire Hilbert space is an eigenstate of the local operator $o_{a,j}$ (except for the base state). In fact, the full wavefunctions are only designed to be eigenstates of the partition operators $O_a$. However, local segments of the base state when perturbed with $(Q^\dagger)^l$ \emph{limited} to perturbing the state only on a set of local sites do appear as eigenstates of the operators $o_{a,j}$. 

Finally, the connection to the broken unitary picture is clear---since $H = \sum_{a}O_a$, time evolution within the scar subspace is determined by $\prod_a e^{-i O_a t}$

%\begin{equation}
%\begin{aligned}
%    e^{-iHt}\sum_{i = 1}c_i(Q^{\dagger})^i\ket{\psi_A} =\\ \prod_{a} e^{-ik_aO_at}\sum_{i = 1}c_i(Q^{\dagger})^i\ket{\psi_A}
%\end{aligned}
%\end{equation}

\subsubsection{Special cases}
We now study some special cases. Consider a Hamiltonian $H = \sum_{j= 1}^{L}h_{j}$ where the operators $h_{j}$ have support onto the sites $j,j+1$. A sensible partition to consider is given by $O_1 = \sum_{j = 1}^{L/2}h_{2j-1}$, $O_2 = \sum_{j = 1}^{L/2}h_{2j}$. Suppose now that the $h_{j}$ terms satisfy the equation
\begin{equation}
\begin{aligned}
h_{j}\left(\sum_{p = j, j+1}b_pQ_p^{\dagger}\right)^{n_j}\ket{\psi_A}_{j,j+1} = \\
(n_j\epsilon + c)\left(\sum_{p = j,j+1}b_pQ_p^{\dagger}\right)^{n_j}\ket{\psi_A}_{j,j+1},
\label{eq:partspecial}
\end{aligned}
\end{equation}
for some constants $c,\epsilon$, meaning that $\epsilon_1(j,n_j) = n_j\epsilon + c$, $\epsilon_2(j,n_j) = n_j\epsilon + c$. It is then clear that $\sum_{q\in P_a}\epsilon_a(q,n_q)|_{\{n_j\}} = l\epsilon + \frac{Lc}{2}$ which only depends on $l$. The states $(Q^{\dagger})^l \ket{\psi_A}$ are then common eigenstates of the operators $O_1$ and $O_2$, and their time evolution is dictated by $U_s(t) = e^{-iO_1t}e^{-iO_2t}$.
\\\\
Another relevant case to consider is whenever the $o_{a,j}$ are such that 
\begin{equation}
\begin{aligned}
    o_{a,j}\left(\sum_{p \in \mathcal{N}_a(j)}b_pQ_p^{\dagger}\right)^{n_j}\ket{\psi_A}_{\mathcal{N}_a(j)} = 0.
\end{aligned}
\end{equation}
When this is the case, the states $(Q^{\dagger})^l\ket{\psi_A}$ are eigenstates of all the $o_{a,j}$ individually. The corresponding partitions can be taken as each $o_{a,j}$ on $\mathcal{N}_a(j)$ individually with $0$ on all other sites. 

\subsubsection{Relation to formulation of Ref.~\cite{MOUDGALYA2020_MPSQuasiparticles}}
The quasiparticle construction in Ref.~\cite{MOUDGALYA2020_MPSQuasiparticles} imposes a different set of restrictions to ensure that the states $\left(Q^\dagger\right)^l\ket{\psi_A}$ are eigenstates of the full Hamiltonian $H$. In particular, excited states are expected to satisfy the conditions 
\begin{equation}\label{eq:sanjayconditions}
\begin{aligned}
\ket{B_jB_{j+1}} &= 0\\
\ket{B_j^2} &= 0\\
h_j(\ket{B_jA_{j+1}} + e^{jk}\ket{A_jB_{j+1}}) &= \\\epsilon(\ket{B_jA_{j+1}} + e^{ik}\ket{A_jB_{j+1}}) &\\
h_{j}\ket{A_jA_{j+1}} &= 0\\
\end{aligned}
\end{equation}
where it was assumed that $b_j = e^{ikj}$.

%where
%\begin{equation}
%\begin{aligned}
%\ket{B_j^2} &= \sum_{m_j}(B^2)^{[m_j]}_j\ket{m_j}\\
%(B^2)^{[m]} &= \sum_{n,l}Q^{\dagger}_{m,l}Q^{\dagger}_{l,n}A^{[n]}\\
%\ket{B_jB_{j+1}} &= \sum_{m_j,m_{j+1}}B_j^{[m_j]}B_{j+1}^{[m_{j+1}]}\ket{m_j,m_{j+1}}\\
%\end{aligned}
%\end{equation}

We can interpret these conditions in terms of the more general construction presented above following from Eq.~(\ref{eq:MPScondition}), and more specifically, Eq.~(\ref{eq:partspecial}).

Assuming, as in Ref.~\cite{MOUDGALYA2020_MPSQuasiparticles} that the operators $h_{j}$ have support on the sites $j,j+1$, one can readily verify that the above conditions imply 
\begin{equation}
\begin{aligned}
    h_{j}\left(\sum_{p = j,j+1}e^{ikp}Q_p^{\dagger}\right)^{n_j}\ket{A_jA_{j+1}} \\= n\epsilon\left(\sum_{p = j,j+1}e^{ikp}Q_p^{\dagger}\right)^{n_j}\ket{A_jA_{j+1}}
\end{aligned}
\end{equation}
Note that this holds even when $n_j>1$ since for $n_j>1$, the resultant state itself vanishes due to the first two conditions of Eq.~(\ref{eq:sanjayconditions}). The partition of $H$ that corresponds to this condition is then \begin{equation}
    O_1 = \sum_{j = 1}^{L/2}h_{2j-1},\text{  } O_2 = \sum_{j = 1}^{L/2}h_{2j}.
\end{equation}
Remarkably, it is easy to show that the AKLT model falls within this restricted scheme~\cite{MOUDGALYA2020_MPSQuasiparticles}, which directly implies (using only local considerations) that the unitary describing the time evolution within the scar subspace is given by $U_s(t) = e^{-iO_1t}e^{-iO_2t}$. As for the Hubbard model as well as the spin-1 XY magnets (single site tower), these models are better understood via Eqs.  (\ref{eq:MPScondition}), (\ref{eq:IntegerCondition}), see App. \ref{app:MPSQuasiparticlesModels}
\\\\
When it comes to generalizations such as multi-site quasiparticles (cases where the $Q_j^{\dagger}$ act on more than one site such that they start overlapping), the situation is more complex. We suspect that the multi-site quasiparticle formalism may also be understood as arising from common eigenstates of simple operators by making use of alternate representations, as observed in the multi-site tower of scar states in the spin-1 XY magnets discussed in Sec. \ref{sec:TwoSiteQPSpinXY}, but we lack proof of such a statement. We defer this issue to future works.

\subsection{Type II rules; extensive number of local eigenoperators}\label{sec:LocalRulesExtensive}
For the second set of rules, let us consider a general Hamiltonian given by 
\begin{equation}
H = \sum_{j=1}^{L} a_jh_j + \sum_{j\in P_v}v_j.
\label{eq:GSMform}
\end{equation}
where periodic boundary conditions are assumed and $P_v$ is some subset of the sites $\{1,2,...,L\}$. 
Our goal is to find a subspace of states $\ket{\psi_s}$ that is described by the `broken' unitary $U_s (t) = e^{-i\sum_{j\in P_v}v_jt}$ for arbitrary coefficients $a_j$. In other words, we must find a subspace of states where the individual terms $h_j$ vanish and whose states are not coupled to the rest of the Hilbert space via the operator $\sum_{j\in P_v}v_j$. Note that the requirement that the $h_j$ vanish when acting on the scar subspace can be thought of as a requirement that all the terms $e^{-i a_j h_j t_*}$ which form part of the broken unitary dynamics equal $\mathbb{1}$ on the scar subspace for some $t_*$; this is necessary to ensure revivals. Since the amplitudes $a_j$ are arbitrary, this is naturally achieved only when the operators $h_j$ annihilate the scar subspace. 

We now ask what conditions enforce such broken unitary dynamics. Naturally, we must first have that $h_j\ket{\psi_s} = 0$ for all $j$. Grouping the $h_j$ and $v_j$ into extensive-local operators $O_1 = \sum_ja_jh_j$, $O_2 = \sum_{j\in P_v}v_j$ leads, per Eq.~(\ref{eq:PolynomialConditions}), to the condition
\begin{align}\label{qe:powerCondition}
[O_1^{n_1},O_2^{n_2}]\ket{\psi_s} = 
O_1^{n_1}O_2^{n_2}\ket{\psi_s} - O_2^{n_2}O_1^{n_1}\ket{\psi_s} = \nonumber\\O_1^{n_1}O_2^{n_2}\ket{\psi_s} = 0.
\end{align}
Note that satisfying Eq.~(\ref{qe:powerCondition}) for $n_1 = 1$ and all integers $n_2$ directly implies satisfaction the condition for higher powers $n_1$. Thus, the necessary and sufficient condition for the above equation to be satisfied is given by 
\begin{equation}
\left( \sum_ja_jh_j \right) \left(\sum_{j\in P_v} v_{j}\right)^{n_2}\ket{\psi_s} = 0.
\end{equation}
Since the above must hold for arbitrary coefficients $a_j$, we can simplify the condition to 
\begin{equation}
h_j\left(\sum_{p\in P_v} v_p\right)^{n_2}\ket{\psi_s} = 0
\end{equation}
which must hold for arbitrary $j$ and integer $n_2$. 

To summarize, a general set of conditions that enforce the broken unitary dynamics are given by 
\begin{align}\label{eq:GenScarConditionGSM}
h_j\ket{\psi_s} &= 0 \; \; \forall \; \; j\nonumber\\
h_j\left(\sum_{p\in P_v} v_p\right)^{n_2}\ket{\psi_s} &= 0 \; \; \forall \; \; j,n_2. 
\end{align}
 
 When these conditions hold for a state $\ket{\psi_s}$, the dynamics under the Hamiltonian is simplified to
\begin{align}
e^{-iHt}\ket{\psi_s} &= e^{-i(\sum_j a_j h_j + \sum_{j\in P_v}v_j)t}\ket{\psi_s} \nonumber\\ 
&= e^{-i\sum_{j\in P_v}v_j t}e^{-i\sum_ja_j h_j t} \ket{\psi_s} \nonumber\\&= e^{-i\sum_{j\in P_v}v_j t}\ket{\psi_s}
\end{align}
as desired. Note that the $\ket{\psi_s}$ are not required to be eigenstates of $H$. 

The conditions of Eq.~(\ref{eq:GenScarConditionGSM}) are so far `global' in the sense that the sum over the terms $v_j$ might include overlapping terms. %We next attempt to simplify these conditions further if $v_i$ satisfies certain properties. 
%
%When the support of $v_i$s overlap for different $i$, the set of conditions in Eq.~(\ref{eq:GenScarConditionGSM}) are rather difficult to verify explicitly since the order into which the $v_j$ appear in the product matters \textcolor{red}{so far we haven't broken Eq 7 down to local relations and the sum over vi is unrestricted. So I don't understand this statemnet?}. 
%
In practice, most QMBS models constructed in the form of Eq.~(\ref{eq:GSMform}) have a sum $\sum_{j\in P_v}v_j$ in which no terms overlap, which greatly simplifies the above conditions. Furthermore, most scar states can be thought of as being generated by the repeated action of a creation operator $Q^{\dagger} = \sum_{j \in P_Q}Q_{j}^{\dagger}$ on some base state $\ket{\psi_A}$, where $P_Q$ is some subset of sites. In the following subsection, we discuss the simple case where both $Q^{\dagger}$ and $\sum_{j \in P_v}v_j$ are operators built out of non-overlapping commuting local operators, and we derive a simplified set of rules by making use of these assumptions. We note that when the $v_j$'s in $\sum_{j \in P_v}v_j$ start overlapping, the simplification of these `global' rules to local ones is not obvious. 

The rules of type II can be interpreted as a generalization of the Shiraishi-Mori construction introduced in Ref. \cite{SHIRAISHI2017_EmbeddedScars}. Recall that the Shiraishi-Mori construction aims at embedding the common null space $\mathcal{S}$ of a set of local operators $P_j$ inside of an otherwise thermalizing spectrum. The eigenstates within this subspace can then be identified as the scar eigenstates. To achieve this embedding, one constructs the Hamiltonian $H_{\text{SM}} = \sum_jP_jK_jP_j + H'$ with the condition $[P_j,H'] = 0$. It can then be verified that $e^{-iH_{\text{SM}}t}\mathcal{S} = e^{-iH't}\mathcal{S}$, see Sec. \ref{sec:ShiraishiMoriContruction} for the  details of the proof. 

\begin{figure}
\centering
\includegraphics[width=0.46\textwidth]{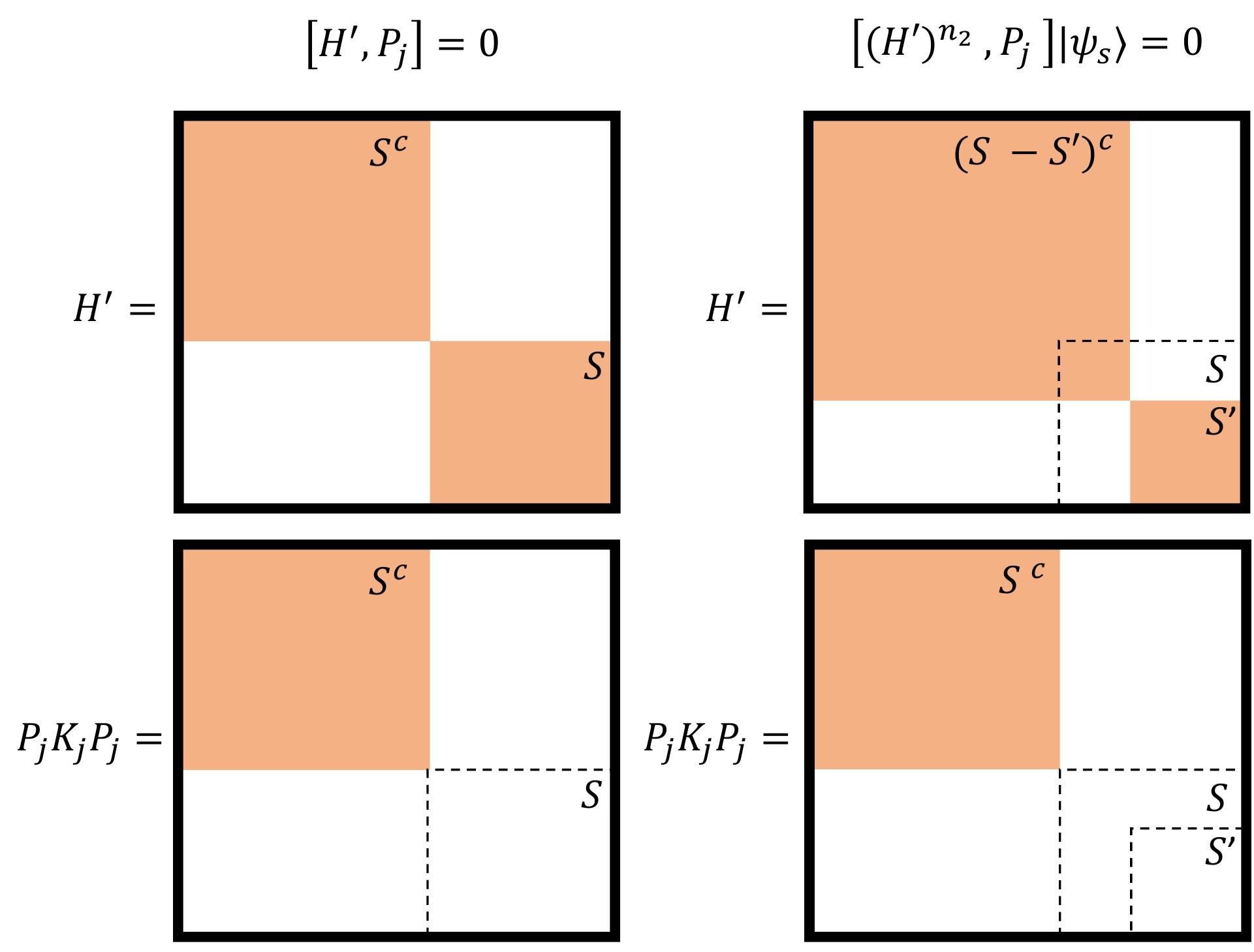}
\caption[]{The usual Shiraishi-Mori condition $[H',P_j] = 0$ requires that the entire common null-space of the $P_j$ evolves in time according to the unitary operator $e^{-iH't}$. $H'$ is then block diagonal, with one block associated with the subspace $\mathcal{S}$, and another block associated with it's complement $\mathcal{S}^c$ as illustrated in the figure. As for the terms $P_j K_j P_j$, they vanish within the subspace $\mathcal{S}$, and have arbitrary matrix elements in it's complement $\mathcal{S}^c$. The Shiraishi-Mori condition further imposes that the $P_j$ operators commute with $H'$ in the subspace $\mathcal{S}^c$. By contrast, when type II rules $[(H')^{n_2},P_j]\ket{\psi_s} = 0$ are satisfied, $H'$ only needs to be block-diagonal with a subspace $\mathcal{S'}$ that can be smaller than the full null-space $\mathcal{S}$ of the $P_j$ operators. Furthermore, it is no more required that the operators $P_j$ commute with $H'$ within $\mathcal{S}^c$. }
\label{fig:ShiraishiMoriTypeII}
\end{figure}

We will now show that the type II rules can be understood as a relaxation of the original Shiraishi-Mori construction and result in a completely analogous dynamical decoupling. The $h_j$ operators introduced in the type II rules can be viewed as the $P_j$ in the Shiraishi-Mori construction, whilst the operator $\sum_{j\in P_v}v_j$ plays the role of $H'$. Let us assume the existence of a state $\ket{\psi_s}$ that satisfies 
\begin{align}
P_j\ket{\psi_s} = 0\nonumber\\
P_j(H')^{n_2}\ket{\psi_s} = 0
\end{align}
for all sites $j$ and integers $n_2 \ge 0$, which can be written compactly as 
\begin{equation}
[P_j,(H')^{n_2}]\ket{\psi_s} = 0
\end{equation} for arbitrary sites $j$ and integers $n_2 \ge 0$. Note that this also implies that the commutation relation $[P_jK_jP_j,(H')^{n_2}]\ket{\psi_s} = 0$ is also satisfied for arbitrary operators $K_j$. Then, 
%by virtue of the type II rules and of the more general condition in Eq. \ref{eq:PolynomialConditions}, 
there must exist a subspace of the Hilbert space $\mathcal{S'}$ characterized by the time evolution operator $e^{-iH't}$ instead of the more complicated time evolution operator $e^{-i\sum_jP_jK_jP_j + H'}$. The type II rules may differ in two ways from the original Shiraishi-Mori condition $[P_j,H'] = 0$. First, it is not a requirement that the full common null-space of the $P_j$ evolves according to the unitary operator $e^{-iH't}$. Mathematically, it may be the case that $S' \subset \mathcal{S} = \{\ket{\psi}: P_j\ket{\psi} = 0\; \forall \;j\}$. Furthermore, $H'$ here is not required to commute with every $P_j$ within the complement of the subspace $\mathcal{S'}$. By contrast, the original Shiraishi-Mori construction requires that $[H',P_j] = 0$ for all $P_j$, so $H'$ and $P_j$ commute over the entire Hilbert space. We illustrate these differences with Fig. \ref{fig:ShiraishiMoriTypeII}.

\subsubsection{Non-overlapping set of $v_j$ and $Q_j^{\dagger}$}
Consider a base state $\ket{\psi_A}$ as well as a creation operator $Q^{\dagger} = \sum_{j \in P_Q}b_jQ_j^{\dagger}$ composed of spatially decoupled operators $Q_j^{\dagger}$, $j \in P_Q$ that pairwise commute, and the $b_j$ are complex numbers. Also, assume that the $v_j$ terms that appear in $H = \sum_{j=1}^{L} a_jh_j + \sum_{j\in P_v}v_j$ are spatially decoupled and pairwise commute. The terms $v_j, Q_j, h_j$ are assumed to have support on a neighborhood $\mathcal{N}_v(j), \mathcal{N}_{Q}(j)$ and $\mathcal{N}_h(j)$, respectively, of  site $j$. We define putative scar states by
\begin{equation}
\ket{\psi_l} \equiv (Q^{\dagger})^l\ket{\psi_A} \; \; \text{for different} \; \; l, 
\end{equation}
and the neighborhood $\mathcal{N}(j)$ as the set of sites on which the $Q_p$ or $v_q$ overlap with $h_j$; more precisely
\begin{align}
\mathcal{N}(j) = \bigcup_{p:\mathcal{N}_Q(p)\cap\mathcal{N}_h(j)\neq\emptyset}\mathcal{N}_Q(p)\bigcup_{q:\mathcal{N}_v(q)\cap\mathcal{N}_h(j)\neq\emptyset}\mathcal{N}_v(q). 
\end{align}
With these conditions in place, note that Eq.~(\ref{eq:GenScarConditionGSM}) can be rewritten as 
%\begin{widetext}
\begin{align}\label{eq:BinomCondition}
h_j\ket{\psi_l} &= 0 \nonumber\\
\sum_{n = 0}^{n_2}\binom{n_2}{n}\left(V - \sum_{q\in \mathcal{N}(j)\cap P_v}v_q\right)^{n_2 - n}\times\nonumber\\h_j\left(\sum_{q\in \mathcal{N}(j)\cap P_v}v_q\right)^{n}\ket{\psi_l} &= 0 \nonumber \\
\end{align}
%\end{widetext}
where $V = \sum_{q\in P_v} v_q$. 

\begin{figure}
\centering
\includegraphics[width=0.46\textwidth]{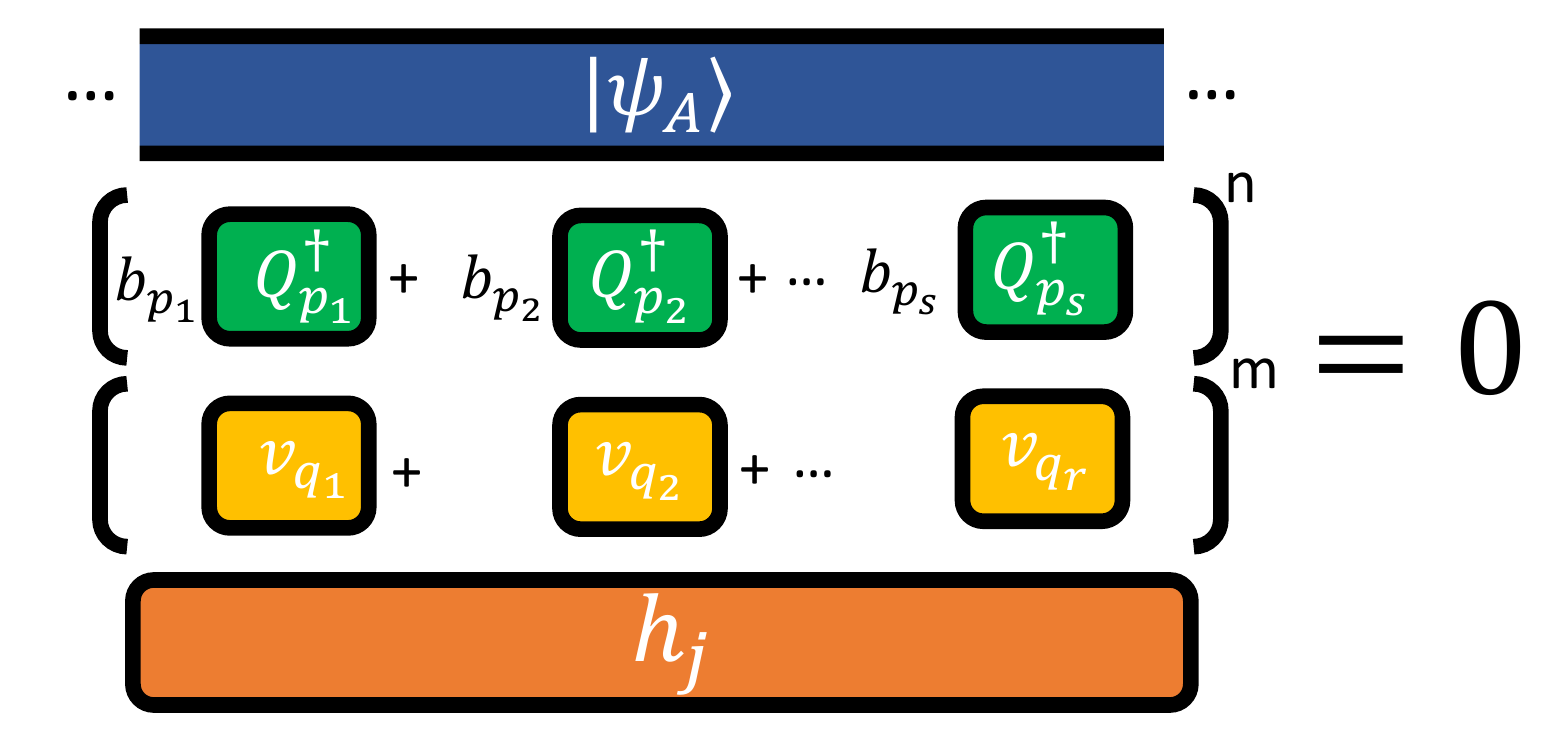}
\caption[]{Visual representation of the rules in Eq. (\ref{eq:localRules}). The action of $h_j$ on the state $\left(\sum_{q\in \mathcal{N}(j)\cap P_v}v_q\right)^{n}\left(\sum_{p\in \mathcal{N}(j)\cap P_Q}b_pQ_p^{\dagger}\right)^{m}\ket{\psi_A}$ vanishes for arbitrary integers $n,m \ge 0$. The $p_i$ indices in the figure are such that $p_i \in \mathcal{N}(j)\cap P_Q$. Similarly, $q_i \in \mathcal{N}(j)\cap P_v$. In practice, one often has that the coefficients $b_{p}$ are momentum factors $b_p = e^{-ikp}$ for some momentum $k$.}
\label{fig:ExtensiveLocalRules}
\end{figure}

The term $\left(V - \sum_{q\in \mathcal{N}(j)\cap P_v}v_q\right)^{(n_2 - n)}$ does not contain any term that acts in the neighborhood $\mathcal{N}(j)$ and when expanded, leads to a sum of non-local operators. Thus, such a term generates a sum of states which differ in global perturbations to the state $\ket{\psi_l}$. For this to be true for arbitrary $n_2,n$ and combinations of sites perturbed, it is natural to demand the local conditions 
\begin{align}
h_j\ket{\psi_l} = 0\nonumber\\
h_j\left(\sum_{q\in \mathcal{N}(j)\cap P_v}v_q\right)^{n}\ket{\psi_l} = 0
\end{align}
to be satisfied for any $n$ and scar state marked by $l$. We may also expand the states $\ket{\psi_l}$ in terms of the quasiparticle operator $Q^\dagger$ acting on the base state. The local conditions we obtain are thus

%$\ket{\psi_l} = (Q^{\dagger})^l\ket{\psi_A}$ and that the $Q_j^{\dagger}$ do not overlap, we may write the conditio $Q^{\dagger}$ after which we finally obtain 

\begin{align}\label{eq:localRules}
h_j\left(\sum_{q\in \mathcal{N}(j)\cap P_v}v_q\right)^{n}\left(\sum_{p\in \mathcal{N}(j)\cap P_Q}b_pQ_p^{\dagger}\right)^{m}\ket{\psi_A} &= 0.
\end{align}
for any site $j$ and arbitrary integers $n \ge 0,m \ge 0$, see Fig. \ref{fig:ExtensiveLocalRules}.
In section \ref{sec:ExplicitModels}, we show that many QMBS models can be understood via this set of conditions. 

It is pertinent to note here distinctions between the rules of type II and the rules of type I. 
The rules of type II are most appropriate when the scar eigenstates are common eigenstates of a single extensive operator and are zero-energy eigenstates of an extensive number of local operators ($\sum_{j\in P_v}v_j$ and the $h_j$ for all $j$, respectively). Crucially, these rules apply even when the state $\ket{\psi_s}$ is not an eigenstate of $H$, the only requirement is that it is part of the scar subspace $\mathcal{S}$. By contrast, the rules of type I only apply to scar states that are eigenstates of the Hamiltonian. They do however have the advantage of applying to cases where the scar states are common eigenstates of multiple global operators. As will be shown in the next sections, it turns out that the rules of type I naturally apply to models like the Spin-1 AKLT model as well as a domain wall conserving model. The rules of type II rather apply to models like the Spin-1 XY magnets and the Hubbard model.
%\textcolor{red}{it would be useful to highlight the differences at this point between local rules 2 and local rules 1. It seems like all models can be studied using local rules 2?}

\subsection{Type III rules; local unitary and Hamiltonian rules}
\begin{figure}
\centering
\includegraphics[width=0.46\textwidth]{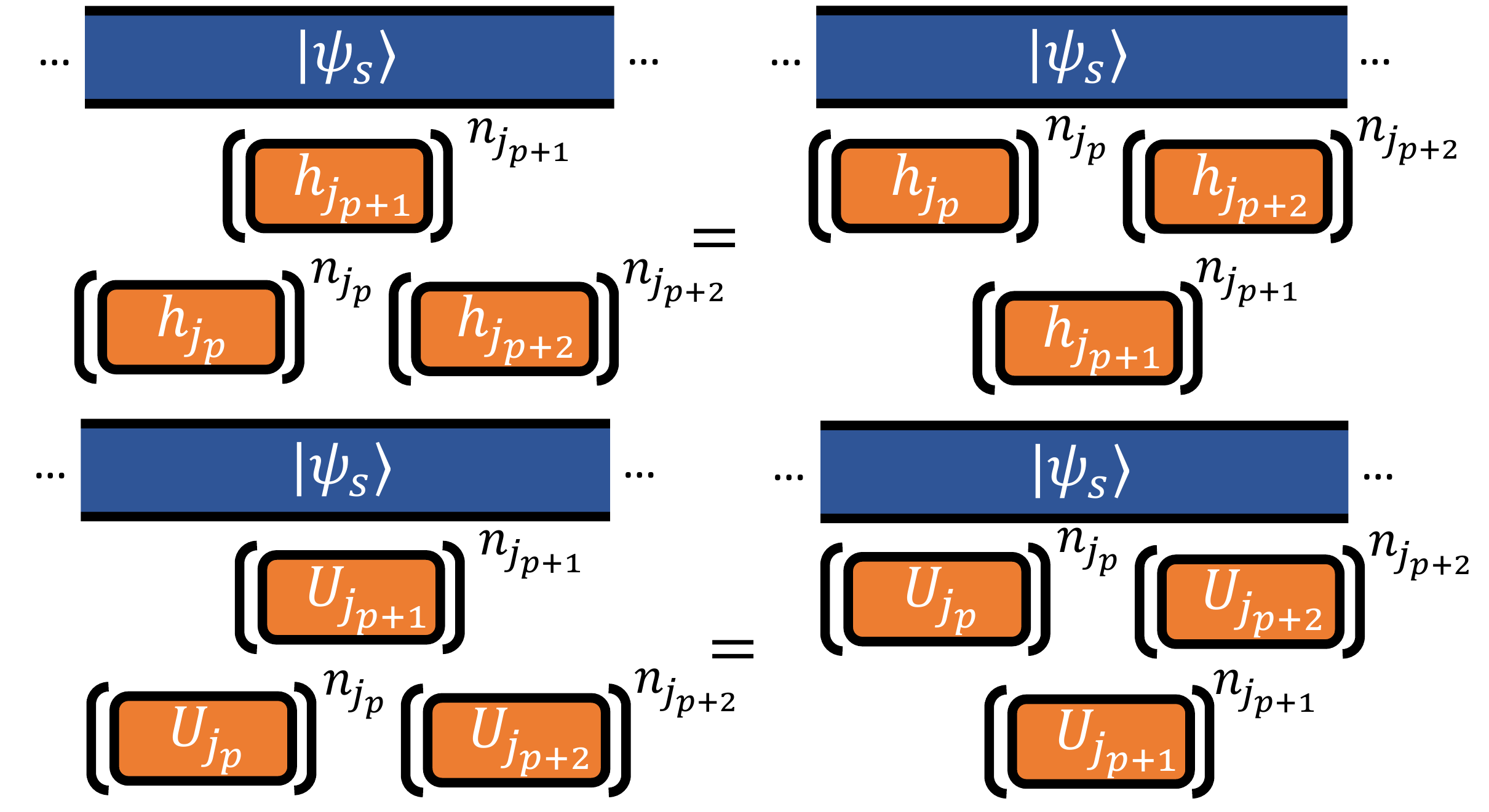}
\caption[]{Visual representation of the local Hamiltonian and unitary rules described in Eqs. (\ref{eq:loclHamiltonianRules}), (\ref{eq:loclUnitaryRules}) respectively. Provided the product of Hamiltonian terms $h_{j_p}^{n_{j_p}}h_{j_{p+2}}^{n_{j_{p+2}}}h_{j_{p+1}}^{n_{j_{p+1}}}$ is equivalent to $h_{j_{p+1}}^{n_{j_{p+1}}}h_{j_p}^{n_{j_p}}h_{j_{p+2}}^{n_{j_{p+2}}}$ when acting on a given scar state $\ket{\psi_s}$, then one obtains that there will exist states that are common eigensates of the operators $h_j$ even if the $h_j$ do not pairwise commute. Similarly, if the same set of rules is satisfied for a unitary operator $U_j$ such that there exists some integer $n$ for which $U_j^n = \mathbb{1}$, then one obtains that $h_j \equiv i\log{(U_j)}$ admits common eigenstates. }
\label{fig:localHURules}
\end{figure}

In a previous work \cite{ROZON2022_FloquetAutomataScars} the present authors considered a set of Hamiltonians given by
\begin{equation}
H = \sum_{j\in P_h}h_j
\end{equation}
where $P_h$ is a subset of the 1D lattice sites $\{1,2,...,L\}$. Periodic boundary condition are assumed and the Hamiltonian terms $h_j$, $j\in P_h$ fail to commute only with their nearest neighbors. Denoting the sites that appear in $P_h$ by $j_1,j_2,...$,  a natural even/odd partition of the Hamiltonian is given by 
\begin{align}
O_1 = \sum_{p=1}^{|P_h|/2}h_{j_{2p - 1}}, \; \; 
O_2 = \sum_{p=1}^{|P_h|/2}h_{j_{2p}}
\end{align}
where $|P_h|$ is the number of sites in the set $P_h$, which is assumed here to be even. All the terms within $O_1$ mutually commute, and similarly for $O_2$. It can then be shown that if a state $\ket{\psi_s}$ satisfies \cite{ROZON2022_FloquetAutomataScars}
\begin{equation}\label{eq:loclHamiltonianRules}
h_{j_p}^{n_{j_p}}h_{j_{p+2}}^{n_{j_{p+2}}}h_{j_{p+1}}^{n_{j_{p+1}}}\ket{\psi_s} = h_{j_{p+1}}^{n_{j_{p+1}}}h_{j_p}^{n_{j_p}}h_{j_{p+2}}^{n_{j_{p+2}}}\ket{\psi_s} 
\end{equation}
for arbitrary integers $n_j$ and integers $p$ (see Fig. \ref{fig:localHURules}), the time evolution of the state $\ket{\psi_s}$ is characterized by the broken unitary $e^{-iO_1t}e^{-iO_2t}$ instead of the more complicated $e^{-iHt}$. 

In fact, it can be shown that Eq.~(\ref{eq:loclHamiltonianRules}) is a sufficient set of conditions to ensure the existence of a common invariant subspace $\mathcal{S}$ on which all the $h_j$ commute pairwise, and which embeds the state $\ket{\psi_s}$. This subspace is spawned by the common eigenstates of the operators $h_j$. However, the common eigenspace $\mathcal{S}$ of the operators $h_j$ can be smaller than the full scar subspace of the model at hand. For instance, in the AKLT model, only two (the extremal states) out of the $L/2$ scar eigenstates (not accounting for SU(2) degeneracy) satisfy the type III rules. All the scar states more generally satisfy type I rules; the complete scar subspace is thus spawned by common eigenstates of only the operators $O_1, O_2$. 

Now, if only a finite set of $h_j^a$ are independent (as is the case in the PXP model), the rules in Eq.~(\ref{eq:loclHamiltonianRules}) are not infinite and can be enumerated. If all of them are satisfied, then a common eigenspace of operators $O_1,O_2$ exists and defines the scar subspace; moreover, the eigenenergies of these states is equidistant, which allows for strong many-body revivals. However, if a large fraction of the rules, but not all of them are satisfied, then the eigenvalues of the scar eigenstates are not expected to be perfectly equidistant; this naturally leads to approximate many-body revivals. We show in Ref. \cite{ROZON2022_FloquetAutomataScars} that this picture appears to apply to models such as the  PXP model which can be shown to satisfy a large fraction of such rules. Another worthy observation is the fact that the above rules can be formulated in terms of local unitary gates $U_{j_p}$. As discussed in Ref. \cite{ROZON2022_FloquetAutomataScars}, provided there exists an integer $n$ such that $U_{j_p}^n = \mathbb{1}$, one can show that provided there exists a state $\ket{\psi_s}$ for which the set of rules
\begin{equation}\label{eq:loclUnitaryRules}
U_{j_p}^{n_{j_p}}U_{j_{p+2}}^{n_{j_{p+2}}}U_{j_{p+1}}^{n_{j_{p+1}}}\ket{\psi_s} = U_{j_{p+1}}^{n_{j_{p+1}}}U_{j_p}^{n_{j_p}}U_{j_{p+2}}^{n_{j_{p+2}}}\ket{\psi_s} 
\end{equation}
are satisfied for arbitrary integers $n_j$ and integers $p$, see Fig. \ref{fig:localHURules}, one obtains the exact same dynamical decoupling as above with $h_{j_p} \equiv i\log(U_{j_p})$. The $\log$ operation is taken to be the principal logarithm of the diagonal matrix containing the eigenvalues of $U_{j,p}$, which is then expressed back in the original basis. This fact is used extensively in \cite{ROZON2022_FloquetAutomataScars} to construct exact as well as approximate QMBS models starting from trivial local unitary gates $U_{j,p}$ that act as a simple permutation of computational basis states. 

\section{Broken unitary dynamics in selected scar models}\label{sec:ExplicitModels}
In this section, we make explicit use of the rules presented above to analyze systems that are well known to host quantum many-body scars. 

\subsection{The Spin-1 AKLT model}
The one dimensional Spin-1 AKLT model \cite{AFFLECK1987_aklt} is given by the Hamiltonian
\begin{align}\label{eq:AKLTHamiltonian}
H_{\text{AKLT}} =\sum_{j=1}^{L}\left(\frac{1}{3}+\frac{1}{2} \vec{S}_{j} \cdot \vec{S}_{j+1}\right.\nonumber\\ \left.+\frac{1}{6}\left(\vec{S}_{j} \cdot \vec{S}_{j+1}\right)^{2}\right).
\end{align}
This model can be rewritten in terms of projectors onto the total spin-2 subspace of pairs of neighboring sites, i.e 
\begin{align}
H_{\text{AKLT}}&=\sum_{j=1}^{L} P_{j}\nonumber\\&=\sum_{j=1}^{L}\left(\sum_{m=-2}^{2}\left|T_{2, m}\right\rangle\left\langle T_{2, m}\right|\right)_{j, j+1}
\end{align} where $P_j$ acts non-trivially on sites $j,j+1$. Let us define the states $\ket{T_{l,m}}$ to be eigenstates of both $(\vec{S}_{j} + \vec{S}_{j+1})^2$ and $S_{j}^z + S_{j+1}^z$ with eigenvalue $l(l+1)$ and $m$ respectively. The state $\ket{m}_j$ are such that $S_j^{z}\ket{m}_j = m\ket{m}_j$
\\\\
The ground state of the Hamiltonian of Eq. (\ref{eq:AKLTHamiltonian}), which we denote by $\ket{\psi_A}$, is given by the unique state that does not have any spin-2 component on any bond $(j, j+1)$, and is easily seen to have eigenenergy 0 since all the local projectors $P_j$ vanish when acting on such a state.  scar eigenstates can be generated by the repeated action of the operator $Q^{\dagger}$ on the ground state~\cite{MOUDGALYA2018_ExactExcitedStates,MOUDGALYA2018_AKLTEntanglement,MARK2020_RestrictedSGA},
\begin{equation}\label{eq:Qp}
\begin{array}{c}
\left|\psi_l\right\rangle = \left(Q^{\dagger}\right)^{l}|\psi_A\rangle, \; \; \text{with} \; \; Q^{\dagger}=\sum_{j=1}^{L}(-1)^{j}(S_{j}^{+})^{2}
\end{array}
\end{equation} 

which produces a tower of sub volume-law entangled states understood as the scar eigenstates of the spin-1 AKLT model. The existence of these eigenstates is closely associated with a spectrum-generating algebra ~\cite{MARK2020_RestrictedSGA} whereby the subspace $W$ of states spawned by the scar states satisfy the condition

\begin{equation}\label{eq:SGACondition}
    \left([H_{\text{AKLT}},Q^{\dagger}] - \epsilon Q^{\dagger} \right) W = 0. 
\end{equation}

Using the above condition, it is easy to see that $(Q^{\dagger})^l\ket{\psi_A}$ are eigenstates of $H_{\text{AKLT}}$ with eigenvalue $E_G + l \epsilon$, and $W$ is spanned by the states $\left(Q^{\dagger} \right)^l \ket{\psi_A}$ for $l = 0, 1, ... L/2$ (and ${(Q^\dagger})^{L/2 + 1} \ket{\psi_A} = 0$).

We now proceed to show that AKLT scar eigenstates are common eigenstates of simple integrable partitions $O_1, O_2$ of the Hamiltonian with $H_{\text{AKLT}} = O_1 + O_2$ and

\begin{align}
    O_1 = \sum_{j=1}^{L/2}P_{2j}, \; \; O_2 = \sum_{j=1}^{L/2}P_{2j-1} 
\end{align}
by making use of condition \ref{eq:MPScondition} and condition \ref{eq:IntegerCondition}. First, it is known that the ground state $\ket{\psi_A}$ of $H_{\text{AKLT}}$ admits the MPS representation $\ket{\psi_A} = \ket{[AA...A]}$ with
\begin{align}
A^{[1]}_j &= \sqrt{\frac{2}{3}}\sigma^+_j, \; \; A^{[0]}_j = -\sqrt{\frac{1}{3}}\sigma^z_j, \; \; A^{[-1]}_j = -\sqrt{\frac{2}{3}}\sigma^-_j\nonumber
\end{align}
where $\sigma$ are Pauli matrices. Next, we examine the action of $P_{2j}$ on local insertions of the quasiparticle operator to arbitrary power $n_{2j}$, $\left(Q_{2j}^{\dagger} - Q_{2j+1}^{\dagger}\right)^{n_{2j}}\ket{\psi_A}_{2j,2j+1}$. The action of $Q_j^{\dagger}$ on $A_j$ is given by the tensor
\begin{align}
B_j^{[m_j]} = \sum_{m_p = -1,0,1}((S_j^+)^2)_{m_j,m_p}A_j^{[m_p]}
\end{align}
leading to 
\begin{align}
B^{[1]}_j = -2\sqrt{\frac{2}{3}}\sigma^-_j, \; \; 
B^{[0]}_j= \boldsymbol{0}, \; \; 
B^{[-1]}_j= \boldsymbol{0}.
\end{align}
From the above, it is clear that $Q_j^{\dagger}B_j = 0$,  $\ket{B_jB_{j+1}} = 0$, and $P_{2j}\ket{\psi_A}_{2j,2j+1} = 0$ by construction; thus, the only non-trivial term to compute corresponds to the case $n_j = 1$. This needs to be checked against Eqs.~(\ref{eq:MPScondition}),(\ref{eq:IntegerCondition}). An explcit calculation can be used to confirm that 
%\begin{widetext}
\begin{align}
P_{2j} & \left(\ket{B_{2j}A_{2j+1}} - \ket{A_{2j}B_{2j + 1}}\right) \nonumber\\ %=\sum_{n,m}(P_{2j,2j+1})_{m_{2j},m_{2j+1},n,m}(B^{[n]}A^{[m]} - A^{[n]}B^{[m]})\ket{m_{2j},m_{2j + 1}} =
&=\left(\ket{B_{2j}A_{2j+1}} - \ket{A_{2j}B_{2j + 1}}\right) 
\end{align}
%\end{widetext}
which implies 
\begin{align}
P_{2j}\left(\sum_{p = 2j,2j+1}(-1)^pQ_{p}^{\dagger}\right)^{n_{2j}}\ket{\psi_A}_{2j,2j+1} =\nonumber \\ n_{2j}\left(\sum_{p = 2j,2j+1}(-1)^pQ_{p}^{\dagger}\right)^{n_{2j}}\ket{\psi_A}_{2j,2j+1}
\end{align}
for any $n_{2j}$. Clearly, this also holds for $P_{2j + 1}$, which per Eq. (\ref{eq:MPScondition}) and Eq. (\ref{eq:IntegerCondition}) shows that the states $(Q^{\dagger})^l\ket{\psi_A}$ are common eigenstates of both $O_1$ and $O_2$ and that the time evolution of any linear combination of the states $(Q^{\dagger})^l\ket{\psi_A}$ evolves according to the unitary operator $e^{-iO_1t}e^{-iO_2t}$.

Note that $O_1$ and $O_2$ are sums of spatially decoupled projectors, and are thus trivially diagonalizable. Their spectrum is composed of positive integer values ranging from 0 to $L/2$. This allows for revivals at some time $t_*$. Specifically,  
\begin{align}
    e^{-iH_{\text{AKLT}}t}\sum_{l = 0}^{L-1}c_l\ket{\psi_l} = e^{-iO_1t}e^{-iO_2t}\sum_{l = 0}^{L-1}c_l\ket{\psi_l},
\end{align} 
for arbitrary coefficient $c_l$, and revivals occur with period $t_* = 2 \pi$ given the fact that the eigenvalues of $O_1$ and $O_2$ are integers. 

One can alternatively corroborate the fact that the states $(Q^{\dagger})^l\ket{\psi_A} = \ket{\psi_l}$ are common eigenstates of $O_1, O_2$ by showing that these simple partitions satisfy the same spectrum generating algebra~\cite{MARK2020_ResctrictedSGA} as $H_{\text{AKLT}}$ itself. Thus, $[O_1,Q^{\dagger}]\ket{\psi_l} = [O_2,Q^{\dagger}]\ket{\psi_l} =  Q^{\dagger}\ket{\psi_l}$, see App. \ref{app:AKLTAlgebraProof} for the calculation details.
%In particular, there exists a subspace $W$ which includes the ground state $\ket{G}$
%, an eigenstate of $H$, $\ket{G} \in W$ with energy $E_G$ (assumed here to be the ground state of the AKLT model, $E_G = 0$), 
%and an operator $Q^{\dagger}$ such that $Q^{\dagger}W \subset W$ and 
%

\subsection{Domain-wall conserving model}
In Ref. \cite{IADECOLA2020_QMBSGlikeModel}, it was shown that the spin-1/2 domain-wall conserving Hamiltonian
\begin{align} H_{\text{DWC}} & =\sum_{j=1}^{L}\left[\lambda\left(\sigma_j^x-\sigma_{j-1}^z \sigma_j^x \sigma_{j+1}^z\right)+\Delta \sigma_j^z+J \sigma_j^z \sigma_{j+1}^z\right] %\nonumber\\ & \equiv H_\lambda+H_z+H_{z z}
\end{align}
hosts a scar tower given by 
\begin{equation}
\ket{\psi_l} = (\mathcal{P}^{\dagger})^l\ket{00...0}
\end{equation}
where the $\sigma^{x/y/z}_j$ are Pauli matrices acting on site $j$ such that $\sigma_j^z\ket{0}_j = -\ket{0}_j$, $\sigma_j^z\ket{1}_j = \ket{1}_j$. Periodic boundary conditions are assumed.

The operator $\mathcal{P}^{\dagger}$ is given by
\begin{equation}
\mathcal{P}^{\dagger}=\sum_{j=1}^{L}(-1)^j P_{j-1}^0 \ket{1}_j\bra{0}_j P_{j+1}^0
\end{equation}
with $P_{j}^0 = \frac{(1 - \sigma_j^z)}{2}$ and is composed of a sum of local terms that flip the spin-1/2 at site $j$ form a $\ket{0}$ to a $\ket{1}$ provided both neighboring spins are in the state $\ket{0}$. The states $\ket{\psi_l}$ can then be viewed as the result of acting with the operator $(\sum_{j= 1}^L(-1)^j\sigma_j^+)^l$ on the state $\ket{00...0}$, and then projecting on the blockade-free subspace (containing no adjacent $\ket{11}_{j,j+1}$ for any site $j$).  

We will now discuss the appropriate partitioning of this Hamiltonian into eigenoperators which satisfy the conditions of Eqs.~(\ref{eq:MPScondition}),(\ref{eq:IntegerCondition}). It is also worth noting that the quasiparticle operator $\mathcal{P}^\dagger$ appears to be constructed out of local terms which spatially overlap. We will show that this operator can be traded with a simpler operator $Q^\dagger$ which is composed of single site operations exclusively, which is a requirement in the derivation of Eqs.~(\ref{eq:MPScondition}),(\ref{eq:IntegerCondition}). 

In particular, we first rewrite the term proportional to $\lambda$ in $H_{\text{DWC}}$ as
\begin{align}
\lambda\left(\sigma_j^x-\sigma_{j-1}^z \sigma_j^x \sigma_{j+1}^z\right) =
(h_L)_{j-1,j,j+1} + (h_R)_{j-1,j,j+1}
\end{align}
where 
\begin{align}
(h_R)_{j} &\equiv \lambda(\ket{110}\bra{100} + \text{h.c})_{j-1.j,j+1} \nonumber\\
(h_L)_{j} &\equiv \lambda(\ket{011}\bra{001} + \text{h.c})_{j-1,j,j+1}.
\end{align}
This decomposition shows that the Hamiltonian terms $\lambda\left(\sigma_j^x-\sigma_{j-1}^z \sigma_j^x \sigma_{j+1}^z\right)$ can be decomposed into two parts, one that extends/reduces a domain wall by one site to the right, which we call $h_R$, and another which extends/reduces it by one site on the left, which we call $h_L$. We now define the following partition 
\begin{align}
o_{1,j} &= (h_L)_{j} + (h_R)_{j+1}, \nonumber\\
O_1 &= \sum_{j=1}^{L}o_{1,j}\nonumber\\
O_2 &= \sum_j\Delta \sigma_j^z\nonumber\\ 
O_3 &= \sum_{j \in \text{even}}J\sigma_j^z\sigma_{j+1}^z\nonumber\\
O_4 &= \sum_{j \in \text{odd}}J\sigma_j^z\sigma_{j+1}^z
\end{align}
such that 
\begin{equation}
H_{\text{DWC}} = O_1 + O_2 + O_3 + O_4.
\end{equation}

In principle, one may consider $O_2 + O_3 + O_4$ as a single eigenoperator since the constituent terms commute on the entire Hilbert space. However, the above decomposition provides a convenient partitioning in terms extensive operators $O_2,O_3,O_4$ that are composed of non-overlapping local operators, in which case the type I rules apply straightforwardly.

To this end, we will now make use of the rules of Eqs. (\ref{eq:MPScondition}),(\ref{eq:IntegerCondition}) to show that the time-evolution characterizing the scar subspace spanned by the states $\ket{\psi_l}$ is given by $U_s(t) = e^{-iO_2t}e^{-iO_3t}e^{-iO_4t}$. 

To do so, first note that the base scar state (the fully polarized state) can be written in MPS notation as 
\begin{equation}
\ket{\psi_A} = \ket{[AA...A]}
\end{equation}
with 
\begin{align}
A^{[1]}_j = \begin{pmatrix} 0 & 0 \\ 0 & 0 \end{pmatrix}, \; \; 
A^{[0]}_j = \begin{pmatrix}
1 & 0 \\
1 & 0
\end{pmatrix}, 
\end{align}
and an excitation, as we show, can be represented by the tensor 
\begin{align}
B^{[1]}_j = \begin{pmatrix}
0 & 1 \\
0 & 0
\end{pmatrix}, \; \; 
B^{[0]}_j = \begin{pmatrix}
0 & 0 \\
0 & 0
\end{pmatrix},
\end{align}
which is created on top of the base state using the map 
\begin{align}
    Q_j^{\dagger} &: A_j \rightarrow B_j, \nonumber \\
    &: B_j \rightarrow 0.  
\end{align}
The quasiparticle creation operator is given by
\begin{equation}
Q^{\dagger} = \sum_j(-1)^jQ_j^{\dagger}.
\end{equation}
Note that one can explicitly check that $A^{[0]}_{j-1}B^{[1]}_jA^{[0]}_{j+1} = A^{[0]}_j$, $B^{[1]}_jB^{[1]}_{j+1} = 0$ and $A^{[0]}_jA^{[0]}_{j+1} = A^{[0]}_j$. With these, it can be shown that 
\begin{equation}
(Q^{\dagger})^l\ket{\psi_A} = (\mathcal{P}^{\dagger})^l\ket{\psi_A}
\end{equation}
which implies that the operator $\mathcal{P}^{\dagger}$, which is composed of local terms acting on three adjacent sites, can be traded for the map $Q^{\dagger}$ composed of single-site maps $Q_j^{\dagger}$. This works due to the property $B^{[1]}_jB^{[1]}_{j+1} = 0$ which ensures that no blockades can be present in the final state, whilst the properties $A^{[0]}_{j-1}B^{[1]}_jA^{[0]}_{j+1} = A^{[0]}_j$ and $A^{[0]}_jA^{[0]}_{j+1} = A^{[0]}_j$ ensure that all the allowed states have the same weight. 

We are now in a position to confirm the conditions of Eqs.~(\ref{eq:MPScondition}),(\ref{eq:IntegerCondition}). Let us start with the terms $o_{1,j}$. It is straightforward to show that
\begin{align}
o_{1,j}\left(\sum_{p = j-1,j,j+1,j+2} (-1)^pQ_p^{\dagger}\right)^{n_j}\ket{\psi_A}_{j-1,j,j+1,j+2}  =  0
\end{align}
for all $n_j$. Indeed, only the case $n_j = 1$ and $n_j = 2$ need to be verified, as for $n_j = 3$ and higher, there are no non-vanishing configurations that don't have adjacent $B^{[1]}$, which all vanish. The case $n_j=1$ and $n_j=2$ are easily verified by direct computation. This shows that the states $\ket{\psi_l}$ are 0 energy eigenstates of all the $o_{1,j}$. For the operator $O_2$, we have that
\begin{align}
\Delta\sigma_j^z\left((-1)^{j}Q_j^{\dagger}\right)^{n_j}\ket{\psi_A}_j = \nonumber\\(-\Delta + 2n_j\Delta)\left((-1)^{j}Q_j^{\dagger}\right)^{n_j}\ket{\psi_A}_j
\end{align}
which shows that the states $\ket{\psi_l}$ are eigenstates of $O_2$, as per Eqs.~(\ref{eq:MPScondition}),(\ref{eq:IntegerCondition}). Finally, we turn to the operators $O_3$ and $O_4$. We can show that 
\begin{align}
J\sigma_j^z\sigma_{j+1}^z\left(\sum_{p = j,j+1} (-1)^pQ_p^{\dagger}\right)^{n_j}\ket{\psi_A}_{j,j+1} = \nonumber \\
(J - 2n_jJ)\left(\sum_{p = j,j+1} (-1)^pQ_p^{\dagger}\right)^{n_j}\ket{\psi_A}_{j,j+1}
\end{align}
which proves that the states $\ket{\psi_l}$ are eigenstates of $O_3$. An analogous result holds for $O_4$ due to the translation symmetry of the scar states (translation by one site simply multiplies the scar states by either a factor of $-1$ or $1$). This completes the proof that the time evolution of the scar subspace is given by the unitary operator $U_s(t) = e^{-iO_2t}e^{-iO_3t}e^{-iO_4t}$. The eigenvalues of the $\ket{\psi_l}$ are given by $2l(\Delta -2J) + L(J -\Delta )$, and perfect revivals occur after a time $t_* = \frac{\pi}{(\Delta -2J)}$
\subsection{Eta pairing and generalizations}\label{sec:GeneralizedHubbard}

In this section, we consider eta pairing scar eigenstates in the Hubbard model~\cite{YANG1989_EtaPairing,ZHANG1990_HubbardPseudoSU2,YANG1990_HubbardSO4} and discuss their interpretation in the broken unitary picture. 
%The Hubbard model is given by 
%\begin{equation}
%\begin{aligned}
%H_{\mathrm{Hub}}=\sum_{\sigma \in\{\uparrow, \downarrow\}}-t \sum_{\left\langle\boldsymbol{r}, \boldsymbol{r}^{\prime}\right\rangle}\left(c_{r, \sigma}^{\dagger} c_{\boldsymbol{r}^{\prime}, \sigma}+h . c\right)\\-\mu \sum_r c_{r, \sigma}^{\dagger} c_{r, \sigma}+U \sum_r \widehat{n}_{\boldsymbol{r}, \uparrow} \widehat{n}_{\boldsymbol{r}, \downarrow}
%\end{aligned}
%\end{equation}

We consider a generalization of the Hubbard model~\cite{MOUDGALYA2020_GeneralizedHubbard} given by 
\begin{align}
H_{\text{Hubbard}} =&-\sum_{\sigma, \sigma^{\prime}} \sum_{\left\langle\boldsymbol{r}, \boldsymbol{r}^{\prime}\right\rangle} \left(t_{\boldsymbol{r}, \boldsymbol{r}^{\prime}}^{\sigma, \sigma^{\prime}} c_{\boldsymbol{r}, \sigma}^{\dagger} c_{\boldsymbol{r}^{\prime}, \sigma^{\prime}}+t_{\boldsymbol{r}^{\prime}, \boldsymbol{r}}^{\sigma^{\prime}, \sigma} c_{\boldsymbol{r}^{\prime}, \sigma^{\prime}}^{\dagger} c_{\boldsymbol{r}, \sigma}\right) \nonumber \\
&-\sum_{\boldsymbol{r}, \sigma} \mu_{\boldsymbol{r}, \sigma} \hat{n}_{\boldsymbol{r}, \sigma}+\sum_{\boldsymbol{r}} U_{\boldsymbol{r}} \hat{n}_{\boldsymbol{r}, \uparrow} \hat{n}_{\boldsymbol{r}, \downarrow},
\end{align}

defined on an arbitrary graph. The usual Hubbard model may be obtained by setting $t_{\boldsymbol{r},\boldsymbol{r}'}^{\sigma,\sigma'} = t\delta_{\sigma,\sigma'}, \mu_{\boldsymbol{r,\sigma}} = \mu, U_{\boldsymbol{r}} = U$.

We want to ask under what choice of hopping amplitudes, chemical potential, and interaction strength one may obtain a spectrum supporting many-body scars. 
%In the common eigenstates approach, the more the number of operators $A, B, ...$, the harder it is to find common eigenstates. Thus, in general, we want to minimize the number of eigenoperators considered.
In this case, it is natural to club the chemical potential and interaction terms into one eigenoperator, say $O_2$, since these terms commute on the entire Hilbert space, and not just over a subspace. The goal of the broken unitary picture is to generate a scar subspace described by a time evolution operator that is much simpler than the original $e^{-iHt}$. The most straightforward way to achieve this here is to require that the hopping terms $o_{1,{\boldsymbol{r},\boldsymbol{r}'}} =-\sum_{\sigma, \sigma^{\prime}} \left(t_{\boldsymbol{r}, \boldsymbol{r}^{\prime}}^{\sigma, \sigma^{\prime}} c_{\boldsymbol{r}, \sigma}^{\dagger} c_{\boldsymbol{r}^{\prime}, \sigma^{\prime}}+t_{\boldsymbol{r}^{\prime}, \boldsymbol{r}}^{\sigma^{\prime}, \sigma} c_{\boldsymbol{r}^{\prime}, \sigma^{\prime}}^{\dagger} c_{\boldsymbol{r}, \sigma}\right)$ act as individual eigenoperators and vanish when acting onto the scar subspace. 
 
For definiteness, we have
\begin{align}
o_{2,\boldsymbol{r},\boldsymbol{r}'} &=  \sum_{\vs{d} = \vs{r}, \vs{r}'}\left(\sum_\sigma \mu_{\boldsymbol{d},\sigma}\hat{n}_{\boldsymbol{d},\sigma} + U_{\boldsymbol{d}} \hat{n}_{\boldsymbol{d}, \uparrow} \hat{n}_{\boldsymbol{d}, \downarrow}  \right), \nonumber\\ 
%&-\mu_{\boldsymbol{r}',\sigma}\hat{n}_{\boldsymbol{r}',\sigma} +  U_{\boldsymbol{r}'} \hat{n}_{\boldsymbol{r}', \uparrow} \hat{n}_{\boldsymbol{r}', \downarrow} \right)\nonumber\\
o_{1,\boldsymbol{r},\boldsymbol{r}'} &=-\sum_{\sigma, \sigma^{\prime}} \left(t_{\boldsymbol{r}, \boldsymbol{r}^{\prime}}^{\sigma, \sigma^{\prime}} c_{\boldsymbol{r}, \sigma}^{\dagger} c_{\boldsymbol{r}^{\prime}, \sigma^{\prime}}+t_{\boldsymbol{r}^{\prime}, \boldsymbol{r}}^{\sigma^{\prime}, \sigma} c_{\boldsymbol{r}^{\prime}, \sigma^{\prime}}^{\dagger} c_{\boldsymbol{r}, \sigma}\right).\nonumber\\
\end{align}
Thus, 
\begin{align}
O_2 &= -\sum_{\boldsymbol{r},\sigma}\mu_{\boldsymbol{r},\sigma}\hat{n}_{\boldsymbol{r},\sigma} + \sum_{\boldsymbol{r}} U_{\boldsymbol{r}} \hat{n}_{\boldsymbol{r}, \uparrow} \hat{n}_{\boldsymbol{r}, \downarrow}\nonumber\\
O_1 &= \sum_{\left \langle \boldsymbol{r},\boldsymbol{r}' \right \rangle}o_{1,\boldsymbol{r},\boldsymbol{r}'}\nonumber\\\nonumber\\
H_{\text{Hubbard}} &= O_1 + O_2
\end{align}

%To obtain recurrences, we expect $e^{-iAt}$ to become trivial for some time $t$ when acting on the scar states $\ket{\psi_l}$.

It is not hard to see that the state $\ket{\psi_A}$ completely devoid of particles is a zero eigenvalue eigenstate of both the hopping terms and the operator $O_2$. Let us now assume the existence of a creation operator $Q^{\dagger} = \sum_{\boldsymbol{r}}Q^{\dagger}_{\boldsymbol{r}}$, where $Q^{\dagger}_{\boldsymbol{r}}$ are single site operators. By making use of Eq.~(\ref{eq:localRules}), we see that if the conditions
\begin{equation}\label{eq:CommConditionHubbardLocal}
\begin{aligned} o_{1,\boldsymbol{r},\boldsymbol{r}'}o_{2,\boldsymbol{r},\boldsymbol{r}'}^{n}(Q_{\vs{r}}^{\dagger} + Q_{\vs{r'}}^{\dagger})^m\ket{\psi_A} = 0,
\end{aligned}
\end{equation}
are satisfied for any integers $n,m \ge 0$, then the model will host a tower of scar states that evolve in time according to the unitary $e^{-iO_2t}$.
For the $Q^{\dagger}_{\vs{r}}$ terms, we consider
\begin{equation}
\begin{aligned}
Q^{\dagger}_{\vs{r}} = q_{\vs{r},\uparrow}c^{\dagger}_{\vs{r},\uparrow} + q_{\vs{r},\downarrow}c^{\dagger}_{\vs{r},\downarrow} + q_{\vs{r},\uparrow\downarrow}c^{\dagger}_{\vs{r},\uparrow}c^{\dagger}_{\vs{r},\downarrow}
\end{aligned}
\end{equation}
which in principle is a mixture of both bosonic and fermionic excitations. We find that the type II rules are only satisfied when the excitations are either fermionic or bosonic if $H_{\text{Hubbard}}$ is assumed to be hermitian. However, solutions where both bosonic and fermionic excitations coexist do exist if $H_{\text{Hubbard}}$ is allowed to be non-hermitian, see App. \ref{app:HubbardConditions} for more details.

It is easy to see that $(Q^{\dagger}_{\boldsymbol{r}} + Q^{\dagger}_{\boldsymbol{r}'})^{l} = 0$ for $l > 2$. The case $l = 0$ (the condition on $\ket{\psi_A}$) is trivially satisfied. The case $q_{\vs{r},\uparrow} = q_{\vs{r},\downarrow} = 0$ with $n = 0$ leads to the set of conditions (see App. \ref{app:HubbardConditions})
\begin{equation}\label{eq:Hubbard0energyconditionText}
\begin{aligned}
q_{\boldsymbol{r},\uparrow\downarrow}t_{\boldsymbol{r}',\boldsymbol{r}}^{\uparrow,\uparrow} + q_{\boldsymbol{r}',\uparrow\downarrow}t_{\boldsymbol{r},\boldsymbol{r}'}^{\downarrow,\downarrow}= 0\\
q_{\boldsymbol{r}',\uparrow\downarrow}t_{\boldsymbol{r},\boldsymbol{r}'}^{\uparrow,\uparrow} + q_{\boldsymbol{r},\uparrow\downarrow}t_{\boldsymbol{r}',\boldsymbol{r}}^{\downarrow,\downarrow} = 0\\
q_{\boldsymbol{r},\uparrow\downarrow}t_{\boldsymbol{r}',\boldsymbol{r}}^{\uparrow,\downarrow} - q_{\boldsymbol{r}',\uparrow\downarrow}t_{\boldsymbol{r},\boldsymbol{r}'}^{\uparrow,\downarrow}=0 \\
q_{\boldsymbol{r}',\uparrow\downarrow}t_{\boldsymbol{r},\boldsymbol{r}'}^{\downarrow,\uparrow} - q_{\boldsymbol{r},\uparrow\downarrow}t_{\boldsymbol{r}',\boldsymbol{r}}^{\downarrow,\uparrow} = 0
\end{aligned}
\end{equation}

%\begin{equation}
%\begin{aligned}
%q_{\boldsymbol{r}} t_{\boldsymbol{r}^{\prime}, \boldsymbol{r}}^{\sigma^{\prime}, \sigma} s_\sigma+q_{\boldsymbol{r}^{\prime}} t_{\boldsymbol{r}, \boldsymbol{r}^{\prime}}^{\bar{\sigma}, \bar{\sigma}^{\prime}} s_{\sigma^{\prime}}=0 \quad \forall \sigma, \sigma^{\prime}, \quad \forall\left\langle\boldsymbol{r}, \boldsymbol{r}^{\prime}\right\rangle, \\ s_\sigma \equiv\left\{\begin{array}{cc}
%+1 & \text { if } \sigma=\uparrow \\
%-1 & \text { if } \sigma=\downarrow
%\end{array}, \quad \bar{\sigma} \equiv\left\{\begin{array}{ll}
%\downarrow & \text { if } \sigma=\uparrow \\
%\uparrow & \text { if } \sigma=\downarrow
%\end{array}\right.\right.
%\end{aligned}
%\end{equation}
%which reduces to 
%\begin{equation}
%\begin{aligned}
%t_{\vs{r},\vs{r'}}^{\sigma,\sigma'}t_{\vs{r'},\vs{r}}^{\sigma',\sigma}= t_{\vs{r},\vs{r'}}^{\bar{\sigma},\bar{\sigma}'}t_{\vs{r'},\vs{r}}^{\bar{\sigma}',\bar{\sigma}}\\
%\frac{q_{\vs{r}}}{q_{\vs{r'}}} = \frac{t_{\vs{r},\vs{r'}}^{\sigma,\sigma'}}{t_{\vs{r'},\vs{r}}^{\bar{\sigma},\bar{\sigma'}}}
%\end{aligned}
%\end{equation}
In this case, the remaining conditions (\ref{eq:CommConditionHubbardLocal})
\begin{equation}
\begin{aligned}
o_{1,\boldsymbol{r},\boldsymbol{r}'}o_{2,\boldsymbol{r},\boldsymbol{r}'}^n(Q_{\boldsymbol{r}}^{\dagger} + Q_{\boldsymbol{r}'}^{\dagger})^m\ket{\psi_A} = 0
\end{aligned}
\end{equation}
for $n>0$ lead to
\begin{equation}\label{eq:LocalEnergyCondition}
(U_{\boldsymbol{r}} - \mu_{\boldsymbol{r},\uparrow} - \mu_{\boldsymbol{r},\downarrow}) = \mathcal{E}
\end{equation}
for an arbitrary scalar $\mathcal{E}$ (see App. \ref{app:HubbardConditions} for more details). The conditions (\ref{eq:Hubbard0energyconditionText}) and (\ref{eq:LocalEnergyCondition}) are completely equivalent to the conditions derived in Ref. \cite{MOUDGALYA2020_GeneralizedHubbard} and were found here using the broken unitary picture.
The case where $q_{\vs{r},\uparrow,\downarrow} = 0$ leads to $(Q^{\dagger}) ^2 = 0$. The only excited state $Q^\dagger\ket{\psi_A}$ then lives in the single-particle sector, which would comprise a trivial example of a scar `tower'.
\\\\
Satisfaction of both Eqs.~(\ref{eq:Hubbard0energyconditionText}) and~(\ref{eq:LocalEnergyCondition}) implies that the unitary describing the time evolution onto the scar subspace reduces to $U_s(t) = e^{-iO_2t}$, as advertised. 
The trivially integrable operator $O_2 = -\sum_{\boldsymbol{r}, \sigma} \mu_{\boldsymbol{r}, \sigma} \hat{n}_{\boldsymbol{r}, \sigma}+\sum_{\boldsymbol{r}, \sigma} U_{\boldsymbol{r}} \hat{n}_{\boldsymbol{r}, \uparrow} \hat{n}_{\boldsymbol{r}, \downarrow}$ has the states $\ket{\psi_l}$ for eigenstates, e.g $O_2\ket{\psi_l} = l\mathcal{E}\ket{\psi_l}$. 
%Furthermore, one has that the few body operators $B_{\boldsymbol{r},\boldsymbol{r}'} = \sum_{\sigma, \sigma^{\prime}} \left(t_{\boldsymbol{r}, \boldsymbol{r}^{\prime}}^{\sigma, \sigma^{\prime}} c_{\boldsymbol{r}, \sigma}^{\dagger} c_{\boldsymbol{r}^{\prime}, \sigma^{\prime}}+t_{\boldsymbol{r}^{\prime}, \boldsymbol{r}}^{\sigma^{\prime}, \sigma} c_{\boldsymbol{r}^{\prime}, \sigma^{\prime}}^{\dagger} c_{\boldsymbol{r}, \sigma}\right)$ satisfy the property $B_{\boldsymbol{r},\boldsymbol{r}'} (\eta^{\dagger})^l\ket{\psi_s} =(\eta^{\dagger})^ lB_{\boldsymbol{r},\boldsymbol{r}'} \ket{\psi_s} = 0 $
%\begin{equation}
%    \begin{aligned}
%H_{\text{Hubbard}}=- \sum_{\left\langle\boldsymbol{r}, \boldsymbol{r}^{\prime}\right\rangle} o_{\boldsymbol{r},\boldsymbol{r}'}+O_1
%\end{aligned}
%\end{equation}
%from which we directly find that the state $\ket{\psi_l}$ are eigenstates of $H_{\text{gen}}$ with eigenvalues $l\mathcal{E}$, Note that the above properties further 
This has the interesting consequence that the time evolution of any linear superposition of the states $\ket{\psi_l}$ 
%\begin{equation}
%\begin{aligned}
%    e^{-iH_{\text{Hubbard}}t}\sum_{l = 1}c_l\ket{\psi_l} = e^{-iAt}\sum_{l = 1}c_l\ket{\psi_l}
%\end{aligned}
%\end{equation}
%shows that the dynamical properties within the scar subspace are fully determined by the simple unitary $U(t) = e^{-iA}(t)$. All the states composed exclusively of pairs of occupied spin up and down states on some given sites are separated in energy by an integer multiple of $\mathcal{E}$ thanks to the condition $U_{\boldsymbol{r}} - \mu_{\boldsymbol{r},\uparrow} - \mu_{\boldsymbol{r},\downarrow} = \mathcal{E}$. The states $\ket{\psi_l}$ are part of this subspace, and thus exhibit exact revivals after a time 
shows exact revival after a time
$t = \frac{2\pi}{\mathcal{E}}$. 
Finally, it is worth noting that further generalizations~\cite{VAFEK2017_ShiftedHubbard} have been discussed, which can also be understood using the scheme outlined above. 

\subsection{Spin-1 XY magnets}
We now study a tower of scar states discussed in Ref.~\cite{SCHECTER2019_Spin1XYScarTower}. The Hamiltonian of interest, dubbed the spin-1 XY model, is given by
\begin{equation}
H_{\text{XY}}=J \sum_{\langle p j\rangle}\left(S_p^x S_j^x+S_p^y S_j^y\right)+h \sum_j S_j^z+D \sum_j\left(S_j^z\right)^2
\end{equation}
where the spin operators act on spin-1 degrees of freedom defined onto some hyper-cubic lattice of dimension $d$. A tower of scar eigenstates was found in Ref.~\cite{SCHECTER2019_Spin1XYScarTower} to be given by
\begin{equation}
\left|\mathcal{\psi}_l\right\rangle\propto \left(Q^{\dagger}\right)^l|\psi_A\rangle
\end{equation}
where the operator $Q^{\dagger}$ is defined as 
\begin{equation}
Q^{\dagger}=\frac{1}{2} \sum_j e^{i \boldsymbol{r}_j \cdot \vs{\pi}}\left(S_j^{+}\right)^2
\end{equation} 
with the vector $\boldsymbol{\pi} = (\pi, \pi, ..., \pi)$. Here, the base state $\ket{\psi_A}$ is the fully polarized state $\otimes_j\ket{-1}_j$, where $S_{j}^z\ket{m}_j = m\ket{m}_j$, and the quasiparticle is a $\pi-$momentum superposition of a single spin flipped from the $\ket{-1}_j$ state to the $\ket{+1}_j$ state. This particular scar tower is found in complete analogy to that in the Hubbard model, and here we simply discuss the partition of $H$ into the relevant eigenoperators of the scar eigenstates. 

We define the operators 
\begin{align}
o_{2,p,j} &= h(S_p^z + S_j^z) + D((S_p^z)^2 + (S_j^z)^2) \nonumber \\
o_{1,p,j} &= \frac{1}{2}\left(S_p^+S_{j}^- + S_p^-S_{j}^+\right) \nonumber \\
O_2 &= h\sum_{j}S_j^z + D\sum_j(S_j^z)^2 \nonumber \\
O_1 &= \sum_{\left \langle p,j \right \rangle}o_{1,p,j}\nonumber \\
Q_{p,j}^{\dagger} &= \frac{e^{i\boldsymbol{r}_p\cdot\boldsymbol{\pi}}}{2}\left(\left(S_p^+\right)^2 -\left(S_j^+\right)^2  \right)
\end{align}
where $p,j$ in $Q_{p,j}^{\dagger}$ are assumed to be nearest neighbors.
Analogously to the Hubbard model, the eigenoperators will be $O_2$ and the $o_{1,p,j}$, where it will be assumed that each $o_{1,p,j}$ vanishes when acting on the scar subspace. Using Eq.~(\ref{eq:GSMform}) 
%with the substitution $v_i = hS_i^z + D(S_i^z)^2$, $h_i = B_{i,j}$, $Q_i^\dagger = \frac{e^{i\vs{r}_i\cdot\vs{\pi}}}{2}S_i^+$
, we see that we are seeking a relation of the form 
\begin{equation}\label{eq:CommSpin1XY}
\begin{aligned}
o_{1,p,j}o_{2,p,j}^n(Q_{p,j}^{\dagger})^m\ket{\psi_A} = 0
\end{aligned}
\end{equation}
for arbitrary integers $n,m \ge 0$. Now note that $(Q_{p,j}^{\dagger})^m$ is $0$ for $m > 2$, and so this condition only needs to be verified for $m = 0,1,2$. A direct calculation then shows that
\begin{align}
o_{2,p,j} (Q_{p,j}^{\dagger})^m\ket{-1,-1}_{p,j} =\nonumber\\ (2D + 2h(2m-1))(Q_{p,j}^{\dagger})^m\ket{-1,-1}_{p,j}
\label{eq:LocalEXY}
\end{align}
and 
\begin{align}
o_{1,p,j}(Q_{p,j}^{\dagger})^m\ket{-1,-1}_{p,j} = 0
\end{align}

from which
it follows that Eq.~(\ref{eq:CommSpin1XY}) is satisfied. 
As a consequence, the unitary describing the time evolution within the scar subspace is simply $e^{-iO_2t}$. The scar states $\ket{\psi_l}$ have eigenvalues $\mathcal{E}_l = h(2l - L^d) + L^dD$ where $d$ is the total dimension of the lattice, and $L^d$ is total number of sites (assuming a hypercubic lattice with $L$ sites along each axis of the hypercube). 
%\textcolor{red}{this is a bit odd; usually we don't have such pieces in the energy...}.
For any linear combination of the states $\ket{\psi_l}$, the revivals are then guaranteed to occur after a time $t_* = \frac{\pi}{h}$.

\subsection{Two-site tower of states in the 1 d Spin-1 XY model}\label{sec:TwoSiteQPSpinXY}
Another tower of scar eigenstates was discovered in Ref. \cite{SCHECTER2019_Spin1XYScarTower} whenever the dimension $d = 1$ and $D = 0$, in which case the scar eigenstates are given explicitly by 
\begin{widetext}
\begin{equation}
\begin{aligned}
\left|\psi_l\right\rangle \propto  \sum_{j_1 \neq j_2 \neq \cdots \neq j_l}(-1)^{j_1+\cdots+j_l}\left(S_{j_1}^{+} S_{j_1+1}^{+}\right) \ldots\left(S_{j_l}^{+} S_{j_l+1}^{+}\right)|\psi_s\rangle
\end{aligned}
\end{equation}
\end{widetext}
where $\ket{\psi_A}$ is again the fully polarized state $\otimes_j\ket{-1}_j$. To be precise, consider the Hamiltonian~\cite{CHATTOPADHYAY2020_SpinOneHalfTwoSiteTowerScars}

\begin{equation}
\begin{aligned}
H_{\text{XY}}' &=\sum_j\left(S_j^x S_{j+1}^x+S_j^y S_{j+1}^y\right)+\epsilon V+h \sum_j S_j^z \\
\end{aligned}
\end{equation}
with 
\begin{equation}
V=\sum_j\left(S_j^{+}\right)^2\left(S_{j+1}^{-}\right)^2+\text { h.c. }
\end{equation}

The local terms $\left(S_j^{+}\right)^2\left(S_{j+1}^{-}\right)^2$ that appear in $V$ all vanish independently when acting on the tower of scar states, since two consecutive sites $j,j+1$ of the states $\ket{\psi_l}$ never admit patterns such as $\ket{-1,1}$ or $\ket{1,-1}$, which are the only states on which these local terms do not vanish. It is also known that the states $\ket{\psi_l}$ are 0-energy eigenstates of the operator $\sum_j\left(S_j^x S_{j+1}^x+S_j^y S_{j+1}^y\right)$, but the proof relies on laborious tensor network calculations~\cite{CHATTOPADHYAY2020_SpinOneHalfTwoSiteTowerScars}. The aim of this section is to show that the effective unitary operator on this scar subspace is given by $U_s (t) = e^{-iht\sum_jS_j^z}$, and that this is a direct consequence of the type II rules. 
\\\\
First, it is shown in App.~\ref{app:MultiSiteTowerProof} that for any state $\ket{\psi}$ defined on the spin-1 lattice, one has that 
\begin{equation}
    e^{-iH_{\text{XY}}'t}\ket{\psi} = Me^{-i(O_1 + O_2)t}K\ket{\psi}
\end{equation}
Here, $M$ and $K$ are time-independent mappings between a spin-1 chain of length $L$ and a related spin-$1/2$ chain of length $2L$. Explicitly, the mapping is given by $M = \otimes_{j = 1}^LM_{2j-1,2j}$ and $K = \otimes_{j = 1}^LK_{2j-1,2j}$ with
\begin{align}
    M_{2j-1,2j} &= \ket{1}_j\bra{\uparrow\uparrow}_{2j-1,2j}+\ket{0}_j(\bra{\uparrow \downarrow} \nonumber + \bra{\downarrow \uparrow })_{2j-1,2j} \\
    & + \ket{-1}_j\bra{\downarrow\downarrow}_{2j-1,2j} \nonumber \\
    K_{2j-1,2j} &= \ket{\uparrow\uparrow}_{2j-1,2j}\bra{1}_j +\frac{(\ket{\uparrow \downarrow} + \ket{\downarrow \uparrow })_{2j-1,2j}}{2}\bra{0}_j\nonumber\\
    & + \ket{\downarrow\downarrow}_{2j-1,2j}\bra{-1}_j.
\end{align}

The partitioning of $H_{\text{XY}}'$ is most easily seen in the spin-$1/2$ representation of $H_{\text{XY}}'$ which we denote by $\tilde{H}^{1/2}_{\text{XY}}$. Here, the Hamiltonian can be partitioned into two terms, $\tilde{H}^{1/2}_{\text{XY}} = O_1 + O_2$ as follows
\begin{align}
    o_{2,j} &= s_j^z\nonumber\\
    O_2 &= h\sum_{j = 1}^{2L}o_{2,j}\nonumber\\
    O_1 &= \sum_{j=1}^{L}o_{1,j}
\end{align}
where $s_j^z$ is such that $s_j^z\ket{\uparrow} = \frac{1}{2}\ket{\uparrow}_j$, $s^z\ket{\downarrow}_j = -\frac{1}{2}\ket{\downarrow}_j$. The explicit form of the operators $o_{1,j}$ is given in Eq.~(\ref{eq:spin1/2Rep}). We now desire to show that the unitary $e^{-i(O_1 + O_2)t}$ reduces to $U_s (t) = e^{-iO_2t}$ when acting on the spin-$1/2 $ representation of the scar states.

First, define the operator $Q^{\dagger}$ given by \cite{CHATTOPADHYAY2020_SpinOneHalfTwoSiteTowerScars}
\begin{equation}
    Q^{\dagger}  = \sum_{j = 1}^{L}(-1)^js_{2j}^{+}s_{2j + 1}^{+}
\end{equation}
where the operators $s_j^{+}$ map the states $\ket{\downarrow}_j$ to the states $\ket{\uparrow}_j$.
\\\\
It is known that the states 
\begin{equation}
    \ket{\phi_l} \propto (Q^{\dagger})^l\ket{\phi_A}
\end{equation}
 are in a one to one correspondence with the spin-$1$ scar states $\ket{\psi_l}$, where $\ket{\phi_A} = \ket{\downarrow,\downarrow,...}$. Indeed, one can show that 
\begin{equation}
    \ket{\psi_l} \propto M\ket{\phi_l}.
\end{equation}
Given the set of sites $D(j) = \{2j-2,2j-1,...,2j+3\}$, we define
\begin{align}
 Q^{\dagger}_{D(j)} &= \nonumber \\
 (-1)^{j} &\left(-s_{2j-2}^+s_{2j-1}^+ + s_{2j}^+s_{2j+1}^+\right. \left.- s_{2j+2}^+s_{2j+3}^+\right).
 \end{align}
In perfect analogy with the calculations done in the Hubbard model and for the single-site quasiparticle tower of states in the Spin-1 XY magnet, one can show that provided
\begin{align}
    o_{1,j}\left(\sum_{p\in D(j)}o_{2,p}\right)^n(Q^{\dagger}_{D(j)})^m\ket{\phi_A} &= 0
\label{eq:Spin1/2XYcommCondition}
\end{align}
for all sites $j$ and integers $n,m \ge 0$, then the time evolution within the scar subspace is fully determined by the unitary operator $U_s(t) = Me^{-iO_2t}K$.
To this end, it is shown in App. \ref{app:MultiSiteTowerProof} that
\begin{equation}
\begin{aligned}
   o_{2,j}(Q^{\dagger}_{D(j)})^{n_j}\ket{\phi_A}_{D(j)}
   \\ =  h(2n_j - 3)(Q_{D(j)}^{\dagger})^{n_j}\ket{\phi_A}_{D(j)}
\end{aligned}
\end{equation}
and that $o_{1,j}(Q_{D(j)}^{\dagger})^{n_j}\ket{\phi_A} = 0$.
\\\\
These properties when taken together directly imply that Eqs.~(\ref{eq:Spin1/2XYcommCondition}) are satisfied, which completes the proof that the two-site quasiparticles tower of scar states can be understood via the broken unitary picture. The scar subspace dynamically evolves according to the unitary $U_s(t) = e^{-iO_2t}$ in the mapped spin-$1/2$ system. To get back the spin-1 representation of the states, one simply makes use of the maps $M$ and $K$. It is straightforward to show that $e^{-ih(\sum_{j=1}^LS_j^z)t} = Me^{-h(\sum_{j=1}^{2L}s_j^z)t}K = Me^{-iO_2t}K$ (see App. \ref{app:MultiSiteTowerProof}), which shows that the time evolution of the scar states is given by the operator $U_s(t) = e^{-i(\sum_{j= 1}^LS_j^z)}$ as advertised. The eigenvalues of $h\sum_{j=1}^LS_j^z$ are separated by an integer multiple of the energy difference $\Delta E = h$. As a consequence, revivals in the scar subspace are guaranteed to occur after a time $t_* = \frac{2\pi}{h}$

\section{Relations to scar construction methods}\label{sec:Methods}
In this section, we show how many general constructions/recipes for QMBS Hamiltonians can be understood in the broken unitary picture introduced in this work. For each method, we first give a brief review of the main ideas, after which we highlight the connections to the broken unitary formalism.

%\subsubsection{Notation}

\subsection{Shiraishi-Mori construction}\label{sec:ShiraishiMoriContruction}
The Shiraishi-Mori (SM) construction~\cite{SHIRAISHI2017_EmbeddedScars} aims at embedding a set of predetermined states $\ket{\psi_s}$ into an otherwise thermalizing spectrum. To do so, one first defines a set of projectors $P_{j}$ which satisfy the property $P_{j}\ket{\psi_s} = 0 \; \forall \; j$. The SM Hamiltonian~\cite{SHIRAISHI2017_EmbeddedScars} is written as
\begin{equation}
    H _{\text{SM}}= \sum_jP_jK_jP_j + H'
\end{equation}
for arbitrary local terms $K_j$ and $H'$ such that $[P_j,H'] = 0$. Due to the arbitrary choice of $K_j$ terms, the Hamiltonian $H_{\text{SM}}$ is in general non-integrable. 

We can understand the presence of scar states in $H_{\text{SM}}$ by examining the matrix elements of the various operators it is made up of. The operators $P_j K_j P_j$ clearly have non-zero matrix elements only between generic/non-scar states. Next, by virtue of the condition $[P_j, H'] = 0 \; \forall \; j$, it is clear that $H'$ is block diagonal with no matrix elements connecting the subspace of scar states to that of generic states. Thus, in the scar subspace, the Hamiltonian is simply $H'$, the dynamics is given by $e^{-iH't}$, and the scar eigenstates are obtained by diagonalizing $H'$ within this subspace. If the subspace is composed of a finite number of low-entanglement states $\ket{\psi_s}$, the scar eigenstates also have low entanglement. Finally, provided $H'$ has a simple spectrum with equidistant eigenvalues, we can expect strong many-body revivals. 

Given the non-zero matrix elements of the operators composing $H_{\text{SM}}$, it is evident that scar eigenstates are common eigenstates of $H'$ (with some eigenvalue) and $\sum_j P_j K_j P_j$, with eigenvalue zero. Thus, scar eigenstates of the SM Hamiltonian $H_{\text{SM}}$ can be viewed as common eigenstates of a natural partition of the model. As it turns out, the type II rules are intimately connected to the Shiraishi-Mori construction and can be used to relax the condition $[H',P_j] = 0$ whilst still guaranteeing the existence of a decoupled subspace of the Hamiltonian that evolves in time according to the unitary operator $e^{-iH't}$. The consequences of this generalization are discussed in Sec. \ref{sec:LocalRulesExtensive}. 

\subsection{The bond algebra method}\label{sec:BondAlgebraMethod}
The bond algebra formalism~\cite{MOUDGALYA2022_BondAlgebra} aims to find all Hamiltonians hosting a predetermined set of scar eigenstates. One defines the commutant algebra $\mathcal{C}_{\text{scar}} = \langle\langle\ket{\epsilon_l}\bra{\epsilon_l}\rangle\rangle$ which includes all projection operators on a chosen set of wavefunctions $\ket{\epsilon_l}$ which are meant to be scar eigenstates of a putative Hamiltonian $H_{\text{BA}}$. Here  $\langle\langle ... \rangle\rangle$ denotes the algebra spanned by the set of operators included within the brackets.  The goal is then to find the associated bond algebra $\mathcal{A}$ (with the possible addition of a degeneracy lifting term $H_0$), which together constitute a basis of all operators $O_a$ that commute with all operators in $\mathcal{C}_{\text{scar}}$. The Hamiltonian hosting the states $\ket{\epsilon_l}$ as scar states is then any arbitrary linear combination of the operators in $\mathcal{A}$ and $H_0$. It is clear that this construction is directly related to the broken unitary picture presented here. 

\begin{figure}
\centering
\includegraphics[width=0.46\textwidth]{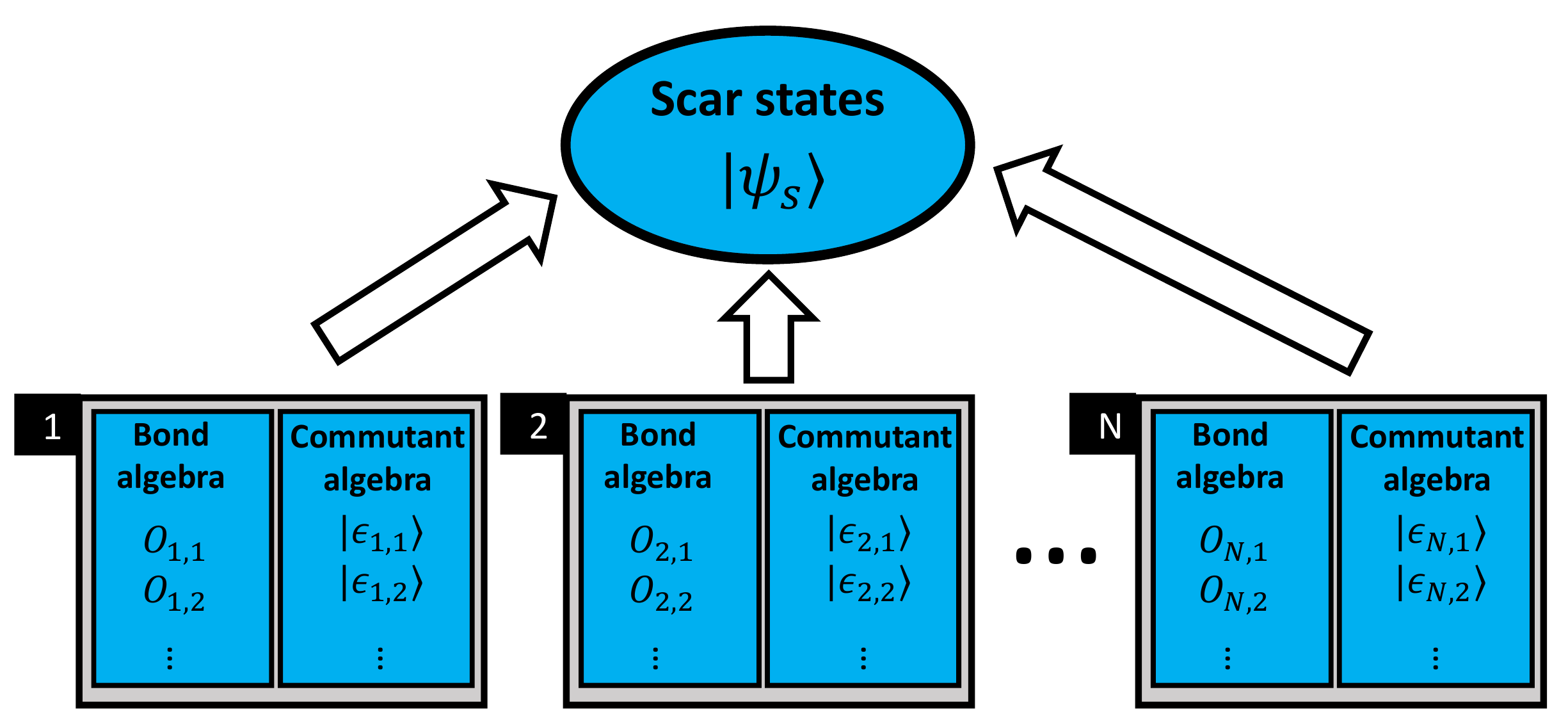}
\caption[]{Multiple commutant algebras, labeled in the figure from 1 to N, span a subspace that contains the specified scar states $\ket{\psi_s}$. The associated bond algebras can be used independently to create QMBS models hosting the scar states $\ket{\psi_s}$. Crucially, the scar states $\ket{\psi_s}$ need not be common eigenstates of the bond algebra operators and can be chosen arbitrarily.}
\label{fig:FromSeedToCommutant}
\end{figure}

The most important distinction between these approaches is that in the broken unitary approach, we do not demand prior knowledge of the states $\ket{\epsilon_l}$. In particular, one starts with some scar states $\ket{\psi_s}$ as well as some operators $O_1,O_2,...$ such that commutator of artbirary polynomials of the operators $O_1,O_2,...$ vanish when acting onto the state(s) $\ket{\psi_s}$. Satisfaction of these commutation relations then directly implies the existence of a subspace $\mathcal{S}$ which contains common eigenstates $\ket{\epsilon_l}$ of the operators $O_1,O_2,...$. The resulting common eigenstates $\ket{\epsilon_l}$ can then be viewed as a commutant algebra and the operators $O_1,O_2,...$ are elements of the corresponding bond algebra. The broken unitary picture can thus be understood as a framework encompassing all possible bond algebras that yield a commutant algebra containing eigenstates $\ket{\epsilon_l}$ from which the scar states $\ket{\psi_s}$ can be constructed; see Fig. \ref{fig:FromSeedToCommutant} for a visual representation. 

%have the states $\ket{\epsilon_l}$ as eigenstates; thus, $\mathcal{A} = \langle\langle O_i : O_i\ket{\epsilon_l} = E_{l,i}\ket{\epsilon_l}\rangle\rangle $. 
%It then follows that arbitrary nested commutators of the $O_j$ vanish when acting on $\ket{\psi_l}$, implying that the time evolution within the scar subspace $\mathcal{T} = \text{span}\{\ket{\psi_l}\}$ is determined by the unitary $U(t) = \prod_je^{-iO_jt}$. 

%By contrast, in the common eigenstates approach,  One key distinction to note here is that with the simple unitary picture, there is no need to know beforehand the scar eigenstates $\ket{\epsilon_l}$, only the seed states $\ket{\psi_l}$ are required, which may be chosen arbitrarily. 

\subsection{Group invariant formalism (GI)}
In the group invariant formalism~\cite{PAKROUSKI2020_GIformalism,PAKROUSKI2021_GIformalismApplication}, one defines an Hamiltonian of the form
\begin{equation}
    H_{\text{GI}} = \sum_a K_a T_a + H_0
\end{equation}
where the $T_a$ are the generators of a Lie group and the $K_a$ are arbitrary operators. $H_0$ satisfies the property $[H_0,C_{G^2}] = WC_{G^2}$ for some arbitrary operator $W$, where $C_{G^2}$ is the Casimir operator of the Lie group. It can be shown that any Hamiltonian of this form hosts a protected subspace $\mathcal{S} = \text{span}\{\ket{\psi_s}:C_{G^2}\ket{\psi_s} = 0\}$, or equivalently, $\mathcal{S}$ is the set of states that are common 0-energy eigenstates of the generators $T_a$. Indeed, the condition $[H_0,C_{G^2}] = WC_{G^2}$ is equivalent to requiring that $H_0\mathcal{S} \subset \mathcal{T}$, see Ref. 
 \cite{MOUDGALYA2022_BondAlgebra} and \cite{PAKROUSKI2020_GIformalism} for more details. Thus, both $\sum_aK_aT_a$ and $H_0$ keep $\mathcal{S}$ invariant and the only operator with non-zero matrix elements within $\mathcal{S}$ is $H_0$. Just like in the Shiraishi-Mori construction, one obtains perfect revivals provided the eigenstates of $H_0$ contained in $\mathcal{S}$ have equidistant eigenvalues.
\\\\
We now connect this formalism to the broken unitary picture. To do so, we start by proving that $(K_aT_a)H_0^{n}\ket{\psi_s} = 0$ for any $n \ge 0$. The condition $C_{G^2}\ket{\psi_s} = 0$ directly implies that $T_a\ket{\psi_s} = 0$. Furthermore, the condition $[H_0,C_{G^2}] = WC_{G^2}$ is equivalent to requiring that $H_0\mathcal{S} \subset \mathcal{T}$, as discussed above. These two facts together directly imply that $(K_aT_a)H_0^{n}\ket{\psi_l} = 0$. We may then choose $O_1 = \sum_aK_aT_a$ and $O_2 = H_0$. Then, we clearly have $[O_1^{n_1},O_2^{n_2}]\ket{\psi_s} = 0$, which implies that the unitary describing the time evolution of the scar subspace $\mathcal{S}$ is $U_s(t) = e^{-iO_2t}e^{-iO_1t}$. $O_1$ vanishes within the scar subspace, and thus the unitary describing the time evolution within $\mathcal{S}$ is given by $U_s(t) = e^{-iH_0t}$. As in the Shiraishi-Mori construction, a natural partition of the Hamiltonian leads to a broken unitary.

\subsection{From tunnels to tower formalism}
In this formalism~\cite{DEA2020_TunnelsToTowers}, one considers Hamiltonians that take the form 
\begin{equation}
    H_{\text{sym}}+ H_{\text{SG}} + H_{\text{A}}
\end{equation}
which is the result of a three-step process. First, one defines a Hamiltonian $H_{\text{sym}}$ that hosts a non-abelian symmetry $G$ [for instance, $\text{SU}(2)$]. This initial Hamiltonian contains multiplets of degenerate eigenstates (for instance, the eigenstates of the operator $H_{\text{sym}} = \sum_j\vec{S}_j\cdot\vec{S}_{j+1}$ with the same Casimir eigenvalue are degenerate, where the Casimir operator is $S_{\text{tot}}^2$). Second, one adds an operator $H_{\text{SG}}$ that breaks the degeneracy within the degenerate multiplets (for instance, by adding $H_{\text{SG}} = \sum_j S_j^z$ to $H_{\text{sym}} = \sum_j\vec{S}_j\cdot\vec{S}_{j+1}$). Since $H_{\text{SG}}$ only lifts the degeneracy within multiplets (it does not mix distinct multiplets), it commutes with $H_{\text{sym}}$. Finally, one chooses a multiplet (or a set of multiplets), which is spanned by some set of states $\ket{\psi_s}$. The $\ket{\psi_s}$ can be built out of common eigenstates of $H_{\text{SG}}$ and $H_{\text{sym}}$. The operator $H_A$ is then chosen such that $H_A\ket{\psi_s} = 0$. The states $\ket{\psi_s}$ then appear in this construction as scar eigenstates since a generic $H_A$ satisfying this property will most likely mix all subspaces except for the subspace spanned by the $\ket{\psi_s}$.
\\\\
This construction can also be interpreted in the broken unitary picture. From the above, we see that the states $\ket{\psi_s}$ are built out of common eigenstates of the three operators $H_{\text{sym}}$, $H_{\text{SG}}$ and $H_A$ with $H_{A}\ket{\psi_s} = 0$. The time evolution operator within the scar subspace is  then $e^{-iH_{\text{SG}}t}e^{-iH_{\text{sym}}t}$. If a single multiplet is considered, $e^{-iH_{\text{sym}}t}$ is just a phase $e^{-iEt}$ where $E$ is the eigenvalue of the states $\ket{\psi_s}$ under the action of $H_{\text{sym}}$, in which case the time evolution on the scar subspace is given by $e^{-iEt}e^{-iH_{\text{SG}}t}$.

%\section{Generalization to multi-site MPS?}
%I may have found a way, need to investigate more.
\section{Discussion and outlook}\label{sec:Discussion}
In this work, we have presented a new way of interpreting the anomalous dynamical features observed in QMBSs models, which we dub the broken unitary picture. The core postulate of this approach is that scar Hamiltonians admit a subspace of the Hilbert space  $\mathcal{S}$ that evolves in time according to a unitary $U_s(t)$ that is much simpler than the original time evolution operator $e^{-iHt}$---in particular, given a suitable partition of the scar Hamiltonian $H = \sum_aO_a$, then $U_s(t)$ corresponds to a unitary operator obtained by breaking the original $e^{-iHt}$ into a product of simpler unitary operators $e^{-iO_at}$, i.e., $U_s(t) = \prod_ae^{-iO_at}$, hence the name of the approach. More often than not, the individual terms $O_a$ are composed of local spatially decoupled terms. When this is the case, the spectrum of the operators $O_a$ becomes trivial which often leads to anomalous dynamical features such as many-body revivals.

The subspace $\mathcal{S}$ is intimately connected to the existence of common eigenstates of non-commuting operators. Indeed, we find that provided there exists at least one state $\ket{\psi_s}$ on which commutators of arbitrary polynomials of the operators $O_a$ vanish directly implies the existence of a subspace $\mathcal{S}$ composed of common eigenstates of the operators $O_a$. It follows that within this subspace, time evolution is dictated by the unitary operator $U_s(t) = \prod_ae^{-iO_at}$.
The resulting common eigenstates usually fail to satisfy the ETH as their properties cannot be accurately predicted using only global symmetries of the Hamiltonian---the subspace $\mathcal{S}$ has more symmetries than the full Hamiltonian. The common eigenstates of the operators $O_a$ can then be identified as the scar eigenstates of the QMBS model at hand. Often, the spectrum of the operators $O_a$ is simple (e.g the $O_a$ are composed of spatially decoupled terms), which is likely to produce common eigenstates $\ket{\psi_l}$ that are separated in energy by some multiple of a constant energy width $\Delta E$. In such cases, one obtains periodic revivals of the full subspace $\mathcal{S}$ after a time period $t_* = \frac{2\pi}{\Delta E}$. 

In practice, calculating the action of commutators of arbitrary polynomials of the operators $O_a$ on some state $\ket{\psi_s}$ is a challenging task. However, we show that conditions for the existence of a scar subspace $\mathcal{S}$ can be greatly simplified provided some assumptions are made about the scar states as well as the QMBS model at hand. To this end, we find three types of strictly local rules that, if satisfied, ensure the existence of a subspace $\mathcal{S}$ described by a broken unitary operator $U_s(t)$. The first set of conditions (type I rules) applies well whenever the subspace $\mathcal{S}$ is composed of common eigenstates of a finite set of extensive operators $O_a = \sum_jo_{a,j}$. The rules are designed to capture scar states $\ket{\psi_l}$ that are built by applying a quasiparticle creation operator repetitively on an MPS base state $\ket{\psi_A}$, that is, $\ket{\psi_l} \equiv (Q^{\dagger})^l\ket{\psi_A}$. Importantly, the rules derived apply to cases in which $Q^{\dagger}$ is composed of single-site operations on the MPS base state $\ket{\psi_A}$. We show using type I rules that the subspace $\mathcal{S}$ spanned by the scar eigenstates of the spin-1 AKLT model is characterized by the broken unitary operator $U_s(t)= e^{-iO_1t}e^{-iO_2t}$, where $O_1(O_2)$ is a sum over decoupled even (odd) bond projectors and $H_{\text{AKLT}} = O_1 + O_2$. Another instance where type I rules apply is in the domain-wall conserving model considered in this work. We show that the quasiparticle excitations in this model can be captured in the MPS language by an operator $Q^{\dagger}$ composed of single-site MPS operations, despite the fact that in the physical basis, the quasiparticle creation operator $\mathcal{P}^{\dagger}$ is composed of local terms that act on three consecutive sites. From there, we find an appropriate partition of $H$ which makes it evident that the type I rules are satisfied. More broadly, these rules generalize the quasiparticle-based MPS construction of scar eigenstates in scar models. 

The second set of rules (type II rules) is most appropriate when describing QMBS models for which an extensive number of local operators $h_j$ vanish when acting on the scar subspace $\mathcal{S}$, in addition to which there is a single extensive local operator $\sum_{j\in P_v}v_j$ that lifts the degeneracy of the states. We show that such rules can be directly applied to the Hubbard model as well as the Spin-1 XY magnets. In the Hubbard model, the hopping terms all individually vanish on the scar subspace whilst the Hubbard $U$ interaction combined with the local chemical potential terms act as the degeneracy lifting term. As for the spin-$1$ XY magnets, the single-site tower of scar states can also be understood via the type II rules. There, the spin hopping terms $\frac{1}{2}(S_i^+S_j^- + S_i^-S_j^+)$ vanish within the scar subspace, while the magnetic terms lift the degeneracy. For the two-site tower of quasiparticles in this model, a similar picture emerges only once it is mapped to a related spin-$1/2$ chain. The type II rules can be interpreted as a direct generalization of the original Shiraishi-Mori (SM) conditions. In the SM case, one requires that the local terms that annihilate the scar states (here these would be the $h_j$) commute with $H'$ (here, this is the degeneracy lifting term). Type II rules relax these conditions by only requiring that the operators $h_j,H'$ commute on a common invariant subspace $\mathcal{S}$; the operators can fail to commute on the full Hilbert space. We note that the Hubbard model and the spin-$1$ XY magnet already serve as examples of this generalization even though these were constructed with the SM procedure in mind. 

Finally, the last set of rules (type III rules), are formulated on a particular set of states, and is most appropriate for constructing scar models in which these chosen states are embedded in a larger scar subspace. Additionally, this set of rules amounts to requiring that the chosen states are common eigenstates of an extensive number of local operators $h_j$. Usually, the full scar subspace of the model, which contains these states, corresponds to the common eigenspace of two operators $O_1 = \sum_{j \in \text{odd}} h_j, O_2 = \sum_{j \in \text{even}} h_j$. We note that this is analogous to the scar subspace found in the AKLT model where the local projectors commute only on the extremal states of the scar subspace, but the entire scar subspace is more generally a common eigenspace of operators $O_1, O_2$ only. On the other hand, in models constructed with the Shiraishi-Mori approach, the entire scar subspace is a common eigenspace of an extensive set of local operators; in this sense, the type III rules allow for both possibilites. One potential difference with the SM approach is that the eigenvalues associated with each $h_j$ need not be equal to $0$. In previous work, the present authors \cite{ROZON2022_FloquetAutomataScars} showed that this last set of rules can be exploited to construct approximate QMBSs. Furthermore, provided the right conditions are met, these rules can also be understood in the context of quantum cellular automata. 

This work was then concluded with a discussion of the relations that exist between the broken unitary picture and the many exact QMBS constructions methods that were introduced in the literature. We show for instance that the SM construction, the bond algebra approach, the group invariant formalism as well as the tunnels to towers formalism can all be interpreted in terms of the broken unitary picture. The connection to all these methods boils down to the simple recipe of identifying the proper partition of the Hamiltonian $H = \sum_aO_a$ that will lead to the emergence of a subspace $\mathcal{S}$ characterized by the simpler unitary $U_s(t) = \prod_ae^{-iO_at}$.

There are many ways in which this work can be further extended. For instance, it is unclear to us yet how the local rules introduced in \cite{MOUDGALYA2020_MPSQuasiparticles} which can be used to build scar eigenstates from a quasiparticle creation operator $Q^{\dagger}$ composed of \emph{overlapping} local $Q_j^{\dagger}$ operators can be recast in terms of the broken unitary picture described in this work. We suspect that in such cases, a mapping to a related model might be required to uncover the appropriate partition, as in the two-site tower of quasiparticle excitations in the Spin-$1$ XY model.

%It has been shown that the existence of a single state $\ket{\psi_s}$ on which the commutator of arbitrary polynomials of the operators $O_a$ vanish implies the existence of a subspace $\mathcal{S}$ that evolves in time according to the unitary $U_s(t) = \prod_ae^{-iO_at}$ (note that the order in which the $O_a$ appear is arbitrary). A remaining open question is under which condition the converse is true, i.e under which condition does the existence of a subspace $\mathcal{S}$ characterized by the unitary $U_s(t) = \prod_ae^{-iO_at}$ (where the order in which the $O_a$ is not important) for arbitrary $t$ directly implies that the states in $\mathcal{S}$ can be written as a superposition of common eigenstates of the $O_a$? A trivial example where this does not work is with a vanishing Hamiltonian $H = 0$. Then, $H$ can be written as $H = K - K$ for arbitrary $K$. The dynamics of any state $\ket{\psi}$ in the Hilbert space can then be written as $e^{-iKt}e^{iKt}\ket{\psi} = e^{iKt}e^{-iKt}\ket{\psi} = e^{-iHt}\ket{\psi}$ for any $t$, but clearly $\ket{\psi}$ is in general not a common eigenstate $K$ and $-K$ since $K$ is arbitrary. This also raises the question of whether there can exist states $\ket{\psi_s}$ characterized by an alternate time evolution operator $U_s(t)$ without the need for the existence of common eigenstates.

Given an explicit listing of all the rules to be satisfied (in general, the set of rules is finite as discussed in \cite{ROZON2022_FloquetAutomataScars}), it would be interesting if a precise relation can be established between the degree to which the rules are satisfied and the decay of many-body revivals, and if some subset of these conditions are more relevant than others in ensuring the largest possible many-body revivals. 

We have shown in this work that the existence of a single state $\ket{\psi_s}$ onto which commutators of arbitrary polynomials of a set of operators $O_a$ vanish is sufficient to ensure the existence of a subspace $\mathcal{S}$ characterized by the unitary operator $U_s(t)$. Since the state $\ket{\psi_s}$ can in practice be arbitrary, this suggests a way of building QMBS models with predetermined scar states. For instance, the type III rules were used in \cite{ROZON2022_FloquetAutomataScars} to construct QMBS Hamiltonians in which the scar states evolve in time according to a related quantum cellular automata. It would be worthwhile to `operationalize' the type I and type II rules developed here to similarly build scar Hamiltonians with predetermined scar states and anomalous dynamical properties.

\section{Acknowledgements}

The authors acknowledge funding support from NSERC, FRQNT, and INTRIQ. The authors also acknowledge useful discussions with Sanjay Moudgalya on this work, and Michael Gullans for related previous collaborations which inspired this work.

\newpage
\appendix
\section{Bipartite spectrum generating algebra in the spin-1 AKLT model}\label{app:AKLTAlgebraProof}
In this section, we show that both the operators 
\begin{align}
    O_1 = \sum_{j=1}^{L/2}P_{2j}, \; \; O_2 = \sum_{j=1}^{L/2}P_{2j-1} 
\end{align}
individually admit a spectrum-generating algebra with the same base state $\ket{\psi_A}$. This then constitutes another proof that the scar eigenstates in the AKLT model are common eigenstates of the operators $O_1,O_2$.

Note that in the AKLT model, we have that 
\begin{align}
\left[O_1, Q^{\dagger}\right] &=\left[\sum_{j=1}^{L/2} P_{2j},\sum_{p=1}^{L}(-1)^{p}(S_{p}^{+})^{2}\right] \nonumber \\
%&= \sum_{j=1,l\in (2j,2j+1)}^{L/2}\left[P_{2j},(-1)^{l}(S_{l}^{+})^{2}\right] %\nonumber \\
&= \sum_{j=1}^{L/2}\left[P_{2j},(S_{2j}^{+})^{2}-(S_{2j+1}^{+})^{2}\right]
\end{align}

%\begin{equation}
 %   [\hat{H},Q^{\dagger}] = 2Q^{\dagger} + A, \quad A\ket{\mathcal{S}_{2n}} = 0
%\end{equation}
%for all n. 
Representing the operator $(S_{2j}^{+})^{2}-(S_{2j+1}^{+})^{2}$ in the $\ket{T_{l,m}}_{2j,2j+1}$ basis (see appendix \ref{app:CompBasisRepOfTotSpin} for more details) yields

\begin{align}
(S_{j}^{+})^{2}- & (S_{j+1}^{+})^{2} = 2 \bigg[-\ket{T_{2,1}}\bra{ T_{1,-1}} +\ket{ T_{1,1}}\bra{T_{2,-1}} \nonumber \\
&-\sqrt{2}\ket{T_{2,2}}\bra{T_{1,0}} + \sqrt{2}\ket{ T_{1,0}}\bra{ T_{2,-2}}\bigg]_{j, j+1}
\end{align}

using which
%the action of the projector can be easily computed, yielding 
%\begin{equation}
%\begin{array}{l}
%{\left[P_{2j, 2j+1},(S_{2j}^{+})^{2}-(S_{2j+1}^{+})^{2}\right]=-2(\ket{T_{2,1}}\bra{ T_{1,-1}}} \\
%+\sqrt{2}\ket{T_{2,2}}\bra{T_{1,0}}+\ket{T_{1,1}}\bra{T_{2,-1}}+\\\sqrt{2}\ket{ T_{1,0}}\bra{ T_{2,-2}})_{2j, 2j+1}
%\end{array}
%\end{equation}
%One can conveniently substract and add back the total spin representation of $(S_{2j}^{+})^{2}-(S_{2j+1}^{+})^{2}$ to this expression to obtain 
one finds
\begin{align}
\bigg[ & P_{2j, 2j+1}, \; (S_{2j}^{+})^{2}-(S_{2j+1}^{+})^{2} \bigg]=(S_{2j}^{+})^{2}-(S_{2j+1}^{+})^{2} \nonumber \\
&-4 \left[ \ket{T_{1,1}}\bra{T_{2,-1}}+\sqrt{2}\ket{ T_{1,0}}\bra{ T_{2,-2}} \right]_{2j, 2j+1}
\end{align}

Identifying $-4(\ket{T_{1,1}}\bra{T_{2,-1}}+\sqrt{2}\ket{ T_{1,0}}\bra{ T_{2,-2}})_{2j, 2j+1}$ with $K_{2j}$, we find
\begin{equation}
\left[O_1, Q^{\dagger}\right] =  Q^{\dagger} + \sum_{j=1}^{L/2}K_{2j}.  
\end{equation}
It thus remains to show that the action of $\sum_{j = 1}^{L/2} K_{2j}$ on $\ket{S_{2n}}$ yields zero. 
This is clearly true for $\ket{\psi_A}$ as it does not contain any spin-2 components on any pairs of sites. By looking at the terms that appear in $(S_{2j}^{+})^{2}-(S_{2j+1}^{+})^{2}$, one can see that if the initial state is a state composed exclusively of total spin-1 and total spin-0 states on any adjacent pairs of spins i.e the ground state, then repeated application of $Q^{\dagger}$ can only produce a state that locally has weight on $\ket{T_{2,1}}$, $\ket{T_{2,2}}$ and $\ket{T_{1,m}}$ states, from which we directly see that $K_{2j}\ket{\psi_l} = 0$. For each state $\ket{\psi_l}$, one can generate $4n$ additional spin rotated states due to the $\text{SU}(2)$ symmetry of the AKLT model.
\\\\
A similar calculation shows that
\begin{equation}
\left[O_2, Q^{\dagger}\right] =  Q^{\dagger} - \sum_{j=1}^{L/2}K_{2j-1},  
\end{equation}
where a similar argument can be used to conclude that only the $Q^\dagger$ piece yields a non-vanishing contribution when acting on $\ket{\psi_l}$. 

Now, since the ground state is a common eigenstate of both $O_1$ and $O_2$, and a spectrum-generating algebra is satisfied for both $O_1$ and $O_2$ individually (with the same base state $\ket{\psi_A}$), one has that all the states $\ket{\psi_l}$ are common eigenstates of both $O_1$ and $O_2$. 
%Note that the above properties imply that the time evolution of any superposition of the scar eigenstates is given as a product of time evolution by $O_1$ and $O_2$, two trivially integrable Hamiltonians, which guarantees revival at some time $t$. Specifically,  
%\begin{equation}
%    e^{-iH_{\text{AKLT}}t}\sum_{l = 0}^{L-1}c_l\ket{\psi_l} = e^{-iO_1t}e^{-iO_2t}\sum_{l = 0}^{L-1}c_l\ket{\psi_l},
%\end{equation} 
%for arbitrary coefficient $c_l$, and revivals occur with period $t = 2 \pi$ given the fact that the eigenvalues of $O_1$ and $O_2$ are integers. %This shows that within the scar subspace, the dynamics is fully determined by the the cellular automaton $\mathcal{T}(t) = e^{-iAt}e^{-iBt}$
\section{Relation between the Spin-1 XY model and a related virtual spin-1/2 chain}\label{app:MultiSiteTowerProof}
In this section, the mapping between the spin-1 XY model to the spin-1/2 spin chain of Sec. \ref{sec:TwoSiteQPSpinXY} and its relation to the type II rules are discussed.
It will be useful to make use of the local map introduced in Ref.~\cite{CHATTOPADHYAY2020_SpinOneHalfTwoSiteTowerScars} given by
\begin{equation}
\begin{aligned}
    M_{2j-1,2j} = \ket{1}_j\bra{\uparrow\uparrow}_{2j-1,2j}+\\\ket{0}_j(\bra{\uparrow \downarrow} + \bra{\downarrow \uparrow })_{2j-1,2j} + \ket{-1}_j\bra{\downarrow\downarrow}_{2j-1,2j},
\end{aligned}
\end{equation}
where it is understood that we are associating any physical site $j$ of the spin-$1$ chain to two consecutive virtual sites $2j-1,2j$ of a spin-$1/2$ chain. The above local map converts virtual degrees of freedom to physical degrees of freedom.
Using the definition of $M_{2j-1,2j}$, one may then construct the map
\begin{equation}
    M = \otimes_{j = 1}^LM_{2j-1,2j}
\end{equation} which connects an entire virtual spin-$1/2$ chain of size $2L$ to a physical spin-1 chain of size $L$.
We now define an inverse local map given by 
\begin{equation}
\begin{aligned}
    K_{2j-1,2j} = \ket{\uparrow\uparrow}_{2j-1,2j}\bra{1}_j+\\\frac{(\ket{\uparrow \downarrow} + \ket{\downarrow \uparrow })_{2j-1,2j}}{2}\bra{0}_j + \ket{\downarrow\downarrow}_{2j-1,2j}\bra{-1}_j
\end{aligned}
\end{equation}
from which we may construct the map 
\begin{equation}
    K = \otimes_{j = 1}^LK_{2j-1,2j}.
\end{equation}
Note that 
\begin{equation}
    K_{2j-1,2j}M_{2j-1,2j} = \mathcal{P}_{2j-1,2j}
\end{equation}
where 
\begin{equation}
\begin{aligned}
  \mathcal{P}_{2j-1,2j} =  
  \ket{\uparrow\uparrow}_{2j-1,2j}\bra{\uparrow\uparrow}_{2j-1,2j}+\\\frac{(\ket{\uparrow \downarrow} + \ket{\downarrow \uparrow })_{2j-1,2j}(\bra{\uparrow \downarrow} + \bra{\downarrow \uparrow })_{2j-1,2j}}{2}\\ + \ket{\downarrow\downarrow}_{2j-1,2j}\bra{\downarrow\downarrow}_{2j-1,2j},
\end{aligned}
\end{equation} which is just the projector onto the triplet subspace of the two adjacent spin-$1/2$ degrees of freedom on the virtual sites $2j-1, 2j$. Furthermore, we define the global projector
\begin{equation}
\mathcal{P} = \otimes_{j = 1}^L\mathcal{P}_{2j-1,2j}.
\end{equation}
Our goal is now to find a spin-1/2 representation of the spin-1 model for which the type II rules apply.
A natural candidate for the appropriate spin-1/2 representation is the operator $KH_{\text{XY}}'M$ with 
\begin{equation}
\begin{aligned}
H_{\text{XY}}' &=\sum_j\left(S_j^x S_{j+1}^x+S_j^y S_{j+1}^y\right)+\epsilon V+h \sum_j S_j^z.
\end{aligned}
\end{equation}
First, we show that $KH_{\text{XY}}'M$ can be rewritten as $\mathcal{P}H_{\text{XY}}^{1/2}$ where $H_{\text{XY}}^{1/2}$ is an effective model acting on the virtual spin-$1/2$ degrees of freedom. By defining
\begin{align}
(H_{\text{S}}')_{j,j+1} &\equiv \left(S_j^x S_{j+1}^x+S_j^y S_{j+1}^y\right) + \nonumber\\ &\epsilon\left(\left(S_j^+\right)^2\left(S_{j+1}^-\right)^2 + \text{h.c}\right),\nonumber\\\nonumber\\
(H_z')_{j,j+1} &\equiv hS_j^z + hS_{j+1}^z,
\end{align}
one then has that 
\begin{align}\label{eq:Spin1/2Projection}
    &KH_{\text{XY}}'M = \nonumber \\ &\sum_{j = 1}^{L}\left(\otimes_{p = 1, p\neq \{j,j+1\}}^{L}\mathcal{P}_{2p-1,2p}\right) K_{2j-1,2j}K_{2j+1,2j+2}\nonumber\\ &\times
    (H_{\text{S}}' + H_z')_{j,j+1}M_{2j-1,2j}M_{2j+1,2j+2}.
\end{align}
Further defining 
\begin{align}
   &(H_{\text{S}}^{1/2})_{j,j+1} \equiv  K_{2j-1,2j}K_{2j+1,2j+2}(H_{\text{S}}')_{j,j+1}\nonumber\\ &\times M_{2j-1,2j}M_{2j+1,2j+2}\nonumber\\\nonumber\\
    &(H_{z}^{1/2})_{j,j+1} \equiv  K_{2j-1,2j}K_{2j+1,2j+2}( H_z')_{j,j+1}\nonumber\\ &\times M_{2j-1,2j}M_{2j+1,2j+2}
\end{align}
and noting that $P_{2j-1,2j}K_{2j-1,2j} = K_{2j-1,2j}$, one may write Eq. (\ref{eq:Spin1/2Projection}) as 
\begin{equation}
    KH_{\text{XY}}'M  = \mathcal{P}H_{\text{XY}}^{1/2}
\end{equation}
with
\begin{equation}
\sum_{j=1}^L(H_{\text{S}}^{1/2} + H_{z}^{1/2})_{j,j+1} = H_{\text{XY}}^{1/2}.
\end{equation}
Interestingly, we note that for any eigenstates $\ket{\alpha}$ of the operator $H_{\text{XY}}^{1/2}$, one has that the state $M\ket{\alpha}$ is an eigenstate of the operator $H_{\text{XY}}'$ with the same eigenvalue. To show this, let us assume that the state $\ket{\alpha}$ is an eigenstate of $H_{\text{XY}}^{1/2}$ with eigenvalue $\alpha$. Then, one has that 
\begin{equation}
\begin{aligned}
    KH_{\text{XY}}M\ket{\alpha} = \\\mathcal{P}H_{\text{XY}}^{1/2}\ket{\alpha} = \mathcal{P}\alpha \ket{\alpha} = K\alpha M\ket{\alpha},
\end{aligned}
\end{equation}
which leads to the relation 
\begin{equation}
    K(H_{\text{XY}}' - \alpha)M\ket{\alpha} = 0
\end{equation}
Now, let us assume that $(H_{\text{XY}}' - \alpha)M\ket{\alpha}$ is non-zero, but that $K(H_{\text{XY}}' - \alpha)M\ket{\alpha}$ vanishes. That directly leads to a contradiction, since provided we write down the state $(H_{\text{XY}}' - \alpha)M\ket{\alpha}$ as a sum of computational basis states 
\begin{equation}
\begin{aligned}
    (H_{\text{XY}}' - \alpha)M\ket{\alpha} = c_1\ket{10(-1)011...} +\\ c_2\ket{110(-1)(-1)0...} + ...,
\end{aligned}
\end{equation}
then the operator $K$ preserves orthogonality of the spin-$1$ computational basis states ($K$ maps orthogonal spin-$1$ states to orthogonal spin-$1/2$ states), and merely re-scales all such terms. Thus, it must be the case that $(H_{\text{XY}}' - \alpha)M\ket{\alpha} = 0$ which shows that the states $M\ket{\alpha}$ are eigenstates of $H_{\text{XY}}'$ with an eigenvalue of $\alpha$. 
\\\\We may also show that provided we found all eigenstates of $H_{\text{XY}}^{1/2}$, then we would have automatically found the entire spectrum of $H_{\text{XY}}'$. To show this, let us prove that provided we have an eigenstate $\ket{\beta}$ of $H_{\text{XY}}'$ with eigenvalues $\beta$, then it must be the case that $K\ket{\beta}$ is an eigenstate of $H_{\text{XY}}^{1/2}$ with the same eigenvalue. First, note that $MK$ is just the identity operator onto the full spin-$1$ Hilbert space. Thus, we have that 
\begin{align}
 \mathcal{P}H_{\text{XY}}^{1/2}K\ket{\beta} &= KH_{\text{XY}}'MK\ket{\beta} = \nonumber\\
   K\beta\ket{\beta} &= \mathcal{P}K\beta\ket{\beta},
\end{align}
where we have used the fact that $\mathcal{P}K\ket{\beta} = K\ket{\beta}$. We have thus obtained
\begin{equation}
    \mathcal{P}(H_{\text{XY}}^{1/2} - \beta)K\ket{\beta} = 0.
\end{equation}
The states $(H_{\text{XY}}^{1/2} - \beta)K\ket{\beta}$ have no components in the null-space of the projector $\mathcal{P}$. By an argument similar to the above, it must then be the case that the state $K\ket{\beta}$ is an eigenstate of $H_{\text{XY}}^{1/2}$ with eigenvalue $\beta$. These considerations mean that we can study the spin-1 model by examining the eigenstates of the model $H_{\text{XY}}^{1/2}$ instead. Indeed, from the above facts, one may show that for any states $\ket{\psi}$ defined onto the physical spin-1 chain, one has that
\begin{equation}
    e^{-iH_{\text{XY}}'t}\ket{\psi} = Me^{-iH_{\text{XY}}^{1/2}t}K\ket{\psi}.
\end{equation}

To make a connection with the type II rules, let us now rewrite $\mathcal{P}H_{\text{XY}}^{1/2}$ using the operator
\begin{widetext}
\begin{align}\label{eq:spin1/2Rep}
\tilde{H}^{1/2}_{\text{XY}} &= \sum_{j=1}^Lo_{1,j}  + h\sum_{j=1}^{2L}o_{2,j}\nonumber\\
o_{1,j} &= (H_{\text{S}}^{1/2})_{j,j+1}-\mathcal{P}_{2j-1,2j}\mathcal{P}_{2j+1,2j+2}(s^+_{2j-1}s^-_{2j+1} + s^+_{2j}s^-_{2j+2} + \text{h.c}) + (s^+_{2j-2}s^-_{2j} + s^+_{2j-1}s^-_{2j+1} + \text{h.c})\nonumber\\
o_{2,j} &= s_j^z.
\end{align}
\end{widetext}
Using $\mathcal{P}_{2j-1,2j}(s_{2j-1}^z + s_{2j}^z) = K_{2j-1,2j}S_j^zM_{2j,1,2j}$, it is clear that $\mathcal{P}\tilde{H}^{1/2}_{\text{XY}}  = \mathcal{P}H_{\text{XY}}^{1/2}$, and thus all the arguments made above for $H_{\text{XY}}^{1/2}$ apply to $\tilde{H}^{1/2}_{\text{XY}} $ as well (note however that $\tilde{H}^{1/2}_{\text{XY}} \neq H_{\text{XY}}^{1/2} $). We are now in a position to apply the type II rules. Here, the operators $o_{1,j}$ play the role of the $h_j$ and locally vanish when acting onto the scar subspace, whilst the $o_{2,j}$ act as the $v_j$ terms. The type II rules then read as 
\begin{align}
    o_{1,j}\left(\sum_{p\in D(j)}o_{2,p}\right)^ n(Q^{\dagger}_{D(j)})^m\ket{\phi_A} &= 0
\end{align}
for arbitrary integers $m,n \ge 0$, where $\ket{\phi_A} = \ket{\downarrow,\downarrow,....}$, $D(j) = \{2j-2,2j-1,...,2j+1\}$ and \begin{align}
 Q^{\dagger}_{D(j)} &= \nonumber \\
 (-1)^{j} &\left(-s_{2j-2}^+s_{2j-1}^+ + s_{2j}^+s_{2j+1}^+\right. \left.- s_{2j+2}^+s_{2j+3}^+\right).
 \end{align} Starting from the result 

\begin{equation}
\begin{aligned}
  (Q_{D(j)}^{\dagger})^{m}\ket{\phi_A}_{D(j)}\\ =  \left\{
    \begin{array}{ll}
        \ket{\downarrow,\downarrow,\downarrow,\downarrow,\downarrow,\downarrow}_{D(j)}\quad m = 0\\\\
         (-1)^{j+1}(\ket{\uparrow,\uparrow,\downarrow,\downarrow,\downarrow,\downarrow} - \ket{\downarrow,\downarrow,\uparrow,\uparrow,\downarrow,\downarrow}  + \\ \ket{\downarrow,\downarrow,\downarrow,\downarrow,\uparrow,\uparrow})_{D(j)}  \quad m = 1 \\\\
         2(-\ket{\uparrow,\uparrow,\uparrow,\uparrow,\downarrow,\downarrow} - \ket{\downarrow,\downarrow,\uparrow,\uparrow,\uparrow,\uparrow}  + \\ \ket{\uparrow,\uparrow,\downarrow,\downarrow,\uparrow,\uparrow})_{D(j)}  \quad m = 2 \\\\
         -6(-1)^{j+1}\ket{\uparrow,\uparrow,\uparrow,\uparrow,\uparrow,\uparrow}_{D(j)} m = 3\\\\
         0 \quad m > 3,
    \end{array}
\right.
\end{aligned}
\end{equation}
a direct calculation can be used to show that 
\begin{equation}
\begin{aligned}
   \left(\sum_{p\in D(j)}o_{2,p}\right)^ n(Q^{\dagger}_{D(j)})^{m}\ket{\phi_A}_{D(j)}
   \\ =  h(2m - 3)(Q_{D(j)}^{\dagger})^{m}\ket{\phi_A}_{D(j)}
\end{aligned}
\end{equation}
and 
\begin{equation}
\begin{aligned}
   o_{1,j}(Q^{\dagger}_{D(j)})^{m}\ket{\phi_A}_{D(j)} = 0
\end{aligned}
\end{equation}
which directly implies that the type II rules are satisfied.

\section{Conditions for quantum scars in the generalized Hubbard model}\label{app:HubbardConditions}
In this section, we derive the set of rules presented in Eq. (\ref{eq:Hubbard0energyconditionText}). First, recall that in the generalized Hubbard model, one has that 
\begin{align}
o_{2,\boldsymbol{r},\boldsymbol{r}'} &=  \sum_{\vs{d} = \vs{r}, \vs{r}'}\left(\sum_\sigma -\mu_{\boldsymbol{d},\sigma}\hat{n}_{\boldsymbol{d},\sigma} + U_{\boldsymbol{d}} \hat{n}_{\boldsymbol{d}, \uparrow} \hat{n}_{\boldsymbol{d}, \downarrow}  \right), \nonumber\\ 
%&-\mu_{\boldsymbol{r}',\sigma}\hat{n}_{\boldsymbol{r}',\sigma} +  U_{\boldsymbol{r}'} \hat{n}_{\boldsymbol{r}', \uparrow} \hat{n}_{\boldsymbol{r}', \downarrow} \right)\nonumber\\
o_{1,\boldsymbol{r},\boldsymbol{r}'} &=-\sum_{\sigma, \sigma^{\prime}} \left(t_{\boldsymbol{r}, \boldsymbol{r}^{\prime}}^{\sigma, \sigma^{\prime}} c_{\boldsymbol{r}, \sigma}^{\dagger} c_{\boldsymbol{r}^{\prime}, \sigma^{\prime}}+t_{\boldsymbol{r}^{\prime}, \boldsymbol{r}}^{\sigma^{\prime}, \sigma} c_{\boldsymbol{r}^{\prime}, \sigma^{\prime}}^{\dagger} c_{\boldsymbol{r}, \sigma}\right).\nonumber\\
\end{align}
Thus, 
\begin{align}
O_2 &= -\sum_{\boldsymbol{r},\sigma}\mu_{\boldsymbol{r},\sigma}\hat{n}_{\boldsymbol{r},\sigma} + \sum_{\boldsymbol{r}} U_{\boldsymbol{r}} \hat{n}_{\boldsymbol{r}, \uparrow} \hat{n}_{\boldsymbol{r}, \downarrow}\nonumber\\
O_1 &= \sum_{\left \langle \boldsymbol{r},\boldsymbol{r}' \right \rangle}o_{1,\boldsymbol{r},\boldsymbol{r}'}\nonumber\\\nonumber\\
H_{\text{Hubbard}} &= O_1 + O_2.
\end{align}
The type II rules read
\begin{equation}\label{eq:AppCommConditionHubbard}
\begin{aligned}
    \left[o_{2,\boldsymbol{r},\boldsymbol{r}'} ^m,o_{1,\boldsymbol{r},\boldsymbol{r}'}\right](Q^{\dagger}_{\boldsymbol{r}} + Q^{\dagger}_{\boldsymbol{r'}})^n\ket{\psi_A} = 0
\end{aligned}
\end{equation}
with 
\begin{equation}
\begin{aligned}
Q^{\dagger}_{\vs{r}} = q_{\vs{r},\uparrow}c^{\dagger}_{\vs{r},\uparrow} + q_{\vs{r},\downarrow}c^{\dagger}_{\vs{r},\downarrow} + q_{\vs{r},\uparrow\downarrow}c^{\dagger}_{\vs{r},\uparrow}c^{\dagger}_{\vs{r},\downarrow}
\end{aligned}
\end{equation}
for arbitrary integers $n,m\ge 0$.
First, we focus on the case where $q_{\vs{r},\uparrow} = q_{\vs{r},\downarrow} = 0$. When $m = 0$, one has to satisfy $o_{1,\boldsymbol{r},\boldsymbol{r}'}(Q^{\dagger}_{\boldsymbol{r}} + Q^{\dagger}_{\boldsymbol{r'}})^n\ket{\psi_A} = 0$. Clearly, for $n = 0,2$, this is satisfied for arbitrary $t_{\boldsymbol{r},\boldsymbol{r'}}^{\sigma,\sigma'}$, $q_{\boldsymbol{r},\uparrow\downarrow}$. For $n =  1$, we obtain
\begin{equation}
\begin{aligned}
o_{1,\boldsymbol{r},\boldsymbol{r'}}(Q^{\dagger}_{\boldsymbol{r}} + Q^{\dagger}_{\boldsymbol{r'}})\ket{0,0,0,0}_{\boldsymbol{r},\boldsymbol{r}'} = \\
(q_{\boldsymbol{r},\uparrow\downarrow}t_{\boldsymbol{r}',\boldsymbol{r}}^{\uparrow,\uparrow} + q_{\boldsymbol{r'},\uparrow\downarrow}t_{\boldsymbol{r},\boldsymbol{r}'}^{\downarrow,\downarrow})\ket{0,\downarrow,\uparrow,0}_{\boldsymbol{r},\boldsymbol{r}'} + \\
(q_{\boldsymbol{r'},\uparrow\downarrow}t_{\boldsymbol{r},\boldsymbol{r}'}^{\uparrow,\uparrow} + q_{\boldsymbol{r},\uparrow\downarrow}t_{\boldsymbol{r}',\boldsymbol{r}}^{\downarrow,\downarrow})\ket{\uparrow,0,0,\downarrow}_{\boldsymbol{r},\boldsymbol{r}'} + \\
(q_{\boldsymbol{r},\uparrow\downarrow}t_{\boldsymbol{r}',\boldsymbol{r}}^{\uparrow,\downarrow} - q_{\boldsymbol{r'},\uparrow\downarrow}t_{\boldsymbol{r},\boldsymbol{r}'}^{\uparrow,\downarrow})\ket{\uparrow,0,\uparrow,0}_{\boldsymbol{r},\boldsymbol{r}'} + \\
(q_{\boldsymbol{r'},\uparrow\downarrow}t_{\boldsymbol{r},\boldsymbol{r}'}^{\downarrow,\uparrow} - q_{\boldsymbol{r},\uparrow\downarrow}t_{\boldsymbol{r}',\boldsymbol{r}}^{\downarrow,\uparrow})\ket{0,\downarrow,0,\downarrow}_{\boldsymbol{r},\boldsymbol{r}'},
\end{aligned}
\end{equation}
where we have made use of the notation \begin{equation}
\begin{aligned}
(c_{\boldsymbol{r},\uparrow}^{\dagger})^{n_1}(c_{\boldsymbol{r},\downarrow}^{\dagger})^{n_2}(c_{\boldsymbol{r}',\uparrow}^{\dagger})^{n_3}(c_{\boldsymbol{r}',\downarrow}^{\dagger})^{n_4}\ket{0,0,0,0}_{\boldsymbol{r},\boldsymbol{r}'} =\\ \ket{f_{\uparrow}(n_1),f_{\downarrow}(n_2),f_{\uparrow}(n_3),f_{\downarrow}(n_4)}\\\\
f_{\sigma}(n) = \sigma \text{ if } n = 1, 0 \text{ otherwise}
\end{aligned}
\end{equation}
and $\ket{0,0,0,0}_{\boldsymbol{r},\boldsymbol{r}'}$ refers to the vacuum state on the two sites $\boldsymbol{r},\boldsymbol{r}'$. This implies that \begin{equation}\label{eq:appHubbardFullStates}
\begin{aligned}
q_{\boldsymbol{r},\uparrow\downarrow}t_{\boldsymbol{r}',\boldsymbol{r}}^{\uparrow,\uparrow} + q_{\boldsymbol{r}',\uparrow\downarrow}t_{\boldsymbol{r},\boldsymbol{r}'}^{\downarrow,\downarrow}= 0\\
q_{\boldsymbol{r}',\uparrow\downarrow}t_{\boldsymbol{r},\boldsymbol{r}'}^{\uparrow,\uparrow} + q_{\boldsymbol{r},\uparrow\downarrow}t_{\boldsymbol{r}',\boldsymbol{r}}^{\downarrow,\downarrow} = 0\\
q_{\boldsymbol{r},\uparrow\downarrow}t_{\boldsymbol{r}',\boldsymbol{r}}^{\uparrow,\downarrow} - q_{\boldsymbol{r}',\uparrow\downarrow}t_{\boldsymbol{r},\boldsymbol{r}'}^{\uparrow,\downarrow}=0 \\
q_{\boldsymbol{r}',\uparrow\downarrow}t_{\boldsymbol{r},\boldsymbol{r}'}^{\downarrow,\uparrow} - q_{\boldsymbol{r},\uparrow\downarrow}t_{\boldsymbol{r}',\boldsymbol{r}}^{\downarrow,\uparrow} = 0
\end{aligned}
\end{equation}
which corresponds to Eq. (\ref{eq:Hubbard0energyconditionText}) which agrees with the conditions originally derived in Ref. \cite{MOUDGALYA2020_GeneralizedHubbard}. Now, whenever $m\neq 0, n = 1$, we have the condition
\begin{equation} o_{1,\boldsymbol{r},\boldsymbol{r}'}o^m_{2,\boldsymbol{r},\boldsymbol{r}'}(Q^{\dagger}_{\boldsymbol{r}} + Q^{\dagger}_{\boldsymbol{r'}})\ket{\psi_A} = 0
\end{equation}
which is the same as Eq. (\ref{eq:appHubbardFullStates}), but with the substitution $q_{\boldsymbol{r},\uparrow\downarrow} \rightarrow q_{\boldsymbol{r},\uparrow\downarrow}(U_{\boldsymbol{r}} - \mu_{\boldsymbol{r},\downarrow} - \mu_{\boldsymbol{r},\uparrow},)^m$ and $q_{\boldsymbol{r}',\uparrow\downarrow} \rightarrow q_{\boldsymbol{r}',\uparrow\downarrow,\uparrow\downarrow}(U_{\boldsymbol{r}'} - \mu_{\boldsymbol{r}',\downarrow} - \mu_{\boldsymbol{r}',\uparrow})^m$. Combined with Eq. (\ref{eq:appHubbardFullStates}), the only way this can vanish is if $(U_{\boldsymbol{r}'} - \mu_{\boldsymbol{r}',\downarrow} - \mu_{\boldsymbol{r}',\uparrow}) = \mathcal{E}$ for some arbitrary constant $\mathcal{E}$.
\subsubsection{Rules for alternate eta pairing operator}
A similar calculation to the above, but now considering the more general local quasiparticle creation operator \begin{equation}
\begin{aligned}
Q^{\dagger}_{\vs{r}} = q_{\vs{r},\uparrow}c^{\dagger}_{\vs{r},\uparrow} + q_{\vs{r},\downarrow}c^{\dagger}_{\vs{r},\downarrow} + q_{\vs{r},\uparrow\downarrow}c^{\dagger}_{\vs{r},\uparrow}c^{\dagger}_{\vs{r},\downarrow}
\end{aligned}
\end{equation}
yields the set of equations
\begin{equation}
\begin{aligned}
q_{\vs{r},\uparrow}t^{\uparrow,\uparrow}_{\vs{r'},\vs{r}} + q_{\vs{r},\downarrow}t^{\uparrow,\downarrow}_{\vs{r'},\vs{r}}  = 0
\\
q_{\vs{r},\downarrow}t^{\downarrow,\downarrow}_{\vs{r'},\vs{r}} + q_{\vs{r},\uparrow}t^{\downarrow,\uparrow}_{\vs{r'},\vs{r}}  = 0
\\
q_{\vs{r'},\uparrow}t^{\uparrow,\uparrow}_{\vs{r},\vs{r'}} + q_{\vs{r'},\downarrow}t^{\uparrow,\downarrow}_{\vs{r},\vs{r'}}  = 0
\\
q_{\vs{r'},\downarrow}t^{\downarrow,\downarrow}_{\vs{r},\vs{r'}} + q_{\vs{r'},\uparrow}t^{\downarrow,\uparrow}_{\vs{r},\vs{r'}}  = 0
\\\\
q_{\vs{r'},\uparrow,\downarrow}q_{\vs{r},\uparrow}t^{\downarrow,\uparrow}_{\vs{r},\vs{r'}} - q_{\vs{r'},\uparrow,\downarrow}q_{\vs{r},\downarrow}t^{\uparrow,\uparrow}_{\vs{r},\vs{r'}}  = 0
\\
q_{\vs{r'},\uparrow,\downarrow}q_{\vs{r},\downarrow}t^{\uparrow,\downarrow}_{\vs{r},\vs{r'}} - q_{\vs{r'},\uparrow,\downarrow}q_{\vs{r},\uparrow}t^{\downarrow,\downarrow}_{\vs{r},\vs{r'}}  = 0
\\
q_{\vs{r},\uparrow,\downarrow}q_{\vs{r'},\uparrow}t^{\downarrow,\uparrow}_{\vs{r'},\vs{r}} - q_{\vs{r},\uparrow,\downarrow}q_{\vs{r'},\downarrow}t^{\uparrow,\uparrow}_{\vs{r'},\vs{r}}  = 0
\\
q_{\vs{r},\uparrow,\downarrow}q_{\vs{r'},\downarrow}t^{\uparrow,\downarrow}_{\vs{r'},\vs{r}} - q_{\vs{r},\uparrow,\downarrow}q_{\vs{r'},\uparrow}t^{\downarrow,\downarrow}_{\vs{r'},\vs{r}}  = 0
\end{aligned}
\end{equation}
together with 
\begin{equation}
\begin{aligned}
q_{\boldsymbol{r},\uparrow\downarrow}t_{\boldsymbol{r}',\boldsymbol{r}}^{\uparrow,\uparrow} + q_{\boldsymbol{r}',\uparrow\downarrow}t_{\boldsymbol{r},\boldsymbol{r}'}^{\downarrow,\downarrow}= 0\\
q_{\boldsymbol{r}',\uparrow\downarrow}t_{\boldsymbol{r},\boldsymbol{r}'}^{\uparrow,\uparrow} + q_{\boldsymbol{r},\uparrow\downarrow}t_{\boldsymbol{r}',\boldsymbol{r}}^{\downarrow,\downarrow} = 0\\
q_{\boldsymbol{r},\uparrow\downarrow}t_{\boldsymbol{r}',\boldsymbol{r}}^{\uparrow,\downarrow} - q_{\boldsymbol{r}',\uparrow\downarrow}t_{\boldsymbol{r},\boldsymbol{r}'}^{\uparrow,\downarrow}=0 \\
q_{\boldsymbol{r}',\uparrow\downarrow}t_{\boldsymbol{r},\boldsymbol{r}'}^{\downarrow,\uparrow} - q_{\boldsymbol{r},\uparrow\downarrow}t_{\boldsymbol{r}',\boldsymbol{r}}^{\downarrow,\uparrow} = 0.
\end{aligned}
\end{equation}
We further have the conditions $(U_{\boldsymbol{r}} - \mu_{\boldsymbol{r},\uparrow} - \mu_{\boldsymbol{r},\downarrow}) = \mathcal{E}$, $\mu_{\boldsymbol{r},\uparrow} = \mu_{\boldsymbol{r},\downarrow}$. Provided we assume that the hopping terms are Hermitian, that is provided we assume that 
$t_{\boldsymbol{r},\boldsymbol{r'}}^{\sigma,\sigma'} = (t_{\boldsymbol{r'},\boldsymbol{r}}^{\sigma',\sigma})^*$, then one can show that this system of equations does not admit non-trivial solutions where both $q_{\boldsymbol{r},\sigma}$ and $q_{\boldsymbol{r},\uparrow\downarrow}$ are simultaneously non-zero (there are no solutions that combine fermionic and bosonic excitations).
However, the above system of equations admits many solutions if we do not require hermiticity. For instance, it easy to show that 
\begin{align}
q_{\boldsymbol{r},\downarrow} = q_{\boldsymbol{r},\uparrow} = q_{\boldsymbol{r},\uparrow\downarrow} = 1
\nonumber\\
t_{\boldsymbol{r},\boldsymbol{r'}}^{\downarrow,\downarrow} =  1, t_{\boldsymbol{r},\boldsymbol{r'}}^{\uparrow,\uparrow}= -1,t_{\boldsymbol{r},\boldsymbol{r'}}^{\uparrow,\downarrow}= 1,t_{\boldsymbol{r},\boldsymbol{r'}}^{\downarrow,\uparrow}= -1\nonumber\\
t_{\boldsymbol{r'},\boldsymbol{r}}^{\downarrow,\downarrow} =  1, t_{\boldsymbol{r'},\boldsymbol{r}}^{\uparrow,\uparrow}= -1,t_{\boldsymbol{r'},\boldsymbol{r}}^{\uparrow,\downarrow}= 1,t_{\boldsymbol{r'},\boldsymbol{r}}^{\downarrow,\uparrow}= -1
\end{align}
is a solution to the above set of equations.

\section{Computational basis representation of the $\ket{T_{l,m}}$ states.}\label{app:CompBasisRepOfTotSpin}
The computational basis representation of the $\ket{T_{l,m}}$ states is given explicitly by
\begin{equation}
\begin{array}{c}
\left|T_{2,-2}\right\rangle=|-1,-1\rangle, \quad\left|T_{2,-1}\right\rangle=\frac{1}{\sqrt{2}}(|0,-1\rangle+|-1,0\rangle), \\
\left|T_{2,0}\right\rangle=\frac{1}{\sqrt{6}}(|1,-1\rangle+2|0,0\rangle+|-1,1\rangle), \\
\left|T_{2,1}\right\rangle=\frac{1}{\sqrt{2}}(|1,0\rangle+|0,1\rangle), \quad\left|T_{2,2}\right\rangle=|1,1\rangle, \\
\left|T_{1,-1}\right\rangle=\frac{1}{\sqrt{2}}(|0,-1\rangle-|-1,0\rangle), \\
\left|T_{1,0}\right\rangle=\frac{1}{\sqrt{2}}(|1,-1\rangle-|-1,1\rangle), \\
\left|T_{1,1}\right\rangle=\frac{1}{\sqrt{2}}(|1,0\rangle-|0,1\rangle), \\
\left|T_{0,0}\right\rangle=\frac{1}{\sqrt{3}}(|1,-1\rangle-|0,0\rangle+|-1,1\rangle) .
\end{array}
\end{equation}
\section{All quantum scars viewed as arising from the type I rules}\label{app:MPSQuasiparticlesModels}
All the quantum scars introduced in this work can be understood by making use of type I rules. 
\subsubsection{Hubbard model}
In the generalized Hubbard model, the local operators $o_{1,\boldsymbol{r},\boldsymbol{r'}}$ all vanish individually onto the scar subspace and can be viewed as individual partitions. The operator $O_2 = -\sum_{\boldsymbol{r},\sigma}\mu_{\boldsymbol{r},\sigma}\hat{n}_{\boldsymbol{r},\sigma} + \sum_{\boldsymbol{r}} U_{\boldsymbol{r}} \hat{n}_{\boldsymbol{r}, \uparrow} \hat{n}_{\boldsymbol{r}, \downarrow} $ is also an eigenoperator. Indeed, provided Eqs. (\ref{eq:Hubbard0energyconditionText}), (\ref{eq:LocalEnergyCondition}) are satisfied, we have that $\left(\sum_\sigma -\mu_{\boldsymbol{r},\sigma}\hat{n}_{\boldsymbol{r},\sigma} + U_{\boldsymbol{r}} \hat{n}_{\boldsymbol{r}, \uparrow} \hat{n}_{\boldsymbol{r}, \downarrow}  \right)(Q^{\dagger}_{\boldsymbol{r}})^n\ket{0,0}_{\boldsymbol{r}} = n\mathcal{E}(Q^{\dagger}_{\boldsymbol{r}})^n\ket{0,0}_{\boldsymbol{r}} $ with $Q_{\boldsymbol{r}}^{\dagger} = q_{\boldsymbol{r},\uparrow\downarrow}c_{\boldsymbol{r},\uparrow}^{\dagger}c_{\boldsymbol{r},\downarrow}^{\dagger}$ and the base state $\ket{0,0,...}$ containing no fermions. This shows that type I rules are satisfied. 
\subsubsection{Spin-1 XY model, single-site tower of quasiparticles}
Recall in this model the operators \begin{align}
o_{1,p,j} &= \frac{1}{2}\left(S_p^+S_{j}^- + S_p^-S_{j}^+\right) \nonumber \\
O_2 &= h\sum_{j}S_j^z + D\sum_j(S_j^z)^2 \nonumber \\
O_1 &= \sum_{\left \langle p,j \right \rangle}o_{1,p,j}\nonumber \\
Q^{\dagger} &= \sum_j\frac{e^{i\boldsymbol{r}_j\cdot\boldsymbol{\pi}}}{2}\left(S_j^+\right)^2.
\end{align}
In the main text, it was shown that $o_{1,p,j}$ annihilates the scar states. Thus, the operators $o_{1,p,j}$ can be understood as individual partitions. The operator $O_1$ is also an eigenoperator. Indeed, we have that $\left(hS_j^z + D(S_j^z)^2\right)\left(\frac{e^{i\boldsymbol{r}_j\cdot\boldsymbol{\pi}}}{2}(S_j^+)^2\right)^n\ket{-1}_j = (D + h(2n - 1))\left(\frac{e^{i\boldsymbol{r}_j\cdot\boldsymbol{\pi}}}{2}(S_j^+)^2\right)^n\ket{-1}_j$ where the base state is the fully polarized state $\ket{-1,-1,...}$, which shows that type I rules are satisfied.
\subsubsection{Spin-1 XY model, two-site tower of quasiparticles}
As shown in App. \ref{app:MultiSiteTowerProof}, by making use of an appropriate mapping of the spin-1 XY model to a related spin-1/2 model, one can find operators that satisfy the type II rules for the two-site tower of quasiparticles. In the spin-1/2 representation, the operators $o_{1,j}$ that are defined in Eq. (\ref{eq:spin1/2Rep}) all vanish individually onto the scar subspace and can thus be understood as individual partitions. In the spin-1/2 representation, the operator $O_2 = \sum_{j=1}^{2L}s_j^z$ is also an eigenoperator. The quasiparticle creation operator in this model is given by $Q^{\dagger} = \sum_{j=1}^L(-1)^j s_{2_j}^+s_{2j+1}^+$ and the base state is $\ket{\psi_A} = \ket{\downarrow,\downarrow,...}$. We then have that $\left(s_{2j}^z + s_{2j + 1}^z\right)\left((-1)^j s_{2_j}^+s_{2j+1}^+\right)^n\ket{\downarrow,\downarrow}_{2j,2j+1} = (2n-1)\left((-1)^j s_{2_j}^+s_{2j+1}^+\right)^n\ket{\downarrow,\downarrow}_{2j,2j+1} $ which shows that type I rules are satisfied.

\newpage
\newpage

\bibliography{VFBrokenUnitary}
\end{document}